\documentclass[12pt,english,floatfix,nofootinbib,superscriptaddress,aps,prd,preprint]{revtex4}
\usepackage[utf8]{inputenc}
\usepackage{float}
\usepackage{array}
\usepackage{lipsum}
\usepackage{graphicx}
\usepackage{amsmath,amsthm,amsfonts,amssymb}
\usepackage{graphicx}
\usepackage[english]{babel}
\usepackage{color}
\usepackage{tensor}
\usepackage{esint}
\usepackage[dvips]{epsfig}
\usepackage[dvips]{graphicx}
\usepackage{float}
\usepackage{units}
\usepackage{textcomp}
\usepackage{mathrsfs}
\usepackage{amsmath}
\usepackage[makeroom]{cancel}
\usepackage{amssymb}
\usepackage{amsbsy}
\usepackage{amsfonts}
\usepackage{amssymb,mathrsfs,xcolor}
\usepackage{esint}
\usepackage{braket}
\usepackage{array}
\usepackage{graphicx}

\usepackage{wasysym}
\usepackage{multirow}
\usepackage{wrapfig}
\usepackage{subfig}

\makeatletter

\makeatletter\usepackage{babel}

\usepackage{hyperref}
\hypersetup{
    colorlinks,
    citecolor=blue,
    filecolor=green,
    linkcolor=purple,
    urlcolor=red,
}

\usepackage{slashed}

\newcommand{\ie}{\begin{equation}}
\newcommand{\fe}{\end{equation}}
\newcommand{\se}{\begin{eqnarray}}
\newcommand{\ff}{\end{eqnarray}}

\begin{document}

\title{Thermal aspects of interacting quantum gases in Lorentz-violating scenarios}


\author{A. A. Ara\'{u}jo Filho}
\email{dilto@fisica.ufc.br}

\affiliation{Universidade Federal do Cear\'a (UFC), Departamento de F\'isica,\\ Campus do Pici,
Fortaleza -- CE, C.P. 6030, 60455-760 -- Brazil.}

\author{J. A. A. S. Reis}
\email{jalfieres@gmail.com}

\affiliation{Universidade Federal do Maranh\~{a}o (UFMA), Departamento de F\'{\i}sica, Campus Universit\'{a}rio do Bacanga, S\~{a}o Lu\'{\i}s -- MA, 65080-805, -- Brazil}

\affiliation{Universidade Estadual do Maranh\~{a}o (UEMA), Departamento de F\'{i}sica,\\ Cidade Universit\'{a}ria Paulo VI, S\~{a}o Lu\'{i}s -- MA, 65055-310, -- Brazil.}


\date{\today}

\begin{abstract}

In this work, we study the interaction of quantum gases in Lorentz-violating scenarios considering both boson and fermion sectors. In the latter case, we investigate the consequences of a system governed by scalar, vector, pseudovector and tensor operators. Besides, we examine the implications of $\left( \hat{k}_{a}\right) ^{\kappa }$ and $\left( \hat{k}_{c}\right) ^{\kappa \xi }$ operators for the boson case and the upper bounds are estimated. To do so, we regard the grand canonical ensemble seeking the so-called partition function, which suffices to provide analytically the calculations of interest, i.e., the mean particle number, the entropy, the mean total energy and the pressure. Furthermore, in low temperature regime, such quantities converge until reaching a similar behavior being in contrast with what is shown in high temperature regime, which brings out the differentiation of their effects. In addition, the particle number, the entropy and the energy exhibit an extensive characteristic even in the presence of Lorentz violation. Also, for the pseudovector and the tensor operators, we notice a remarkable feature due to the breaking process of spin degeneracy: the system turns out to have greater energy and particle number for the spin-down particles in comparison with the spin-up ones. Finaly, we propose two feasible applications to corroborate our results: \textit{phosphorene} and \textit{spin precession.}

\end{abstract}

\maketitle

\section{Introduction}

The well-known Lorentz symmetry is an equivalence of observation as a result of Special Relativity. This entails that the physical laws keep the same for all observers as long as the condition of inertial frames is ensured. Being the association of both rotational and boost symmetries, the Lorentz invariance is a fundamental feature when one regards the General Relativity and the Standard Model of particle physics. On the other hand, if one considers a violation of such condition, one will generally produce either directional or velocity dependences modifying, therefore, the dynamics of particles and waves \cite{STR1,STR2,STR3,STR4,STR5,STR6,STR7}.

Generically, any symmetry breaking process brings about unusual consequences, which can exhibit some fingerprints of new physical phenomena. Specially when the Lorentz symmetry is broken it leads to various particularities \cite{liberati2013,tasson2014,hees2016} being possibly feasible in quantum gravity \cite{rovelli2004}. Moreover, models involving closed-string theories \cite{New1,New2,New3,New4,New5}, loop quantum gravity \cite{New6,New7}, noncommutative spacetimes \cite{New8,New9}, spacetime foam models \cite{New10,New11}, (chiral) field theories defined on spacetimes with nontrivial topologies \cite{New12,New13,New14,New15}, Ho\v{r}ava-Lifshitz gravity \cite{New16} and cosmology \cite{sv1,sv2} are also based on the assumption that Lorentz invariance is no longer maintained. In this sense, in order to entirely characterize the effects due to Lorentz symmetry violation, one requires the obtainment of a reasonable theory that gives a dynamical characteristic to the system.

Thereby, there exists a widespread theoretical framework to support such approach, the Standard Model Extension (SME) \cite{k1,k2,k4,k5,k7}. In a general manner, it describes violations of CPT and Lorentz symmetries concerning both General Relativity and the Standard Model at attainable energies \cite{k2,k1,k7,kostelecky2004,mewes2019}. The Lorentz-violating operators are rather tensor terms coupled with physical fields that acquire a nonzero vacuum expectation value \cite{adailton2}. This latter feature gives rise to a violation of Lorentz symmetry when particle frames are taken into account and a preservation of its invariance when observer frames are assumed though \cite{kostelecky2001}. In this manner, this theoretical background gave the viability for many works involving the fermion sector \cite{f1,f2,f3,f4,f5,f6,f7} and electromagnetic CPT-odd \cite{e1,e2,e3,e4,e5,e6,e7,e8,e9} as well as CPT-even coefficients \cite{c1,c2,c3,c4}.

In addition, the connection established from Lorentz violation and theories including higher-dimensional operators has gained much attention through the last years. Within such approach, we can have operators with higher mass dimensions concerning higher-derivative terms for instance. The nonminimal version of SME has the advantage to hold indefinite numbers of such contributions \cite{kostelecky2009electro,kostelecky2012neutrinos,kostelecky2013fermions,kostelecky2016Penning,kostelecky2019gauge} in contrast with its minimal version. In this context, there are many works whose theoretical properties were studied involving nonrenormalizable operators \cite{d1,d2,d3,schreck2014,cuzinatto2011,casana2018,adailton,anacleto2018,borges2019}.

Although there are some works in the literature looking toward to investigate the thermodynamic aspects of distinct systems with Lorentz violation \cite{t1,t2,t3,t4,t5,t6,t7,t8,Colladay1t}, up to date, there is a lack of studies considering relativistic interacting quantum gases governed by higher-dimensional operators.
In such a way, we pioneer present a model in order to provide such derivation. Here, we focus on the following quantities of interest: particle number, entropy, mean energy and pressure. For doing so, we utilize the so-called grand canonical partition function and the grand canonical potential. With them, all the following evaluations could be carried out in high temperature regime.

In this work, we consider both fermion and boson sectors to proceed our calculations. Additionally, not being restricted to the case involving Lorentz symmetry breaking, the present model, which regards the treatment of the energy of an arbitrary quantum state, can lead to further analyses for different scenarios. It is worth mentioning that it is only possible to accomplish this as long as the operators, which modify the kinematics, be written in terms of momenta only.

This paper is organized as follows: in Secs. \ref{sec:fermionSector} and \ref{sec:scalarSector}, we provide a general discussion concerning the mathematical background highlighting the main aspects of fermion and boson sectors with Lorentz violation. In Sec. \ref{sec:ThermalModel}, we establish our model seeking to perform the derivation of the respective thermal quantities. In Secs. \ref{sec:Therm-state-quantities}, \ref{sec:fermionsInterecting} and \ref{sec:BosonsInterecting}, we focus on studying some particular operators for both fermions and bosons. In Sec. \ref{sec:Results}, we outline the principal features encountered in this manuscript and we analyse the magnitude of Lorentz-violating coefficients estimating the upper bounds for the bosonic case. In Sec. \ref{applications}, we propose two viable applications in order to corroborate our results: \textit{phosphorene} and \textit{spin precession}. In Sec. \ref{sec:conclusion}, we present our remarks and conclusions. Finally, we present a crucial prove considering the thermodynamic limit in Appendix \ref{sec:Therm-Limit} and Tables in Appendix \ref{sec:num-cal}.

\pagebreak

\section{SME Fermion Sector}\label{sec:fermionSector}

Initially, we introduce the Lagrange density for both minimal and nonminimal fermion sectors as being%
\begin{equation}
\mathcal{L}=\frac{1}{2}\bar{\psi}\left( \gamma ^{\mu }\mathrm{i}\partial
_{\mu }-m_{\psi }\mathrm{I}_{4}+\mathcal{\hat{Q}}\right) \psi+\mathrm{H.c.},
\label{eq:SME-L-fermion}
\end{equation}%
where $\psi $ is a Dirac spinor, $\bar{\psi}\equiv \psi ^{\dag }\gamma ^{0}$ is the conjugate Dirac spinor and $m_{\psi }$ is the fermion mass. Moreover, Lorentz-violating contributions are all contained in $\mathcal{\hat{Q}}$, which is a $4\times 4$ matrix $\in SL(2,\mathbb{C})$ lying in spinor space $\mathbb{C}^{2}$. In nonminimal SME, $\mathcal{%
\hat{Q}}$ can be regarded as an expansion in either derivative $\partial _{\mu }$ or momenta $p_{\mu }=\mathrm{i}\partial _{\mu }$ operators. Besides, in the spinor space $\mathcal{\hat{Q}}$ can be decomposed into the 16 Dirac bilinear terms, namely
\begin{equation}
\hat{\mathcal{Q}}=\hat{\mathcal{S}}\mathrm{I}_{4}+\hat{\mathcal{P}}\gamma
^{5}+\hat{\mathcal{V}}^{\mu }\gamma _{\mu }+\hat{\mathcal{A}}^{\mu }\gamma
^{5}\gamma _{\mu }+\frac{1}{2}\hat{\mathcal{T}}^{\mu \nu }\sigma _{\mu \nu
}\,,  \label{eq:Q-decomposition}
\end{equation}%
where the scalar, pseudoscalar, vector, pseudovector and tensor operators are written in momentum space as follows
\begin{subequations}
\begin{align}
\hat{\mathcal{S}}=& \sum_{d=3}^{\infty }\mathcal{S}^{(d)\alpha _{1}\alpha
_{2}\ldots \alpha _{d-3}}p_{\alpha _{1}}p_{\alpha _{2}}\ldots p_{\alpha
_{d-3}},  \label{eq:Operators} \\
\hat{\mathcal{P}}=& \sum_{d=3}^{\infty }\mathcal{P}^{(d)\alpha _{1}\alpha
_{2}\ldots \alpha _{d-3}}p_{\alpha _{1}}p_{\alpha _{2}}\ldots p_{\alpha
_{d-3}}, \\
\hat{\mathcal{V}}^{\mu }=& \sum_{d=3}^{\infty }\mathcal{V}^{(d)\mu \alpha
_{1}\alpha _{2}\ldots \alpha _{d-3}}p_{\alpha _{1}}p_{\alpha _{2}}\ldots
p_{\alpha _{d-3}}, \\
\hat{\mathcal{A}}^{\mu }=& \sum_{d=3}^{\infty }\mathcal{A}^{(d)\mu \alpha
_{1}\alpha _{2}\ldots \alpha _{d-3}}p_{\alpha _{1}}p_{\alpha _{2}}\ldots
p_{\alpha _{d-3}}, \\
\hat{\mathcal{T}}^{\mu \nu }=& \sum_{d=3}^{\infty }\mathcal{T}^{(d)\mu \nu
\alpha _{1}\alpha _{2}\ldots \alpha _{d-3}}p_{\alpha _{1}}p_{\alpha
_{2}}\ldots p_{\alpha _{d-3}}, \label{coeffi3a}
\end{align}%
These decompositions were first proposed in Ref. \cite{kostelecky2013fermions}. Here, we also point out that, having mass dimension $4-d$, the controlling coefficients $\mathcal{S}^{(d)\alpha
_{1}\ldots \alpha _{d-3}}$, $\ldots $, $\mathcal{T}^{(d)\mu \nu \alpha
_{1}\ldots \alpha _{d-3}}$ are spacetime independent in order to maintain the conservation of energy and momentum.

It is possible to obtain the dispersion relations directly from the determinant of the Dirac operator. According to \cite{kostelecky2013fermions}, the first-order dispersion relation (for particle modes) related to the modified Dirac equation coming from Lagrangian (\ref{eq:SME-L-fermion}) is
\end{subequations}
\begin{equation}
E\approx E_{0}-\frac{m_{\psi }\hat{\mathcal{S}}+p\cdot \mathcal{\hat{V}}}{%
E_{0}}\pm \frac{\mathcal{Y}}{E_{0}},  \label{eq:EQDispersion-GP}
\end{equation}%
where $E_{0}=\pm\sqrt{m_{\psi }^{2}+\boldsymbol{p}^{2}}$ and $\mathcal{Y}$ is given by the following expression%
\begin{equation}
\mathcal{Y}^{2}=\left( p\cdot \hat{\mathcal{A}}\right) ^{2}-m_{\psi }^{2}%
\hat{\mathcal{A}}^{2}-2m_{\psi }p\cdot \tilde{\hat{\mathcal{T}}}\cdot \hat{%
\mathcal{A}}+p\cdot \tilde{\hat{\mathcal{T}}}\cdot \tilde{\hat{\mathcal{T}}}%
\cdot p.  \label{Pro.Eq2}
\end{equation}

Here, we see that there exist two possible configurations for energy $E_{0}$. It suffices to describe both particles and antiparticles modes depending on the sign, i.e., positive for particles and negative for antiparticles respectively. Also, we observe that terms $\hat{\mathcal{S}}$, $\hat{\mathcal{V}}$ and $\mathcal{Y}$ displayed above depend on the 4-momentum, that at the leading order in Lorentz violation may be considered as $p^{\mu} \approx (E_{0},- {\bf{p}})$ on the right hand side in Eq. (\ref{eq:EQDispersion-GP}). Thereby, in the existence of Lorentz violation, it is verified that such expression can have four non-degenerate solutions for each ${\bf{p}}$.

Additionally, an important remark which is worth taking into account is the case when the spin degeneracy of a Dirac fermion is broken. Thereby, being in contrast with the aspects encountered in scalar and vector cases, pseudovector and tensor operators no longer maintain the spin degeneracy as a consequence of nonzero $\mathcal{Y}$ term. Moreover, it is worth pointing out that such degeneracy between either particles or antiparticles is no longer preserved when any of these Lorentz-violating operators have nonzero CPT-odd components. Besides, on the other hand, the pseudoscalar operator plays no role at the relevant order dispersion relation. 

From Eq. (\ref{eq:EQDispersion-GP}), we can represent several deviations from the traditional Lorentz covariant approach concerning massive fermions. Many of them lie in anisotropy, dispersion, and birefringence cases, being analogous to those effects that arouse in the nonminimal photon sector of the SME \cite{kostelecky2007,kostelecky2009,cambiaso2012,mewes2012}. Next, we explore the consequences of scalar field in the context of Lorentz violation.


\section{Lorentz-Violating Scalar Fields}\label{sec:scalarSector}

Recently, in the literature it was proposed conjectures about Riemann-Finsler geometries ascribed to Lorentz-violating field theories \cite{EDWARDS2018}. In this sense, knowing how this novelty can be associated with the study of thermodynamic properties of interacting quantum gases is a remarkable question to be investigated. Looking toward to accomplish this, we initially introduce the respective model. Analogously to what was done in Ref. \cite{EDWARDS2018}, we regard a complex scalar field $\phi \left( x^{\mu
}\right) $ of mass $m$ for $n=4$. The effective quadratic Lagrangian for the scalar field is given by%
\begin{equation}
\mathcal{L}=\partial ^{\mu }\phi ^{\dagger }\partial _{\mu }\phi -m\phi
^{\dagger }\phi -\frac{1}{2}\left( \mathrm{i}\phi ^{\dagger }\left( \hat{k}%
_{a}\right) ^{\mu }\partial _{\mu }\phi +\mathrm{h.c.}\right) +\partial
_{\mu }\phi ^{\dagger }\left( \hat{k}_{c}\right) ^{\mu \nu }\partial _{\nu
}\phi ,  \label{eq:SME-L-Scalar}
\end{equation}%
where $\left( \hat{k}_{a}\right) ^{\mu }$ and $\left( \hat{k}_{c}\right)
^{\mu \nu }$ are operators constructed as series of even powers of the partial spacetime derivatives $\partial _{\mu }$. In this way, since Lorentz violation is presumed to contain its effects around the Planck scale, both $\left( \hat{k}_{a}\right) ^{\mu }$ and $\left( \hat{k}_{c}\right)^{\mu \nu }$ may be assumed to bring out only perturbations for ordinary physics. Again, we assume that these operators are independent quantities of spacetime position seeking to maintain the translational invariance, which accounts for the conservation of energy and momentum. The operators have the following structure in momentum space%
\begin{eqnarray}
\left( \hat{k}_{a}\right) ^{\kappa } &=&\sum_{d\geq 3}\left( k_{a}\right)
^{\left( d\right) \kappa \alpha _{1}\ldots \alpha _{\left( d-4\right)
}}p_{\alpha _{1}}\ldots p_{\alpha _{\left( d-3\right) }}, \label{coeffi21} \\
\left( \hat{k}_{c}\right) ^{\kappa \xi } &=&\sum_{d\geq 4}\left(
k_{c}\right) ^{\left( d\right) \kappa \xi \alpha _{1}\ldots \alpha _{\left(
d-4\right) }}p_{\alpha _{1}}\ldots p_{\alpha _{\left( d-4\right) }}. \label{coeffi2}
\end{eqnarray}%
Moreover, the respective dispersion relation for the theory encountered in Eq. (\ref{eq:SME-L-Scalar}) is given by
\begin{equation}
p^{2}-m^{2}-\left( \hat{k}_{a}\right) ^{\kappa }p_{k}+\left( \hat{k}%
_{c}\right) ^{\kappa \xi }p_{\kappa }p_{\xi }=0,
\label{eq:Dispersion-equation-scalar}
\end{equation}%
whose first order dispersion relation is written as %
\begin{equation}
E\approx E_{0}+\frac{1}{2}\frac{\left( \hat{k}_{a}\right) ^{\mu }p_{\mu }}{%
E_{0}}-\frac{\left( \hat{k}_{c}\right)^{\mu \nu }p_{\mu }p_{\nu }}{%
E_{0}}.  \label{eq:Dispersion-relatio-scalar}
\end{equation}
With all these features, it is verified that the boson sector has noteworthy properties and applications. Moreover, in Refs. \cite{passos2008,gomes201}, the authors studied the scalar field in different scenarios as well. In Ref. \cite{EDWARDS2018}, the authors proposed the correspondence between Riemann-Finsler geometries and effective field theories when spin-independent Lorentz violation is taken into account. Nevertheless, it is encountered a gap the literature looking toward to investigate the respective thermodynamic quantities for such case. In this way, we primarily present a model to derive the thermal quantities of interest, i.e., particle number, entropy, mean energy and pressure. For obtaining them, we utilize the so-called grand canonical partition function as well as the grand canonical potential. With these, the following procedure can be fully carried out.

\section{Thermodynamic model}\label{sec:ThermalModel}

At the beginning, let us start our discussion considering a general quantum state $\varphi$ of a free quantum gas, which is entirely determined when one specifies the occupation numbers, i.e, $\{n_{1},n_{2},...,n_{r},...\}$, regarding a discrete quantum state $r$ with energy $\epsilon_{r}$. In this sense, the sum in $\varphi$ must be carried out over all quantum states taking into account the restriction
\ie
\sum_{r}n_{r}=N,
\fe
where $N$ is the particle number. Here, since there exist modifications in the relativistic particle dispersion relations due to Lorentz-violating terms, we can use the advantage of separating the energy of the quantum state $\varphi$ as follows
\ie
E_{\varphi}^{free}=\sum_{r}n_{r}\epsilon _{r}+\sum_{r}n_{r}\delta _{r},
\label{eq:GDEquation}
\fe
where $\epsilon _{r}=\sqrt{\boldsymbol{p}_{r}^{2}+m^{2}}$ is the usual relativistic energy and $\delta _{r}$ refers to Lorentz-violating contribution term that may assume a specific form depending on the case under consideration. Besides, it is worth pointing out that parameter $\delta _{r}$ is in general a function of the 3-momentum $\boldsymbol{p}_{r}$ that modifies the relativistic kinetic term only\footnote{Here, we can still use the background coming from the standard Statistical Mechanics.}. These and other features will be treated in what follows. Now, we derive the so-called grand canonical partition function, which is given by 
\ie
\mathcal{Z}\left( T,V,z\right) =\sum_{N=0}^{\infty }z^{N}Z\left(
T,V,N\right),
\fe
where $z=\exp \left( \beta \mu \right)$ is the fugacity of the system and the usual canonical partition function is written as 
\ie
Z\left( T,V,N\right) =\sum_{{\varphi}}\exp \left( -\beta E_{{\varphi}}\right).
\fe
Here, using the general energy $E_{\varphi}^{free}$, we get the following grand canonical partition function%
\ie
\mathcal{Z}\left( T,V,\mu \right) =\sum_{\left\{ n_{1},n_{2},\ldots \right\}
=0}^{\left\{ \infty /1\right\} }\exp \left\{ -\beta \left[
\sum_{r}n_{r}\left( \epsilon _{r}+\delta _{r}-\mu \right) +U\left(
V,n\right) \right] \right\} ,  \label{eq:g-partition-function}
\fe
and
\ie
z^{N}=\exp \left\{ N\beta \mu \right\} =\exp \left\{ \beta \sum_{r}n_{r}\mu
\right\},
\fe
where we have considered $U\left( V,n\right)$ as being the interaction energy that depends only on the particle density $n$ and the volume $V$. As we shall argue below, such type of interaction can be obtained through the well-know mean field approximation which can give us the advantage of obtaining analytical results. Moreover, these ones will allow us to identify how such type of interaction modifies the thermodynamic properties of the system. It is worth to mention that the interaction term is a monotonically increasing function of the density particle. In other words, if we increase the density $n$, the particles come closer to each other and the interactions between them are expected to increase. More so, the opposite behavior happens otherwise: when $n$ decreases, $U\left( V,n\right)$ must decrease.

It is important to notice that the upper summation index in Eq. (\ref%
{eq:g-partition-function}), namely $\left\{ \infty /1\right\} $, indicates that despite having infinitely many bosons in the quantum state $r$, rather for the fermion case, only one fermion is allowed due to the Pauli exclusion principle. As an example, we regard the upper index “$\infty$” for bosons and “$1$” for fermions. With this notation, we are able to treat both of them simultaneously without losing generality. Moreover, it is also convenient to assume that the total interaction energy can be
written as $U\left( V,n\right) =Vu\left( n\right)$, where $n=N/V$ is the particle density. Such density-dependent interaction has applications in nuclear \cite{khoa1997nuclear} and elementary particle physics \cite{barnett1996,eidelman2004,lalazissis2005new}. Assuming this decomposition, we have%
\ie
\mathcal{Z}\left( T,V,\mu \right) =\sum_{\left\{ n_{1},n_{2},\ldots \right\}
=0}^{\left\{ \infty /1\right\} }\exp \left\{ -\beta \left[
\sum_{r}n_{r}\left( \epsilon _{r}+\delta _{r}-\mu \right) +Vu\left( n\right) %
\right] \right\},  \label{eq:g-partition-function-1}
\fe
and, for further evaluation of the above expression, it is
necessary that the exponential function be decomposed into factors of the form $%
\exp \left\{ -\beta n_{r}\left( \ldots \right) \right\}$. It is noteworthy to mention that this assumption can be ensured if and only if the term $Vu\left( n\right)$ is linear in $\sum_{r}n_{r}=N$. However, for most cases this is not a straightforward procedure; $Vu\left( n\right)$ is a difficult function of $N$ that has to be determined afterwards when the interaction is particularized. In this way, there is a traditional form in the literature to linearize $Vu\left(
n\right)$ as a function of $N$, which is via Taylor series. We 
expand $u\left( n\right)$ around the mean value of the particle number density $\bar{n}$, that is%
\ie
u\left( n\right) =u\left( \bar{n}\right) +u^{\prime }\left( \bar{n}\right)
\left( n-\bar{n}\right) +\ldots.   \label{eq:Taylor_u}
\fe
Here, we choose only these two terms, since the other ones may be overlooked due to the fact that fluctuations $(n-%
\bar{n})$, close to the mean value of the particle number density, turn out to be tiny when one regards the thermodynamic limit (see Appendix \ref{sec:Therm-Limit} for further details). 

Indeed, having splitted $U\left( V,n\right)$ into different terms, namely $U\left( V,\bar{n}\right) =Vu\left( \bar{n}\right)$ and $Vu^{\prime
}\left( \bar{n}\right) \bar{n}=u^{\prime }\left( \bar{n}\right) \bar{N}$ as well as into $
V u^{\prime}\left( \bar{n}\right) n=u^{\prime}\left( \bar{n}\right) N$, we are properly able to write down the total energy of the quantum state $\varphi$ as follows
\ie
E_{\varphi}=\sum_{r}n_{r}\epsilon _{r}+\sum_{r}n_{r}\delta
_{r}+\sum_{r}n_{r}u^{\prime }\left( \bar{n}\right) +U\left( V,\bar{n}\right)
-u^{\prime }\left( \bar{n}\right) \bar{N},  \label{eq:Total_energy}
\fe
where, up to these two latter terms, the energy of a particle in the quantum state $r$ would be simply $\epsilon_{r}+\delta _{r}+u^{\prime}\left(\bar{n}\right)$. Certainly, it is worth mentioning that the mean energy $u^{\prime}\left( \bar{n}\right)$ of the respective ensemble of particles arouses from the fact that there exists interaction among them. Furthermore, we could suppose that such potential energy would come from a specified mean field\footnote{Here, we are mentioning that the mean field approximation it is just one way to get such potential. However, our results are quite general and they are not restricted to the assumption of the validity of the \textit{molecular field approximation}. Moreover, it is worth pointing out that such an approach is widely used to obtain many interactions in condensed-matter physics. Nevertheless, it is not the unique way to do so. Also, from the point of view of numerical results, the mean field approximation turns out to be convenient since we can find a lot of interesting results in the literature using the same procedure, as already mentioned in the manuscript.} at the position of the particle, and consequently Eq. (\ref{eq:Taylor_u}) would precisely address the widespread \textit{molecular field approximation} which is massively used in condensed matter physics \cite{humphries1972,klein1969,wojtowicz,ter1962molecular,araujo2017,silva2018}. Now, Eq. (\ref{eq:g-partition-function-1}) can be rewritten in a straightforward manner as follows
\begin{eqnarray}
\mathcal{Z}\left( T,V,\mu \right)  &=&\exp \left\{ -\beta \left[ U\left( V,\bar{n}\right) -u^{\prime }\left(
\bar{n}\right) \bar{N}\right] \right\}   \notag \\
&&\times \prod_{r=1}^{\infty }\left( \sum_{n_{r}=0}^{\left\{ \infty
/1\right\} }\exp \left\{ -\beta \left[ \epsilon _{r}+\delta
_{r}+u^{\prime }\left( \bar{n}\right) -\mu \right] n_{r}\right\} \right) .
\label{eq:Modified_GCP}
\end{eqnarray}%
Here, one may notice that in the latter line of the above expression there exists a sum in brackets which turns out to have merely two available values, namely $n_{r}=0,1$, if one considers either fermions or simply a geometric series when bosons are taken into account. In this way, we obtain
\begin{eqnarray}
\mathcal{Z}\left( T,V,\mu \right)  &=&\exp \left\{ -\beta \left[ U\left( V,%
\bar{n}\right) -u^{\prime }\left( \bar{n}\right) \bar{N}\right] \right\}
\notag \\
&&\times \prod_{r=1}^{\infty }\left\{
\begin{array}{c}
1+\exp \left[ -\beta \left( \epsilon _{r}+\delta _{r}+u^{\prime }\left( \bar{%
n}\right) -\mu \right) \right], \text{      Fermions} \\
\left( 1-\exp \left[ -\beta \left( \epsilon _{r}+\delta _{r}+u^{\prime
}\left( \bar{n}\right) -\mu \right) \right] \right) ^{-1}, \text{  Bosons}%
\end{array}%
\right. ,
\end{eqnarray}%
and, seeking for a generalized notation which can fit both of them, we provide the notation $\chi =+1$ for fermions and $\chi =-1$ for bosons respectively. Thus, we can rewrite such expression in a suitable way
\begin{eqnarray}
\mathcal{Z}\left( T,V,\mu \right)  &=&\exp \left\{ -\beta \left[ U\left( V,%
\bar{n}\right) -u^{\prime }\left( \bar{n}\right) \bar{N}\right] \right\}
\notag \\
&&\times \prod_{r=1}^{\infty }\left( 1+\chi \exp \left[ -\beta \left(
\epsilon _{r}+\delta _{r}+u^{\prime }\left( \bar{n}\right) -\mu \right) %
\right] \right) ^{\chi },
\end{eqnarray}%
which yields the so-called grand canonical potential
\begin{eqnarray}
\Phi  &=&-T\ln \mathcal{Z} \notag\\
&=&-T\chi \sum_{r}\ln \left( 1+\chi \exp \left[ -\beta \left( \epsilon
_{r}+\delta _{r}+u^{\prime }\left( \bar{n}\right) -\mu \right) \right]
\right) +U\left( V,\bar{n}\right) -u^{\prime }\left( \bar{n}\right) \bar{N}.
\label{eq:GCP-new}
\end{eqnarray}
As we shall see, the grand canonical potential $\Phi$ will play a crucial role in deriving the following thermodynamic functions. Next, from these preliminaries, we will develop the thermal quantities, i.e., mean particle number, entropy, mean total energy as well as pressure, highlighting the contribution of parameters that account for Lorentz violation. In addition, it is important to notice that the derivation of such quantities is fully carried out in an analytical manner.

\section{Thermodynamic state quantities}\label{sec:Therm-state-quantities}

Certainly, whenever one considers the study of thermodynamic properties of a given system, one seeks the derivation of the main thermal parameters as a traditional procedure. Thereby, we devote our analysis to provide the development of mean particle number, entropy, mean total energy and finally pressure. As a matter of fact, the terms that account for the violation of Lorentz symmetry will be inferred for different cases. Moreover, if one takes the limit when such terms vanish, one will recover the usual case being in agreement with the literature \cite{greiner2012,salinas1997,pathria32} so that it corroborates our results. Now, we start off with the mean particle number given by
\begin{eqnarray}
\bar{N} &=&-\left. \frac{\partial \Phi }{\partial \mu }\right\vert _{T,V},
\notag \\
&=&-V\left. \frac{\partial u\left( \bar{n}\right) }{\partial \mu }%
\right\vert _{T,V}+\bar{N}\left. \frac{\partial u^{\prime }\left( \bar{n}%
\right) }{\partial \mu }\right\vert _{T,V}+u^{\prime }\left( \bar{n}\right)
\left. \frac{\partial \bar{N}}{\partial \mu }\right\vert _{T,V}+  \notag \\
&&+T\chi \sum_{r}\frac{\chi \exp \left[ -\beta \left( \epsilon _{r}+\delta
_{r}+u^{\prime }\left( \bar{n}\right) -\mu \right) \right] }{1+\chi \exp %
\left[ -\beta \left( \epsilon _{r}+\delta _{r}+u^{\prime }\left( \bar{n}%
\right) -\mu \right) \right] }\beta \left( 1-\left. \frac{\partial u^{\prime
}\left( \bar{n}\right) }{\partial \mu }\right\vert _{T,V}\right),
\end{eqnarray}%
and, from it, we obtain%
\begin{eqnarray}
\bar{N}\left( 1-\left. \frac{\partial u\left( \bar{n}\right) }{\partial \mu }%
\right\vert _{T,V}\right)  &=&-V\left. \frac{\partial u\left( \bar{n}\right)
}{\partial \mu }\right\vert _{T,V}+u^{\prime }\left( \bar{n}\right) \left.
\frac{\partial \bar{N}}{\partial \mu }\right\vert _{T,V}  \notag \\
&&+\chi ^{2}\beta T\left( 1-\left. \frac{\partial u^{\prime }\left( \bar{n}%
\right) }{\partial \mu }\right\vert _{T,V}\right)\notag\\
&&\times\sum_{r}\frac{1}{\exp %
\left[ \beta \left( \epsilon _{r}+\delta _{r}+u^{\prime }\left( \bar{n}%
\right) -\mu \right) \right] +\chi }.  \label{eq:Mean-N-1}
\end{eqnarray}%
Here, since $u$ depends only on the particle number density, we obtain directly $du=\frac{du}{dn}dn$. In this sense,
\begin{equation}
\left. \frac{\partial u\left( \bar{n}\right) }{\partial \mu }\right\vert
_{T,V}=u^{\prime }\left( \bar{n}\right) \left. \frac{\partial \bar{n}}{%
\partial \mu }\right\vert _{T,V}=\frac{u^{\prime }\left( \bar{n}\right) }{V}%
\left. \frac{\partial \bar{N}}{\partial \mu }\right\vert _{T,V},
\label{eq:relation-u}
\end{equation}%
and it is worth pointing out that the first two terms in Eq. (\ref{eq:Mean-N-1}) cancel out each other. Regarding natural units, i.e., $\chi ^{2}=\beta T=1$, then%
\begin{equation}
\bar{N}=\sum_{r}\frac{1}{\exp \left[ \beta \left( \epsilon _{r}+\delta
_{r}+u^{\prime }\left( \bar{n}\right) -\mu \right) \right] +\chi }.
\label{eq:Mean-number-N}
\end{equation}%
We obtain such expression due to the fact that the brackets do not vanish in Eq. (\ref{eq:Mean-N-1}), since $%
u\left( n\right)$ can be an arbitrary function. Besides, the mean occupation number can be immediately written as 
\begin{equation}
\bar{n}_{r}=\frac{1}{\exp \left[ \beta \left( \epsilon _{r}+\delta
_{r}+u^{\prime }\left( \bar{n}\right) -\mu \right) \right] +\chi }.
\label{eq:Mean-number-N-new}
\end{equation}%
Here, when compared to the usual case, Lorentz-violating parameters only modify the single-particle energy. Additionally, another further analysis can be done in such direction. Since Lorentz violation will be treated perturbatively, we can expand $\bar{n}_{r}$ at first order in $\delta _{r}$,
namely%
\begin{equation}
\bar{n}_{r}\approx \frac{1}{\exp \left[ \beta \left( \epsilon _{r}+u^{\prime
}\left( \bar{n}\right) -\mu \right) \right] +\chi }-\delta _{r}\frac{\beta
\exp \left[ -\beta \left( \epsilon _{r}+u^{\prime }\left( \bar{n}\right)
-\mu \right) \right] }{\left( \exp \left[ \beta \left( \epsilon
_{r}+u^{\prime }\left( \bar{n}\right) -\mu \right) \right] +\chi \right) ^{2}%
},
\end{equation}%
and, seeking a shorter expression, we define the following useful quantity%
\begin{equation*}
\mathcal{\bar{N}}_{r}\equiv \frac{1}{\exp \left[ \beta \left( \epsilon
_{r}+u^{\prime }\left( \bar{n}\right) -\mu \right) \right] +\chi }.
\end{equation*}%
As a result, the mean occupation number can be written as
\begin{equation}
\bar{n}_{r}\approx \mathcal{\bar{N}}_{r}-\delta _{r}\beta \mathcal{\bar{N}}%
_{r}^{2}\exp \left[ -\beta \left( \epsilon _{r}+u^{\prime }\left( \bar{n}%
\right) -\mu \right) \right] .  \label{eq:first-order-occ-number}
\end{equation}%
yielding therefore the mean particle number
\begin{equation}
\bar{N}\approx \sum_{r}\mathcal{\bar{N}}_{r}-\sum_{r}\delta _{r}\mathcal{%
\bar{N}}_{r}^{2}\beta \exp \left[ -\beta \left( \epsilon _{r}+u^{\prime
}\left( \bar{n}\right) -\mu \right) \right],
\label{eq:first-order-N-number}
\end{equation}%
where the second term represents the deviation from the standard result highlighting the contribution due to the Lorentz violation. We can also notice that the interaction modifies the mean particle number since the term $u^{\prime
}\left( \bar{n}\right)$ is present in Eq. (\ref{eq:Mean-number-N-new}). This modification is directly related to the fact that we chose a interaction energy that is a function of the particle density.

Furthermore, looking toward to bring out the development to the calculation of the entropy, we proceed as follows
\begin{eqnarray}
S &=&-\left. \frac{\partial \Phi }{\partial T}\right\vert _{\mu ,V}  \notag
\\
&=&-V\left. \frac{\partial u\left( \bar{n}\right) }{\partial T}\right\vert
_{\mu ,V}+\bar{N}\left. \frac{\partial u^{\prime }\left( \bar{n}\right) }{%
\partial T}\right\vert _{\mu ,V}+u^{\prime }\left( \bar{n}\right) \left.
\frac{\partial \bar{N}}{\partial T}\right\vert _{\mu ,V}  \notag \\
&&+\chi \sum_{r}\ln \left( 1+\chi \exp \left[ -\beta \left( \epsilon
_{r}+\delta _{r}+u^{\prime }\left( \bar{n}\right) -\mu \right) \right]
\right)   \notag \\
&&+\chi ^{2}T\sum_{r}\frac{\left( \epsilon _{r}+\delta _{r}+u^{\prime
}\left( \bar{n}\right) -\mu \right) \left( -\frac{d\beta }{dT}\right) -\beta
\left. \frac{\partial u^{\prime }\left( \bar{n}\right) }{\partial T}%
\right\vert _{\mu ,V}}{\exp \left[ \beta \left( \epsilon _{r}+\delta
_{r}+u^{\prime }\left( \bar{n}\right) -\mu \right) \right] +\chi },
\end{eqnarray}%
and, in agreement with what happened to Eq. (\ref{eq:relation-u}), we also realize that the first and third terms cancel out each other. Now, regarding Eq. (\ref{eq:Mean-number-N}) in the numerator of the last sum, it follows that the second term cancels out the last term in the sum. Here, using the fact that $-\frac{d\beta }{dT}=\frac{1}{T^{2}}$, the result can be
written as
\begin{eqnarray}
S &=&\chi \sum_{r}\ln \left( 1+\chi \exp \left[ -\beta \left( \epsilon
_{r}+\delta _{r}+u^{\prime }\left( \bar{n}\right) -\mu \right) \right]
\right)   \notag \\
&&+\frac{1}{T}\sum_{r}\bar{n}_{r}\left( \epsilon _{r}+\delta _{r}+u^{\prime
}\left( \bar{n}\right) -\mu \right) .  \label{eq:Entropy}
\end{eqnarray}%
Expanding the above expression up to the first order in $\delta _{r}$, we yield%
\begin{eqnarray}
S &\approx &\chi \sum_{r}\ln \left( 1+\chi \exp \left[ -\beta \left(
\epsilon _{r}+u^{\prime }\left( \bar{n}\right) -\mu \right) \right] \right) +%
\frac{1}{T}\sum_{r}\left( \epsilon _{r}+u^{\prime }\left( \bar{n}\right)
-\mu \right) \mathcal{\bar{N}}_{r}  \notag \\
&&-\sum_{r}\delta _{r}\beta \mathcal{\bar{N}}_{r}+\frac{1}{T}\sum_{r}\delta
_{r}\mathcal{\bar{N}}_{r}\left\{ 1-\left( \epsilon _{r}+u^{\prime }\left(
\bar{n}\right) -\mu \right) \beta \exp \left[ \beta \left( \epsilon
_{r}+u^{\prime }\left( \bar{n}\right) -\mu \right) \right] \mathcal{\bar{N}}%
_{r}\right\}, \label{eq:first-order-Entropy}
\end{eqnarray}%
and, in particular, the last two terms represent the deviation from the standard result, exhibiting clearly the contribution of Lorentz-violating parameters. Following these approaches, we derive the mean total energy
\begin{eqnarray}
\bar{E} &=&\left. \frac{\partial \left( \beta \Phi \right) }{\partial \beta }%
\right\vert _{z,V},  \notag \\
&=&\left. \frac{\partial }{\partial \beta }\left[ \beta Vu\left( \bar{n}%
\right) -\beta \bar{N}u^{\prime }\left( \bar{n}\right) \right] \right\vert
_{z,V}  \notag \\
&&-\chi \sum_{r}\frac{\chi z\exp \left[ -\beta \left( \epsilon _{r}+\delta
_{r}+u^{\prime }\left( \bar{n}\right) \right) \right] }{1+\chi z\exp \left[
-\beta \left( \epsilon _{r}+\delta _{r}+u^{\prime }\left( \bar{n}\right)
\right) \right] }\left( -\epsilon _{r}-\delta _{r}-\left. \frac{\partial }{%
\partial \beta }\left[ \beta u^{\prime }\left( \bar{n}\right) \right]
\right\vert _{z,V}\right) ,  \notag \\
&=&U\left( V,\bar{n}\right) +\beta V\left. \frac{\partial u\left( \bar{n}%
\right) }{\partial \beta }\right\vert _{z,V}-\bar{N}\left. \frac{\partial %
\left[ \beta u^{\prime }\left( \bar{n}\right) \right] }{\partial \beta }%
\right\vert _{z,V}-\beta u^{\prime }\left( \bar{n}\right) \left. \frac{%
\partial \bar{N}}{\partial \beta }\right\vert _{z,V}  \notag \\
&&+\sum_{r}\frac{\epsilon _{r}+\delta _{r}}{z^{-1}\exp \left[ \beta \left(
\epsilon _{r}+\delta _{r}+u^{\prime }\left( \bar{n}\right) \right) \right]
+\chi }+\bar{N}\left. \frac{\partial \left[ \beta u^{\prime }\left( \bar{n}%
\right) \right] }{\partial \beta }\right\vert _{z,V}.  \label{eq:mean-energy}
\end{eqnarray}%
It is important to mention that because of%
\begin{equation}
\left. \frac{\partial u\left( \bar{n}\right) }{\partial \beta }\right\vert
_{z,V}=u^{\prime }\left( \bar{n}\right) \left. \frac{\partial \bar{n}}{%
\partial \beta }\right\vert _{z,V}=\frac{u^{\prime }\left( \bar{n}\right) }{V%
}\left. \frac{\partial \bar{N}}{\partial \beta }\right\vert _{z,V},
\end{equation}%
the second term cancels out the fourth in Eq. (\ref{eq:mean-energy}) and, therefore we get
\begin{equation}
\bar{E}=\sum_{r}\bar{n}_{r}\left( \epsilon _{r}+\delta _{r}\right) +U\left(
V,\bar{n}\right) .  \label{eq:Energy}
\end{equation}%
As expected, the energy is the average of the kinetic term plus the interactions energy. As an analogous procedure, we accomplish the expansion of the mean energy at first order in $\delta _{r}$. This yields
\begin{equation}
\bar{E}\approx \sum_{r}\mathcal{\bar{N}}_{r}\epsilon _{r}+U\left( V,\bar{n}%
\right) +\sum_{r}\delta _{r}\mathcal{\bar{N}}_{r}\left\{ 1-\epsilon
_{r}\beta \exp \left[ \beta \left( \epsilon _{r}+u^{\prime }\left( \bar{n}%
\right) -\mu \right) \right] \mathcal{\bar{N}}_{r}\right\},
\label{eq:first-order-mean-energy}
\end{equation}%
with the identification that the third term is the Lorentz-violating contribution that differs, thus, from the usual result. Finally, after all these features, we also provide the derivation of pressure as follows%
\begin{eqnarray}
\mathcal{P} &=&-\left. \frac{\partial \Phi }{\partial V}\right\vert _{\mu ,T}  \notag
\\
&=&-u\left( \bar{n}\right) +u^{\prime }\left( \bar{n}\right) \left. \frac{%
\partial \bar{N}}{\partial V}\right\vert _{\mu ,T}+\frac{\chi T}{V}%
\sum_{r}\ln \left( 1+\chi \exp \left[ -\beta \left( \epsilon _{r}+\delta
_{r}+u^{\prime }\left( \bar{n}\right) -\mu \right) \right] \right)  \notag \\
&=&-\frac{\Phi }{V},
\label{eq:pressurePP}
\end{eqnarray}%
where we have used Eq. (\ref{eq:Mean-number-N}), and the essential consideration that the respective particle number density is independent of the volume; as a result, $u\left(
\bar{n}\right) $ and $u^{\prime }\left( \bar{n}\right)$ do not
depend on the volume as well, as shown in Appendix \ref{sec:Therm-Limit}. From the Eq. (\ref{eq:pressurePP}), we can also realize how interaction plays an important role on the pressure. The first term in Eq. (\ref{eq:pressurePP}), - $u\left( \bar{n}\right)$, for instance, is responsible for reduce the pressure of the system, while the second one, $u^{\prime }\left( \bar{n}\right)$, plays the rule of increasing it. Now, as we did before, let us expand the pressure at first order in
$\delta _{r}$. In doing so, we get%
\begin{equation}
\mathcal{P} \approx -\frac{\Phi _{\text{\textrm{Standard}}}}{V}-\frac{T}{V}%
\sum_{r}\delta _{r}\beta \mathcal{\bar{N}}_{r},
\end{equation}%
where conveniently we define
\begin{equation}
\Phi _{\text{\textrm{Standard}}}=-\chi T\sum_{r}\ln \left( 1+\chi \exp \left[
-\beta \left( \epsilon _{r}+u^{\prime }\left( \bar{n}\right) -\mu \right) %
\right] \right) .
\end{equation}
Moreover, it is worth to point out that in Refs. \cite{anacleto2018,adailton,adailton3}, in the context of Lorentz violation, the authors also accomplished a similar analysis to the one encountered in this work. However, they brought out rather the advantage of using the accessible states of the system in order to derive the respective thermodynamic functions for higher-derivative electrodynamics. Besides, considering other viewpoints, the thermal quantities were calculated as well taking into account the canonical ensemble from its respective partition function \cite{oliveira2020,oliveira2019,pacheco2014,casana2008}. In the next sections, we will provide an analysis concerning different Lorentz-violating operators in order to verify their contributions when interacting quantum gases are regarded. Particularly, we examine both fermion and boson sectors respectively. For the first sector, we contemplate scalar, vector, pseudovector and tensor operators. On the other hand, for the latter one, we explore both $\left( \hat{k}_{a}\right) ^{\kappa }$ and $\left( \hat{k}_{c}\right) ^{\kappa \xi }$ operators to complete our discussion. We also point out that such analysis will bring out the behavior of the thermodynamic functions that we have calculated so far. Furthermore, as we could verify, the mathematical structure of those functions are quite complicated to explain in a phenomenological manner. Nevertheless, we overtake this situation performing a numerical analysis where we display the plots for the thermodynamic functions under consideration.

\pagebreak

\section{Interacting fermions}\label{sec:fermionsInterecting}

In this section, we devoted our attention to study fermions concerning the minimal sector of the Standard Model Extension. In doing so, we seek this analysis for knowing how such coupling terms affect the interaction energy when a quantum gas is taken into account. In order to carry out such investigation, we assume scalar, vector, pseudovector and tensor operators respectively. Moreover, we summarize them in Table \ref{Tab:Fermion-Delta} for the sake of bringing about a better comprehension to the reader. Now, we are on the verge of showing some interesting dispersion relations that can be used from the background established in the previous sections. Thereby, the general dispersion relation has the following structure for both fermions and bosons

\begin{equation}\label{eq:general_DR-Delta}
E\approx E_{0}+\frac{\delta _{r}}{E_{0}}.
\end{equation}
Particularly, the subsequent specifications of $\delta_{r}$ are displayed in Table \ref{Tab:Fermion-Delta}. In the last column on the right hand, we have used the definition $\bar{p}_{\alpha }\equiv \left(E_{0},-\boldsymbol{p}\right) $ so that it allows to obtain the first order dispersion relation in a covariant way.

\begin{table}[h]
  \centering
\begin{tabular}{c|c|c}
\hline
Operator & $\delta _{r}$ & Definition \\ \hline\hline
Scalar & $-m_{\psi }\hat{\mathcal{S}}$ & $\hat{\mathcal{S}}=\mathcal{S}%
^{\left( d\right) \alpha _{1}\ldots \alpha _{\left( d-3\right) }}\bar{p}%
_{\alpha _{1}}\ldots \bar{p}_{\alpha _{\left( d-3\right) }}$ \\ \hline
Vector & $-\bar{p}\cdot \mathcal{\hat{V}}$ & $\mathcal{\hat{V}}^{\kappa }=%
\mathcal{V}^{\left( d\right) \kappa \alpha _{1}\ldots \alpha _{\left(
d-3\right) }}\bar{p}_{\alpha _{1}}\ldots \bar{p}_{\alpha _{\left( d-3\right)
}}$ \\ \hline
Pseudovector & $\pm \sqrt{\left( \bar{p}\cdot \hat{\mathcal{A}}\right)
^{2}-m_{\psi }^{2}\hat{\mathcal{A}}^{2}}$ & $\hat{\mathcal{A}}^{\kappa }=%
\mathcal{A}^{\left( d\right) \kappa \alpha _{1}\ldots \alpha _{\left(
d-3\right) }}\bar{p}_{\alpha _{1}}\ldots \bar{p}_{\alpha _{\left( d-3\right)
}}$ \\ \hline
Tensor & $\pm \sqrt{p\cdot \tilde{\hat{\mathcal{T}}}\cdot \tilde{\hat{%
\mathcal{T}}}\cdot p.}$ & $\hat{\mathcal{T}}^{\kappa \xi }=\mathcal{T}%
^{\left( d\right) \kappa \xi \alpha _{1}\ldots \alpha _{\left( d-3\right) }}%
\bar{p}_{\alpha _{1}}\ldots \bar{p}_{\alpha _{\left( d-3\right) }}$%
\\\hline
\end{tabular}
\caption{This table summarizes four particular cases of $\delta_{r}$, namely, scalar, vector, pseudovector and tensor operators for the fermion sector. Their respective definitions are shown as well.}\label{Tab:Fermion-Delta}
\end{table}

In order to carry out the study of the main thermal properties discussed in section \ref{sec:Therm-state-quantities}, we need to perform the thermodynamic limit of Eqs. (\ref%
{eq:first-order-N-number}), (\ref{eq:first-order-mean-energy}) and (\ref%
{eq:first-order-Entropy}) as developed in detail in appendix \ref{sec:Therm-Limit}. Since such limits give rise to complicated integrals in momentum space, we provide numerical calculations for some particular backgrounds. Furthermore, with the purpose of offering a better arrangement to this paper, we display all numerical outputs in Appendix \ref{sec:num-cal}.

Now, we start with a simple configuration for the scalar operator, namely $\mathcal{S}^{(4)0}p_{0}=\mathcal{S}^{(4)0}E_{0}$. This configuration belongs to the minimal SME and yields the following dispersion relation%
\begin{equation*}
E\approx E_{0}-m_{\psi }\mathcal{S}^{(4)0}.
\label{eq:scalarconff}
\end{equation*}
Now, with the dispersion relation above, we can replace the results encountered in the Eqs. (\ref{eq:first-order-N-number}), (\ref{eq:first-order-mean-energy}) as well as  (\ref{eq:first-order-Entropy}), and perform integrations over the momenta. In Table \ref{Tab:VScalar}, we exhibit numerical evaluations for different values of $\beta$.

The second case taken into account has a vector operator whose non null controlling coefficient is $\mathcal{V}^{(4)33}$. With it, we can obtain the following dispersion relation, namely

\begin{equation}
E\approx E_{0}-\frac{\mathcal{V}^{(4)33}p_{3}p_{3}}{E_{0}}.
\label{eq:vectorconff}
\end{equation}
Now, we need to perform numerical integrations in order to find out the contribution to vector operator. The respective results are displayed in Table \ref{Tab:VVector}.

Another interesting case is the one related to pseudovector operators. We choose $\mathcal{A}^{(3)0}$ as the only non null controlling coefficient, which yields the following dispersion relation%

\begin{equation}
E\approx E_{0}\pm \frac{\sqrt{\left( E_{0}\mathcal{A}^{(3)0}\right)
^{2}-m_{\psi }^{2}\left( \mathcal{A}^{(3)0}\right) ^{2}}}{E_{0}}.
\label{eq:pseudoconff}
\end{equation}
Here, we notice that we have two different values to be considered. By inserting this result in the expression involving mean particle number, energy and entropy, we get the results shown in Table \ref{Tab:VPseudovector}.

On the other hand, we will treat the tensor operator whose non null controlling coefficient chosen is $\tilde{\hat{\mathcal{T}}}^{\left( 4\right)010}$. This configuration gives rise to the dispersion relation given below

\begin{equation}
E\approx E_{0}\pm E_{0}\tilde{\hat{\mathcal{T}}}^{\left( 4\right) 010}
\label{eq:tensorconff}
\end{equation}
Likewise, we need to perform numerical integrations for the sake of obtaining the information about the quantities of interest. Table \ref{Tab:Vtensor} shows how the thermodynamic quantities change with $\beta$.

\begin{figure}[tbh]
\centering
\subfloat [Scalar operator]{
  \includegraphics[width=8cm,height=5cm]{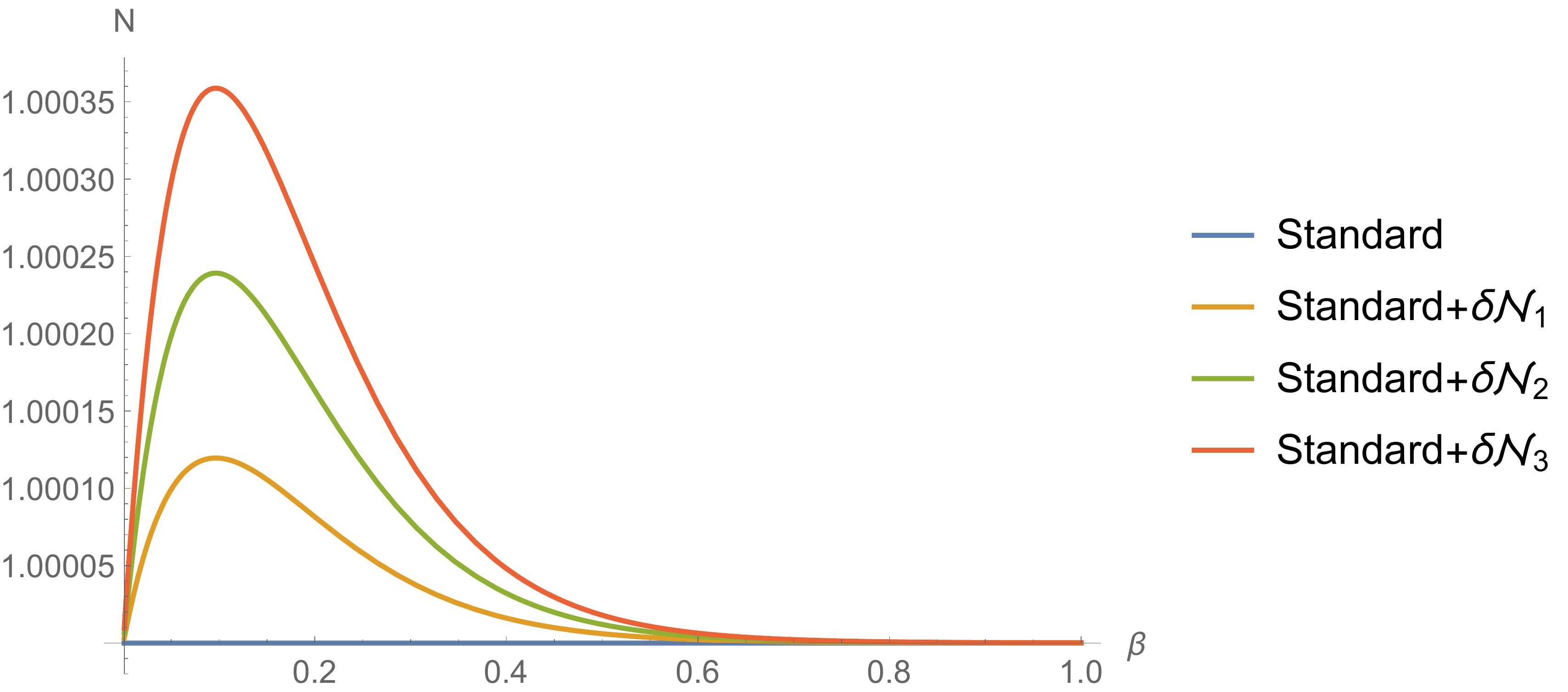}
  \label{fig:ScalaN}}
\subfloat[Vector operator]{
  \includegraphics[width=8cm,height=5cm]{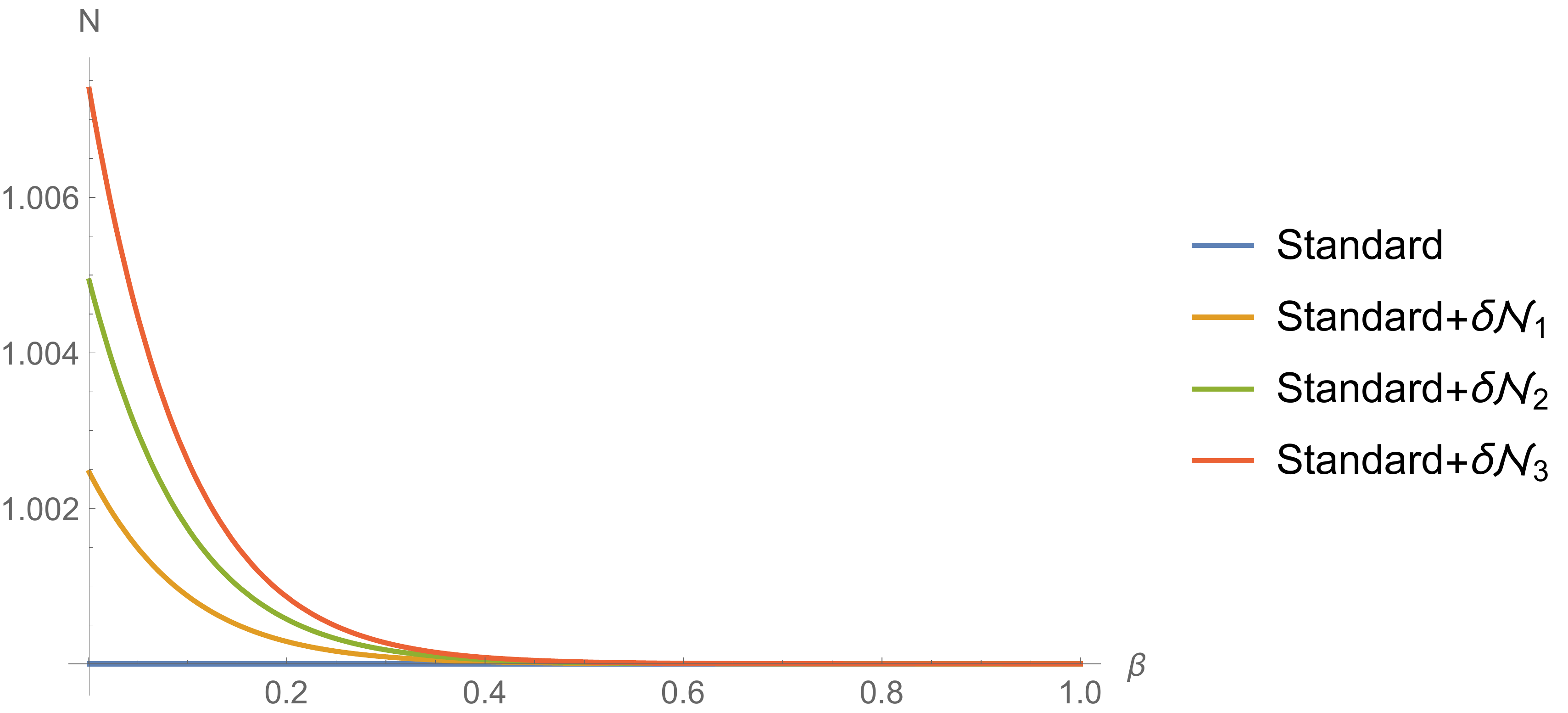}
  \label{fig:VectorN}}
\label{fig:BackgroundFields}
\subfloat[Pseudovector operator]{
  \includegraphics[width=8cm,height=5cm]{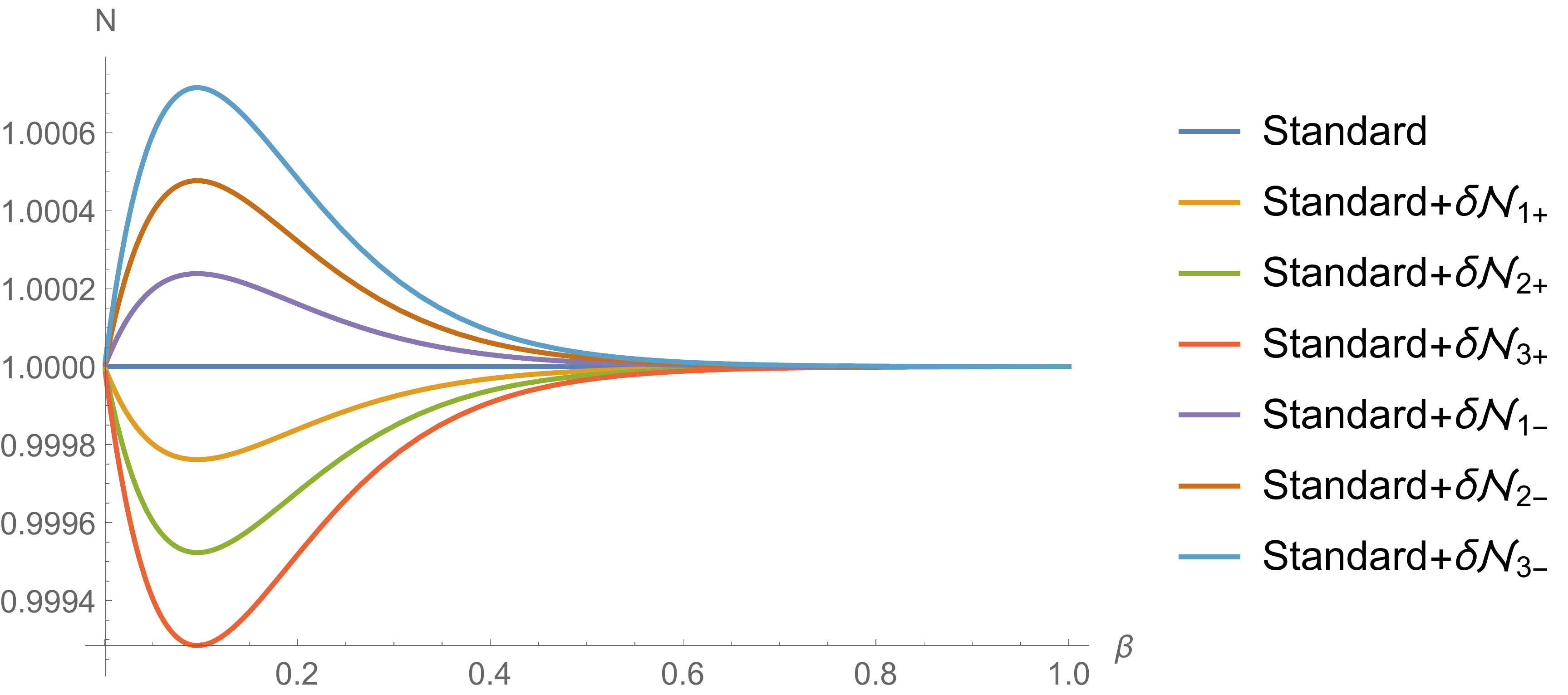}
  \label{fig:PseudovectorN}}
\subfloat[Tensor operator]{
  \includegraphics[width=8cm,height=5cm]{pseudovectorplotn.pdf}
  \label{fig:TensorN}}
\caption{This figure shows a comparison between mean particle number for different background configurations. These plots are normalized by the standard result, which means that the value 1 is the behavior regardless Lorentz violation. In order to  clarify the notation, we should notice that $\delta\mathcal{N}_{i}$ means the deviation from the standard result using the values $10^{-3}$, $10^{-5}$ and $10^{-7}$ for the controlling coefficient respectively.
}
\label{fig:PlotsNumber}
\end{figure}

Although the previous Tables are quite useful to afford quantitatively our results, they do not give us a reasonable visualization of what happens. In this sense, seeking to overtake this situation, we display the plots for the mean particle number for all dispersion relations treated in this section. In Fig. \ref{fig:PlotsNumber}, it is interesting to notice that we have two different solutions whenever we contemplate pseudovector and tensor operators. This occurs because these operators break the spin degeneracy. In other words, it turns out that spin-up $\left( +\right)$ particles propagate differently if compared with spin-down $\left( -\right) $ ones. This feature leads to a remarkable consequence: the mean particle number of a full interacting quantum gas of spin-down particles becomes greater in comparison with spin-up ones when we consider high temperature regime, as shown in Fig. \ref{fig:PseudovectorN} and \ref{fig:TensorN} . In addition, the other equation of state will exhibit the same property.

\begin{figure}[tbh]
\centering
\includegraphics[width=8cm,height=5cm]{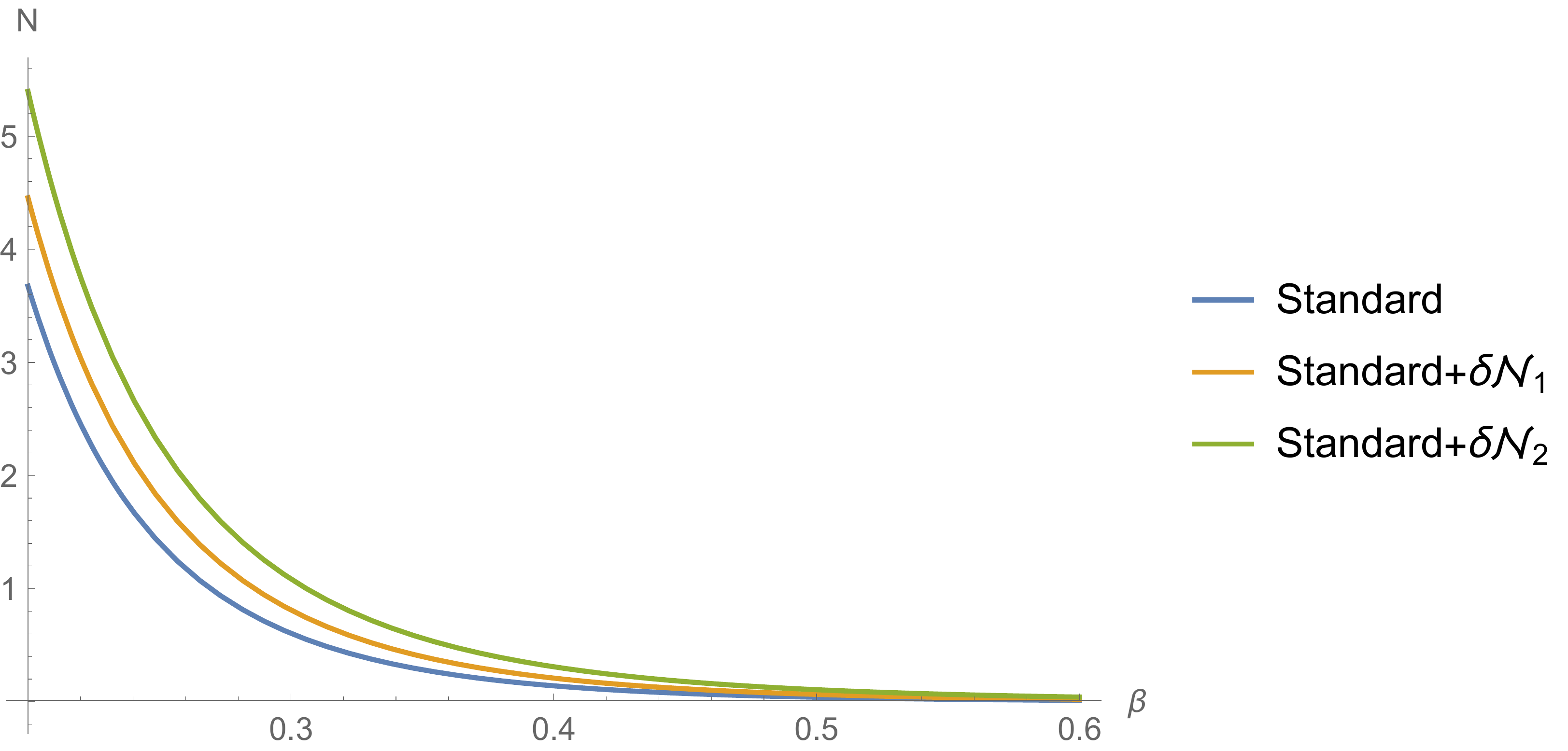}
\includegraphics[width=8cm,height=5cm]{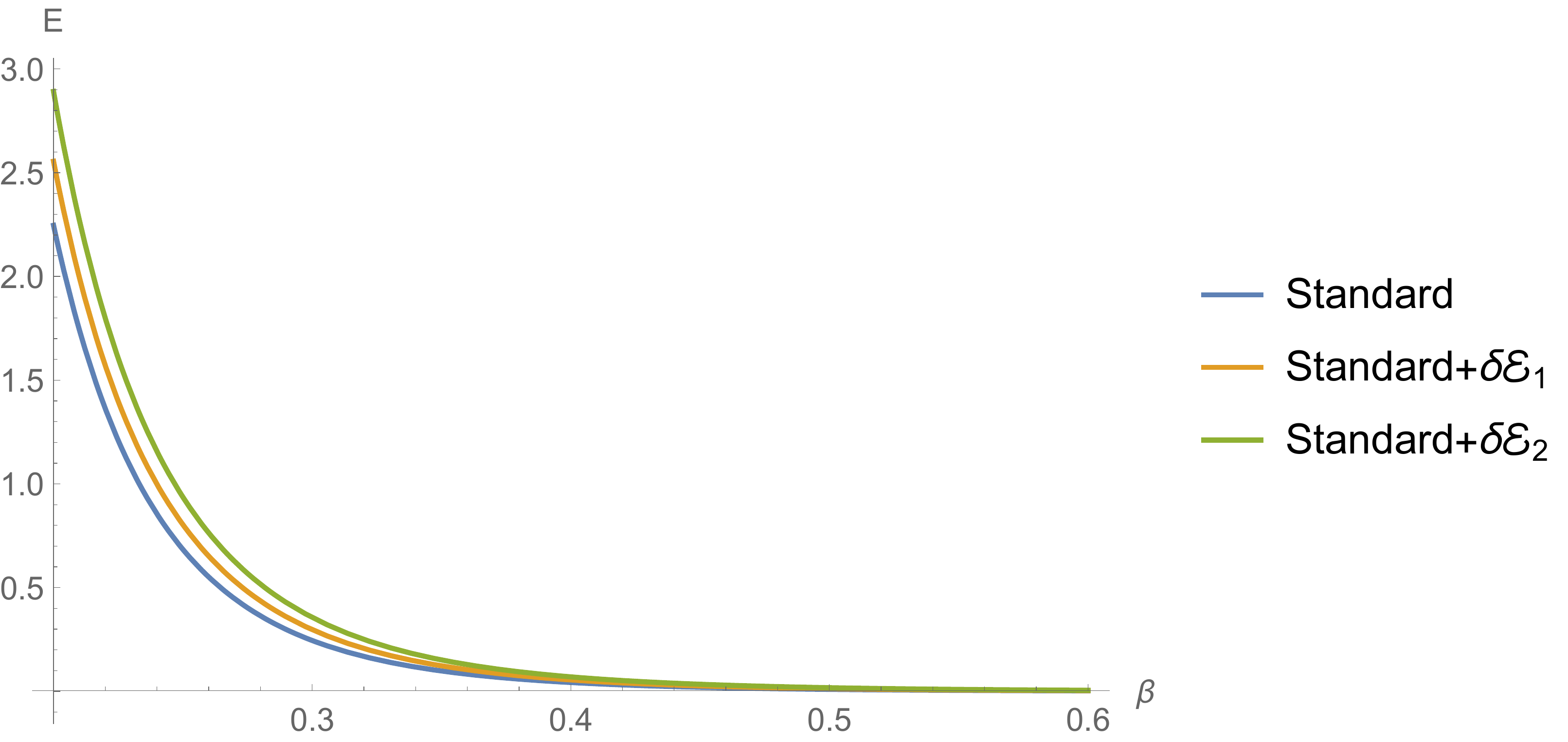}
\includegraphics[width=8cm,height=5cm]{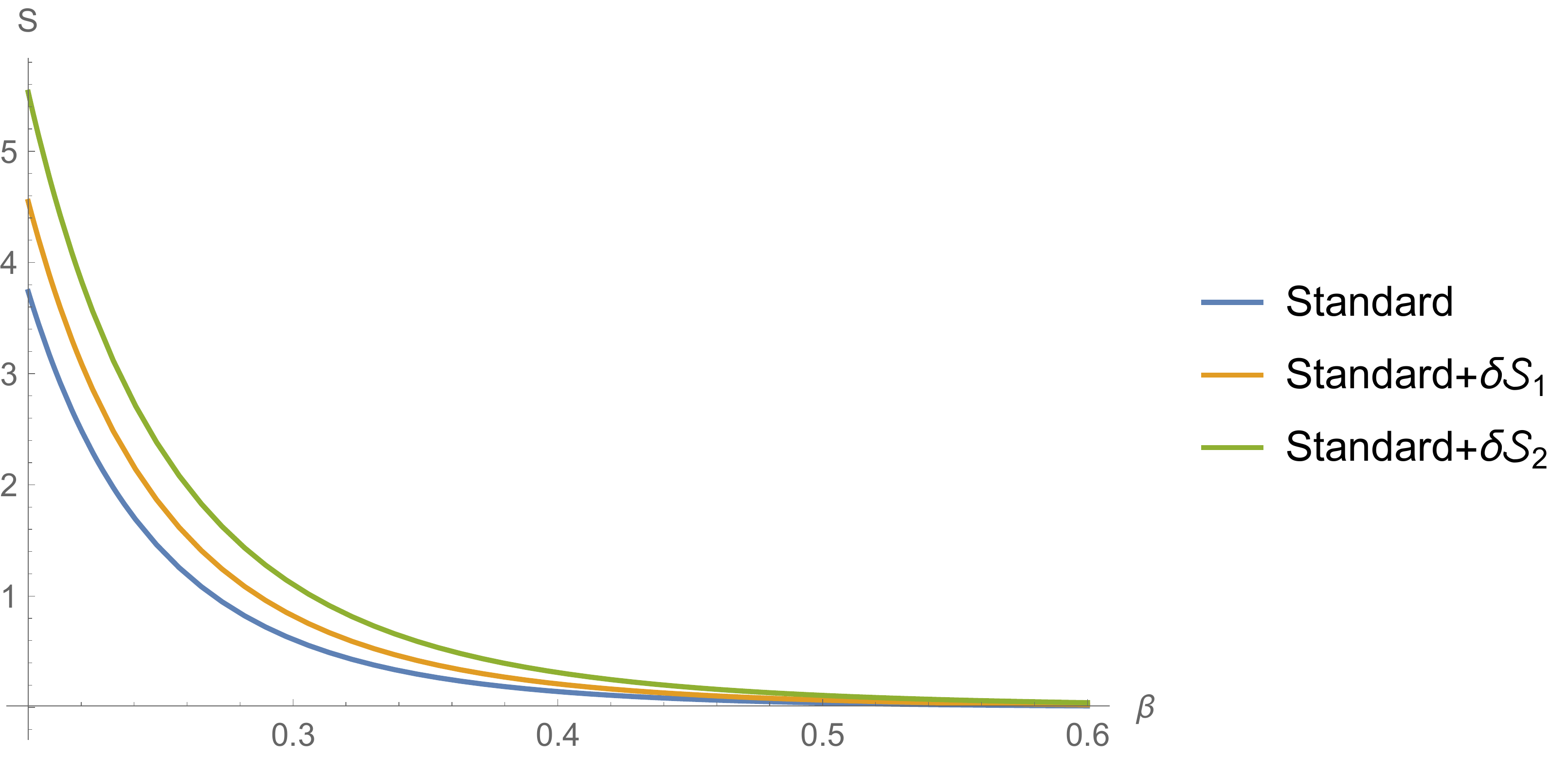}
\includegraphics[width=8cm,height=5cm]{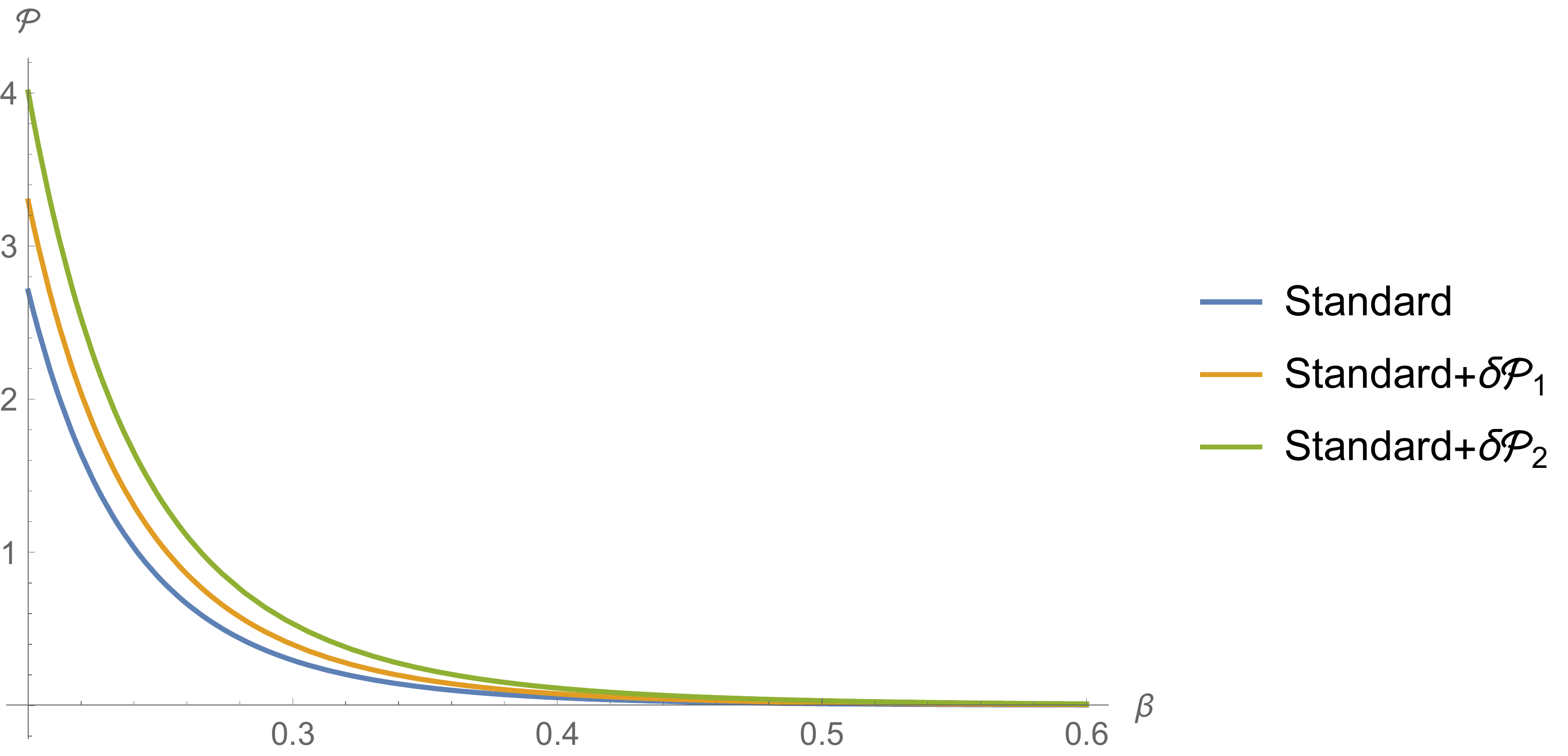}
\caption{This figure shows the general behavior of spin-degenerate case for fermions, where the same notation as in Fig. \ref{fig:PlotsNumber} is used. As one can see, the deviation from the standard results become relevant for high energy regime, which means small $\beta$. Here, it is worth to remember that temperature is given in \textrm{eV}.
}
\label{fig:GPlotsSP}
\end{figure}

\begin{figure}[b]
\centering
\includegraphics[width=8cm,height=5cm]{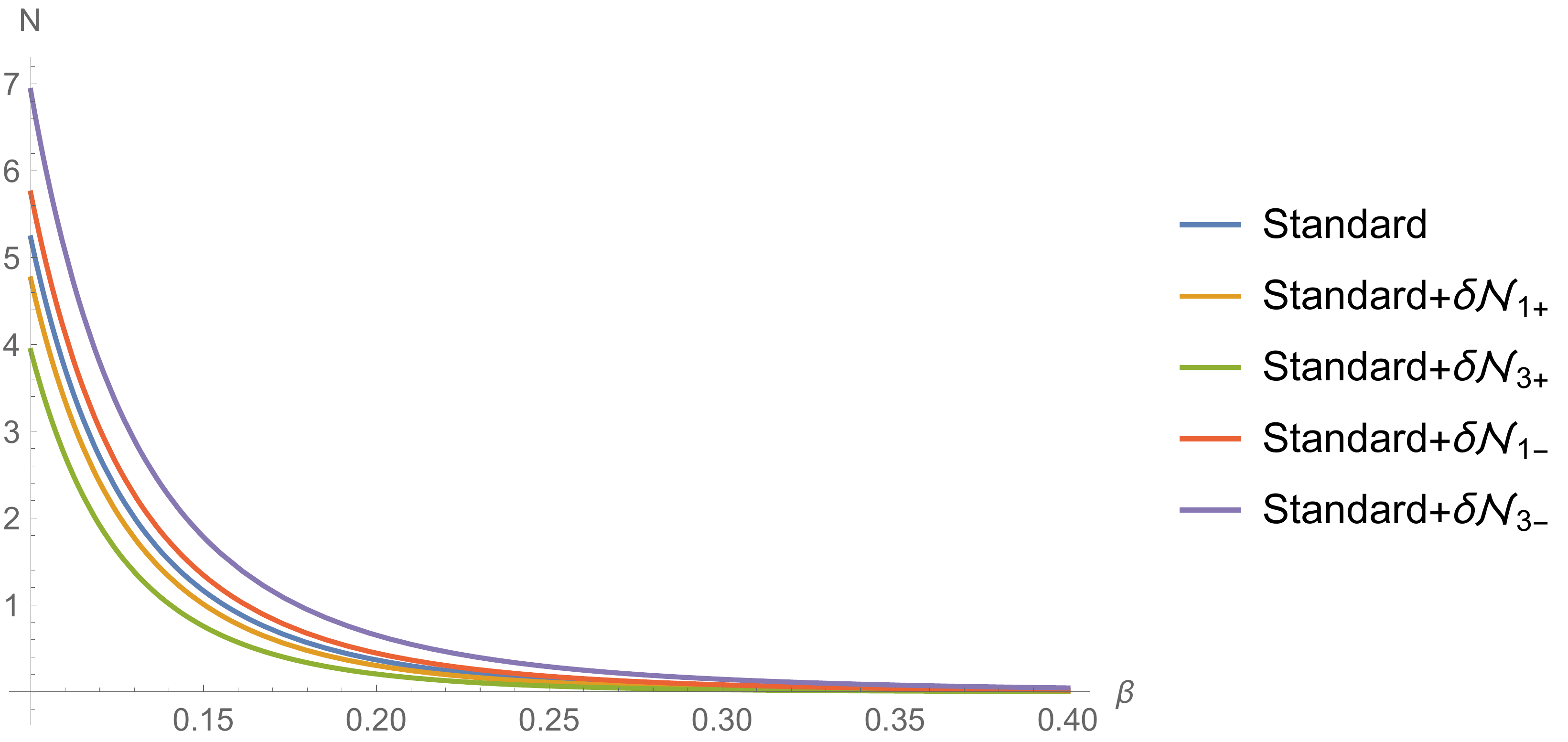}
\includegraphics[width=8cm,height=5cm]{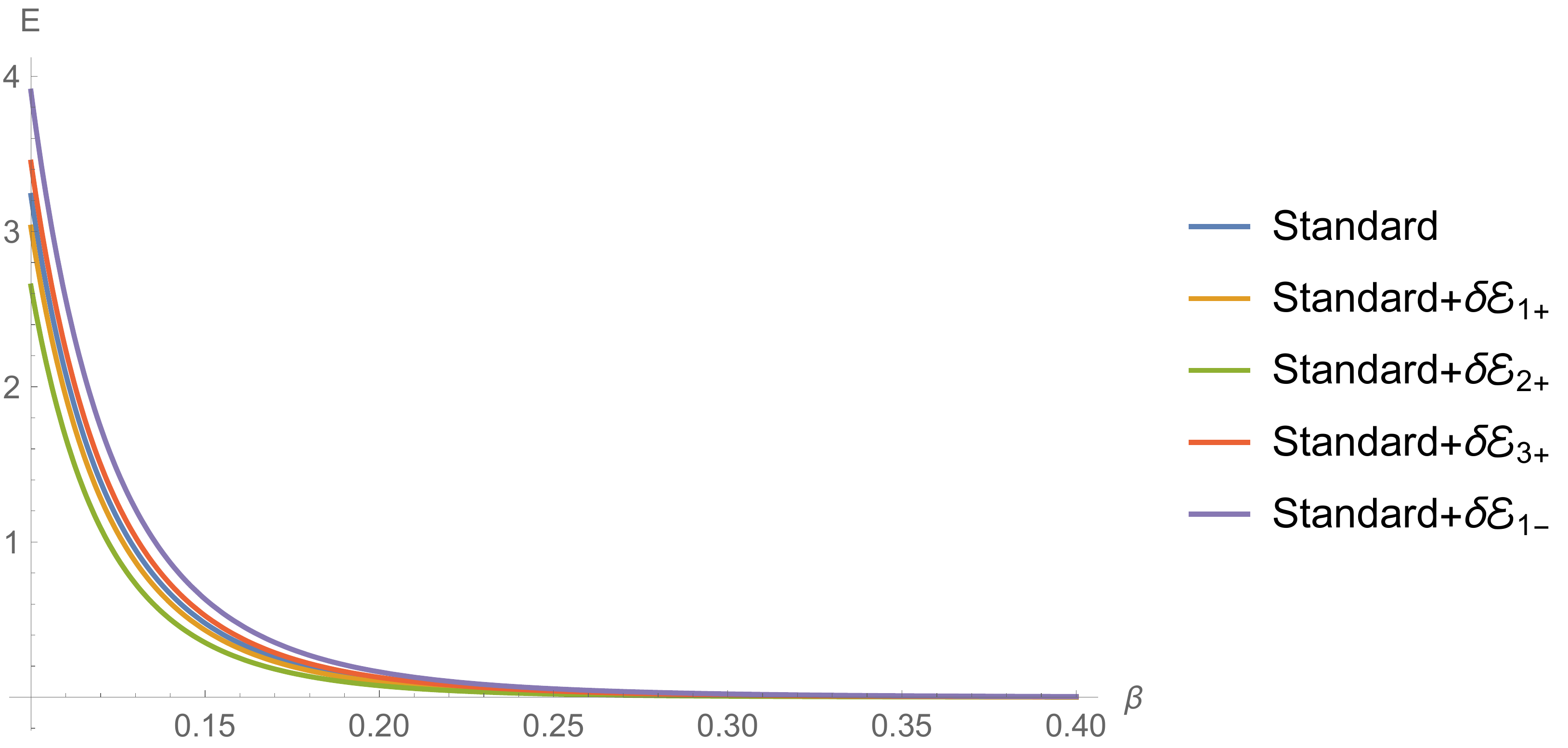}
\includegraphics[width=8cm,height=5cm]{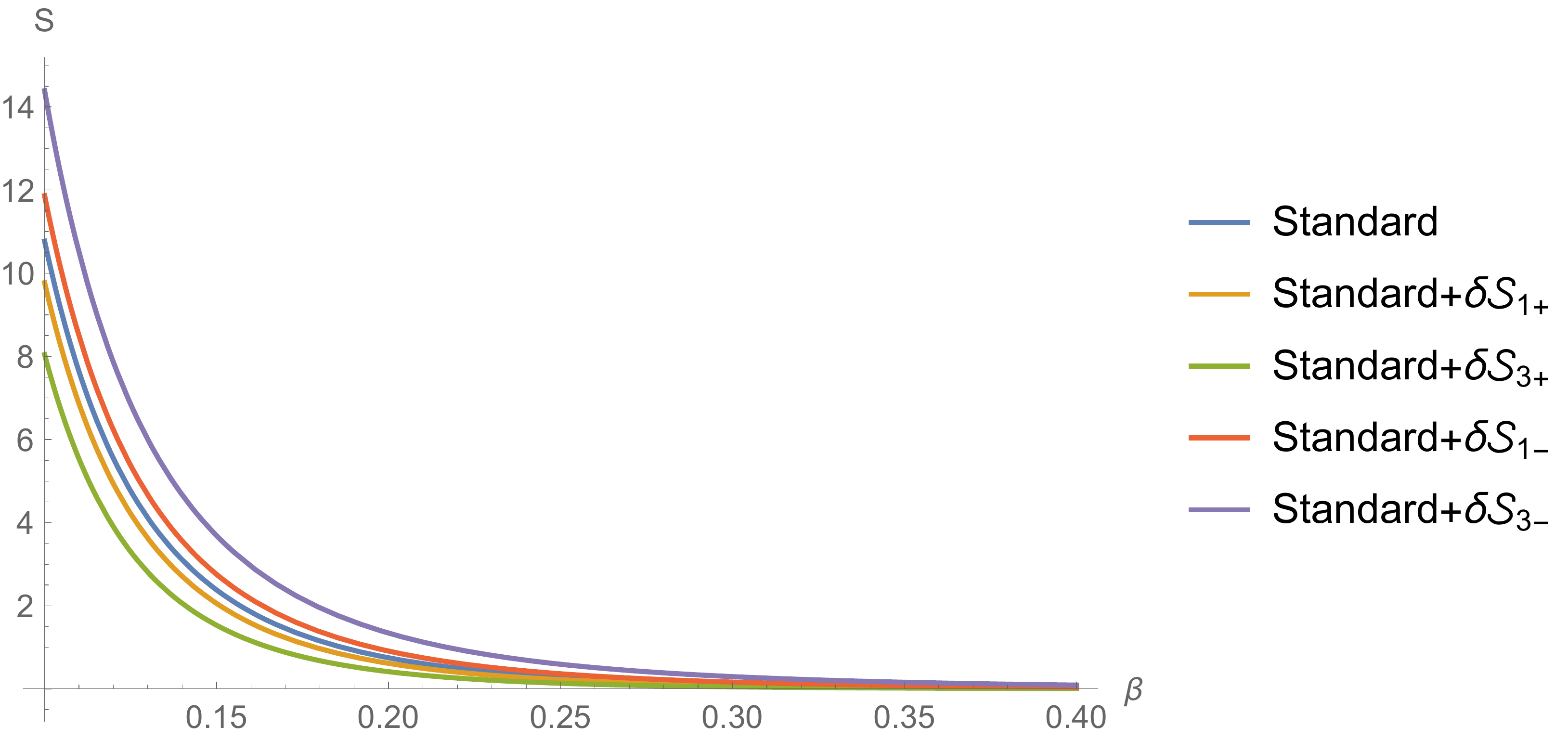}
\includegraphics[width=8cm,height=5cm]{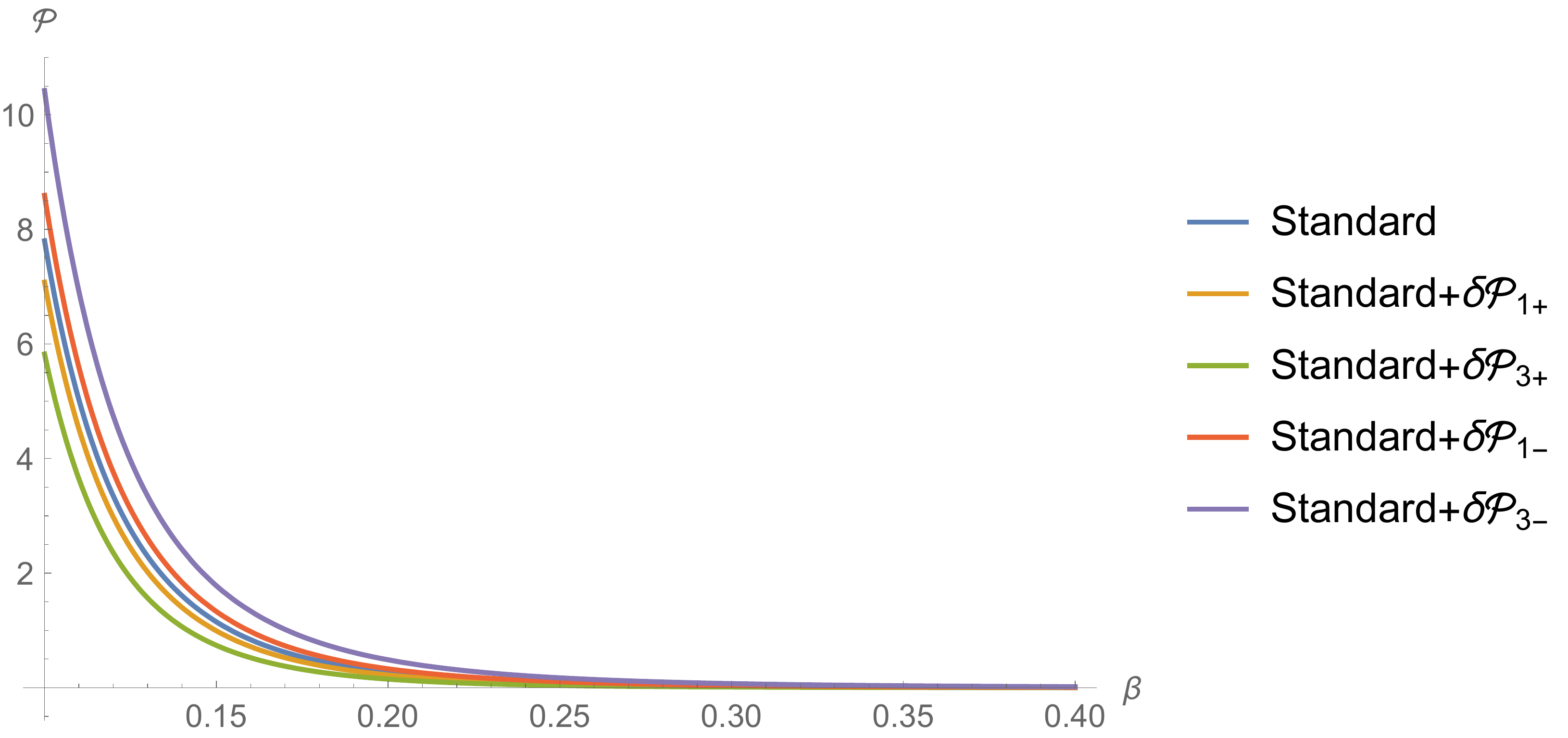}
\caption{These plots exhibit the general behavior to spin-nondegenerate case for fermions.}
\label{fig:GPlotsSNP}
\end{figure}

Another important feature to mention is that independently of the configuration that we have chosen for the spin-degenerate case, the global behavior remains the same as we can check in Fig. \ref{fig:GPlotsSP}. Analogously, the same behavior is shown for the spin-nondegenerate case exhibited in Fig. \ref{fig:GPlotsSNP}.

\section{Interacting bosons}\label{sec:BosonsInterecting}

Now, we perform a similar analysis that was done for fermions. In Table \ref{Tab:Bosons-Delta}, we summarize some definitions used from now on.
\begin{table}[h!]
  \centering
\begin{tabular}{c|c|c}
\hline
Operator & $\delta _{r}$ & Definition \\ \hline\hline
Vector & $\frac{1}{2}\left( \hat{k}_{a}\right) ^{\mu }\bar{p}_{\mu }$ & $%
\left( \hat{k}_{a}\right) ^{\kappa }=\left( k_{a}\right) ^{\left( d\right)
\kappa \alpha _{1}\ldots \alpha _{\left( d-4\right) }}\bar{p}_{\alpha
_{1}}\ldots \bar{p}_{\alpha _{\left( d-3\right) }}$ \\ \hline
Tensor & $-\left( \hat{k}_{c}\right) ^{\mu \nu }\bar{p}_{\mu }\bar{p}_{\nu }$
& $\left( \hat{k}_{c}\right) ^{\kappa \xi }=\left( k_{c}\right) ^{\left(
d\right) \kappa \xi \alpha _{1}\ldots \alpha _{\left( d-4\right) }}\bar{p}%
_{\alpha _{1}}\ldots \bar{p}_{\alpha _{\left( d-4\right) }}$%
\\\hline
\end{tabular}
\caption{This table summarizes two particular cases of $\delta_{r}$, i.e., vector and tensor operators for the boson sector. The respective definitions are exhibited as well.}\label{Tab:Bosons-Delta}
\end{table}
The first case considered is related to a vector coupling, whose non-null controlling coefficient is $\left( \hat{k}_{a}\right) ^{0}$. The dispersion relation associated with it is given below

\begin{equation}
E\approx E_{0}+\frac{1}{2}\left( \hat{k}_{a}\right) ^{0}.
\label{eq:k-a}
\end{equation}%
Here, we can perform a numerical study for such configuration. In this sense, we proceed analogously to what was accomplished in Sec. \ref{sec:fermionsInterecting}. In doing so, we obtain the results displayed in Table \ref{Tab:allboson1}.

\begin{figure}[!htb]
\centering
\subfloat [$\hat{k}_{a}$ operator]{
  \includegraphics[width=8cm,height=5cm]{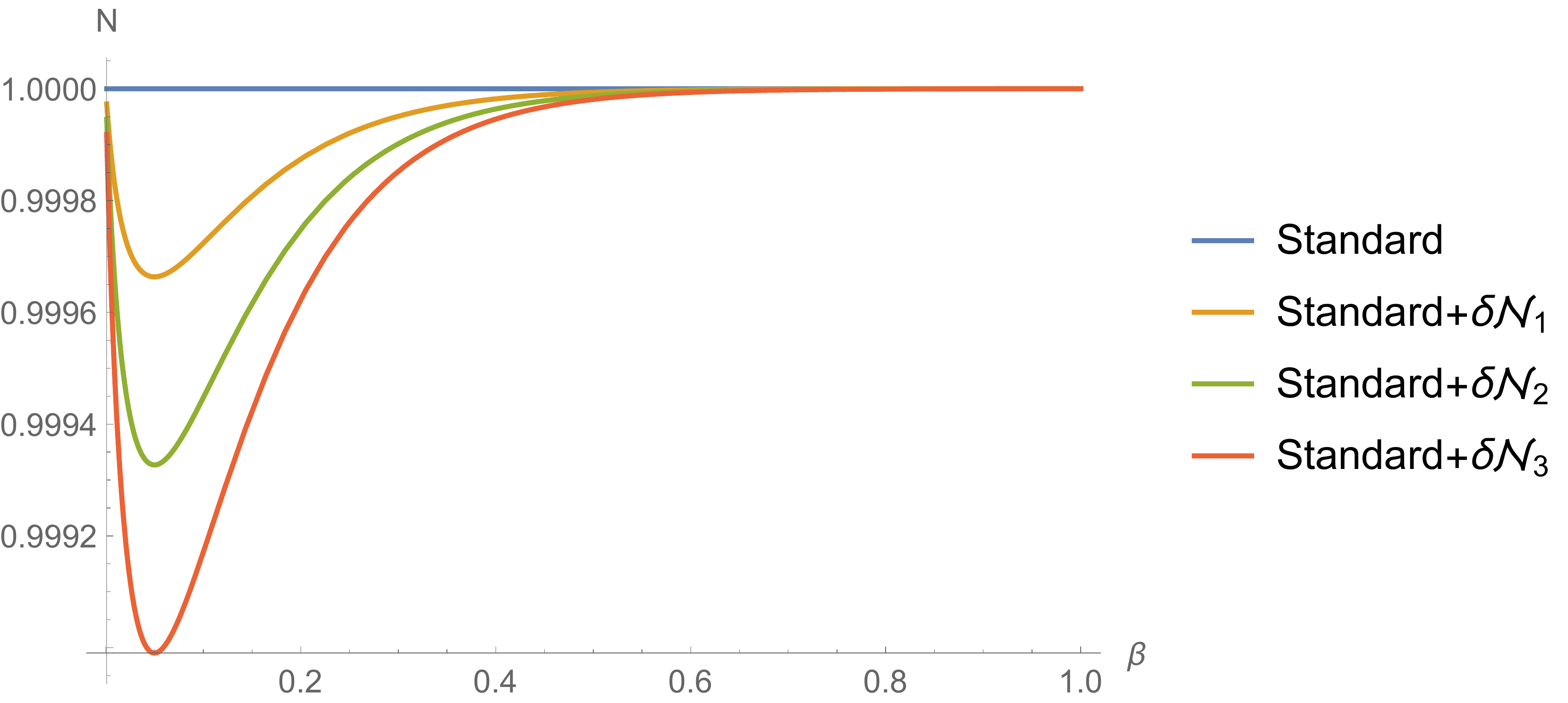}
  \label{fig:BosonC1}}
\subfloat[$\hat{k}_{c}$ operator]{
  \includegraphics[width=8cm,height=5cm]{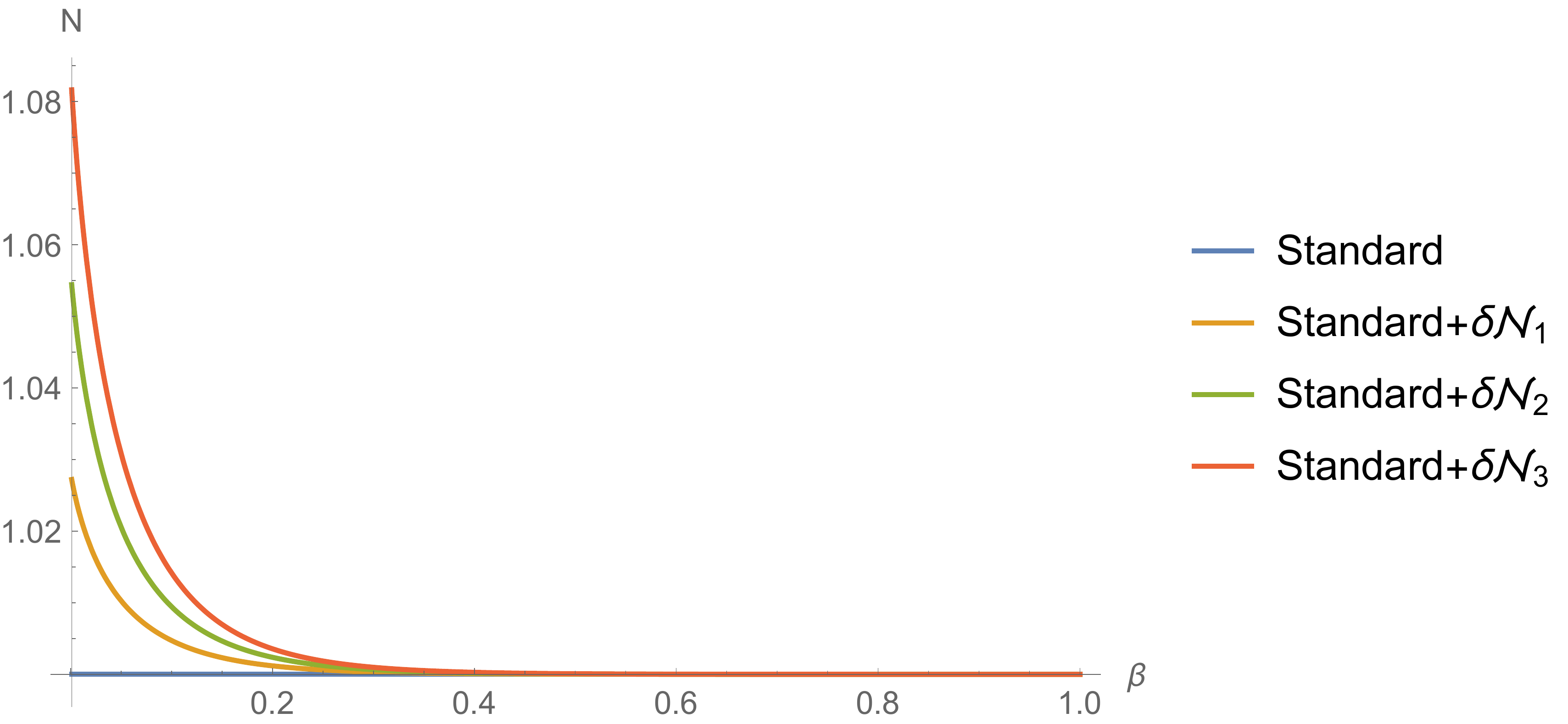}
  \label{fig:BosonC2}}
  
\caption{This figure shows a comparison between particle number to different background configurations for bosons. Here, the normalization is also applied.}
 \label{fig:BosonCase}
\end{figure}

Furthermore, the second case is the modification due to a symmetric tensor operator. We choose the non-null controlling coefficient $\left( \hat{k}_{c}\right)^{00}$, which gives rise to the following dispersion relation shown below 

\begin{equation}
E\approx E_{0}-\left( \hat{k}_{c}\right) ^{00}E_{0}.
\label{eq:k-c}
\end{equation}
With it, we can perform again a numerical analysis that produces the respective values displayed in Table \ref{Tab:allboson2}. Besides, we also show in Fig. \ref{fig:BosonCase} the behavior of the particle number for both cases considered above.

\section{Results and discussions}\label{sec:Results}

We focused on displaying the main aspects encountered in the study of relativistic interacting quantum gases for both fermion and boson sectors when Lorentz violation was taken into account. We provided some discussions based on the graphics exhibited throughout this manuscript. In a complementary way, some tables were shown as well in order to offer the reader a better comprehension of our results in Appendix \ref{sec:num-cal}. Here, we pointed out some noteworthy remarks as proceeded.

Initially, we verified that only at high temperature regime, when Lorentz violation was considered, there existed a change in the results in comparison with the standard ones. Besides, if we had considered an opposite regime, the same behavior presented in the usual case would be expected (without considering Lorentz violation). Furthermore, interesting phenomena occurred when we took into account the analysis of the particle number as we could see in Fig. \ref{fig:PlotsNumber}. We realized that for the spin-degenerate operators the mean particle number aroused when the temperature increased. The same occurred when Lorentz-violating coefficients increased. Now, with the nondegenerate spin operators for a fermion gas, we obtained an intriguing result. For spin-down particles, the mean particle number raised when the temperature reached high values. However, if instead  the spin-up particles were regarded, the mean particle number would reach values below to the usual one at the same temperature regime. Also, in this context of nondegenerate spin operators, we acquired another remarkable feature. As a result, the spin-up modes had lower energy values in contrast with those presented in spin-down cases, as seen in Fig. \ref{fig:GPlotsSNP}. It is also worth to mention that these differences do not imply a disbalance between spin-up and spin-down particles; it means that if we have a gas with spin-down particles, we will find a different result encountered in spin-up ones. On the other hand, if we have a gas made of a mixture of both, the effects turn out to be suppressed; this is because, in average, we have the same amount of spin-up and spin-down modes.

In general, the behavior of all quantities converged to the standard values when low temperature regime was regarded. Nevertheless, they differed from each other in the case of high temperature scenario. Here, it is important to mention that these behaviors were expected, since the Lorentz-violating coefficients are suppressed in low temperature cases. In a complementary manner, the tables were displayed exhibiting the first order corrections, which aroused from the deviations of the standard result. As a matter of fact, if we took the non-relativistic and the non-interacting limits of our calculations, we would recover the well-established results in the literature \cite{Colladay1t}.

Moreover, concerning boson modes, we saw that the mean particle number for the vector case decreased, whereas increased for the tensor case at high temperatures. This could be checked by examining the plot displayed in Fig. \ref{fig:BosonCase}. A similar analysis was also accomplished in Tables \ref{Tab:allboson1} and \ref{Tab:allboson1} to energy, entropy and pressure. Definitely, for our case, we noticed that the dispersion relation emerged from a genuine scalar theory. On the contrary, in Ref. \cite{Colladay1t}, a spin-0 boson gas was described by combining two fermions into a singlet representation of the spin group.

Furthermore, we analyzed the fermion case in the minimal SME regarding a system having instead scalar, vector, pseudovector and tensor operators. Thereby, we observed that the pseudoscalar operator played no role in the leading-order dispersion relation. From the discussion above, it emerges a straightforward question: in which manner the parameters of nonminimal SME could modify those respective thermal properties? We do not provide the answer to this question because such investigation lies beyond the scope of the current work. Nonetheless, we shall address it in an upcoming manuscript.

Here, since we were dealing with Lorentz and CPT violations, it is worth pointing out that we can use the previous descriptions to address a study considering rather the analysis of antiparticles. It is well known that the degeneracy between particles and antiparticles is broken when any of these operators have nonzero CPT-odd components. Then, we expect that there can be modifications of all properties studied previously for the antiparticle case. Nevertheless, such analysis lies beyond the scope of the current work and will be addressed in an upcoming one. Here we also would like to emphasize that the Eqs. (\ref{eq:Mean-number-N}), (\ref{eq:Entropy}), (\ref{eq:Energy}) and (\ref{eq:pressurePP}) were valid for different types of kinematic modifications\footnote{It is important to mention that the only requirement is that $\delta_{r}$ can be written in terms of momenta.}, like the ones that come from the \textit{Very Special Relativity} \cite{VSR}, for instance.

Now, we estimate the magnitude of Lorentz-violating background. To do that, we can use the experimental data from \cite{Data1}. In this reference, in the context of pions, we find the fluctuations for both particle density and mean energy. Here, we chose $u\left( n\right) =\bar{n}$, which is the simplest interaction energy function to estimate our results. Particularly, we use the following respective values to the temperature $T=115~\mathrm{MeV}$, the chemical potential $\mu=134.9~\mathrm{MeV}$ and the pion mass $m_{\pi}=139~\mathrm{MeV}$. To get the bounds, we first calculate the value of the thermodynamics functions that comes from our boson model considering the background coefficients as free variables. Then, we are able to compare our outputs with the available data\footnote{Here we point out that all the results below were calculated numerically. So, to avoid a plethora of numerical terms, we decide to omit them and show just the important results.}. In this direction, we can estimate the upper bound from the fluctuations of the thermodynamic functions. Thereby, we look toward to obtain the relative mean-square fluctuation in the particle density \cite{pathria32}:%
\begin{equation}
\frac{\overline{\left( \Delta n\right) ^{2}}}{\bar{n}^{2}}=\frac{kT}{\bar{N}%
^{2}}\left( \frac{\partial \bar{N}}{\partial \mu }\right) _{T,V}.
\end{equation}%
Inserting Eq.~(\ref{eq:Mean-number-N}) and the dispersion relation (\ref{eq:k-a}) in the above equation and comparing the results with the experimental data present in \cite{Data1}, we get the following upper bound for the vector background $\left( \hat{k}_{a}\right) ^{0}$
\begin{equation}
    |\left( \hat{k}_{a}\right) ^{0}| < 1.2~\times~10^{-10} .
\end{equation}%

Now, we shall examine fluctuations for the energy of a system modified by the dispersion relation (\ref{eq:k-c}). Following the usual procedure, we obtain%
\begin{equation}
\overline{\left( \Delta E\right) ^{2}}=kT^{2}\left( \frac{\partial U}{%
\partial T}\right) _{z,V}.
\end{equation}
Inserting now Eq.~(\ref{eq:mean-energy}) into the above equation and comparing it with the available data in \cite{Data1}, we obtain the following upper bound for the configuration $\left( \hat{k}_{c}\right) ^{00}$, namely,
\begin{equation}
 |\left( \hat{k}_{c}\right) ^{00}| < 1.8~\times~10^{-8}. 
\end{equation}

As we can see from both bounds obtained so far, the contribution that arises due to Lorentz-violating coefficients are very slight. Such contribution becomes even smaller when one takes into account a quadratic interaction energy $u\left( n\right) =\bar{n}^2$.



\section{Applications}\label{applications}

Here, we propose some feasible applications for our model concerning the quantum gases developed in this manuscript. It is important to highlight that the only requirement established is that the term $\delta_{r}$, in Eq. (\ref{eq:GDEquation}), be dependent only on the momenta. In this way, we can apply our model to address such applications. Particularly, we look forward to two different scenarios involving \textit{phosphorene}, and \textit{spin precession}. 

At the beginning, when one considers the measurements of some experimental physics, inevitable one stumbles upon the observable operators. In our case, following the ideas shown in Ref. \cite{kostelecky2013fermions}, we consider the Hamiltonian operator in order to perform the following investigations. In doing so, we implement the decomposition of the Hamiltonian in terms of the spherical harmonics. With this, we can perform a coherent categorization of those previous coefficients which influenced the dynamic of particle modes studied so far. Moreover, it is opportune because rotation violations are a key signature of Lorentz violation.

Nevertheless, an exact expression to Hamiltonian is a challenge since we are dealing with higher-order derivative operators. To overtake this situation, we consider the Hamiltonian as being
\ie
\mathcal{H}=\mathcal{H}_{0} + \delta \mathcal{H},
\fe
where the nonpertubatibed form of the Hamiltonian is 
\ie
\mathcal{H}_{0}=\gamma_{0}({\bf{p \cdot \gamma}} + m_{\psi})
\fe
and its respective perturbated version is written as
\ie
\begin{split}
\delta \mathcal{H} = \frac{1}{E_{0}} &\left[  m_{\psi} \hat{\mathcal{S}} \gamma_{5} - E_{0} \hat{\mathcal{V}}^{0} -\hat{\mathcal{V}}^{j}p^{j}\gamma_{5} + \hat{\mathcal{A}}^{0} p^{j}\gamma^{j}\gamma_{0} + m_{\psi}\hat{\mathcal{A}}^{j}\gamma^{j}\gamma_{0}\gamma_{5} + \frac{\hat{\mathcal{A}}^{j}p^{j}p^{k}\gamma^{k}\gamma_{0}\gamma_{5}}{E_{0}+m_{\psi}} \right.\\
& \left.
+ i p^{j}\hat{\mathcal{T}}^{0k}\gamma^{j}\gamma^{k} + i\hat{\mathcal{T}}^{0j}p^{j} - E_{0} \tilde{\hat{\mathcal{T}}}^{0j}\gamma_{0} + \frac{\tilde{\hat{\mathcal{T}}}^{0j}p^{j}p^{k}\gamma^{k}\gamma_{0}}{E_{0}+ m_{\psi}}         \right]
\end{split}.
\fe
It is important to note that the, from the above expression, we shall only consider the leading order, namely, $\delta \mathfrak{h}$. In this way, we can obtain therefore
\ie
\begin{split}
\delta \mathfrak{h} &= \mathfrak{h}_{a} + \mathfrak{h}_{c} + \mathfrak{h}_{g} + \mathfrak{h}_{H}, \\
& = \frac{1}{E_{0}} \left( \hat{a}^{\nu}_{\text{eff}} - \hat{c}^{\nu}_{\text{eff}} - \tilde{\hat{g}}^{\mu\nu}_{\text{eff}}\tau_{\mu} +  \tilde{\hat{H}}^{\mu\nu}_{\text{eff}}\tau_{\mu} \right)p_{\nu},
\end{split}
\fe
where the explicit definitions of $\hat{a}^{\nu}_{\text{eff}}$, $\hat{c}^{\nu}_{\text{eff}}$, $\tilde{\hat{g}}^{\mu\nu}_{\text{eff}}$ and  $\tilde{\hat{H}}^{\mu\nu}_{\text{eff}}$ can properly be encountered in Ref. \cite{kostelecky2013fermions}. In the subsection \ref{spin-precession}, we take into account the advantage of using such approach to address an application involving the Lamor-like precession. As we shall see in what follows, the components $\mathfrak{h}_{g}$ and $\mathfrak{h}_{H}$ gives rise to a new phenomenon due to the Lorentz violation, the birefringence.


\subsection{Phosphorene} \label{phosphorene}

In this case, we can encompass both particle modes, namely, fermions (electrons) and bosons (phonons), in order to address our possible applications. Given the existence of some well-known approximations, the electrons of a metal can be assumed to be a gas, as they are effectively free particles \cite{lee1988development,araujo2017,aa1,aa2,aa3,silva2018}. Therefore, our model, concerning the theory of ensembles, fits pretty well to this case. 

Moreover, there can exist a promising aspect to be investigated in condensed matter physics, which is the thermal properties of anisotropic systems considering Lorentz violation.  Such anisotropy may reveal new phenomena which might be in principle confronted with experimental physics.  In  the case  of \textit{phosphorene},  one can  assume electrons  to have an  anisotropic effective mass \cite{cunha2020}, which raises many possibilities:  if we confine these electrons in a box, collisions  with  the  walls  will  depend  on  the  angle  that  the  wall  makes  with  anisotropy direction.  More so, \textit{phonons} behave likewise in such systems.

Therefore, our proposal is to investigate how the coefficients, which triggers the Lorentz violation pointed out in Eqs. (\ref{eq:Operators} - \ref{coeffi3a}), (\ref{coeffi21}) and (\ref{coeffi2}), can disturb the system in order to uncover some fingerprints of any signal of the Lorentz violation in the context of phosphorene.

\subsection{Spin precession}\label{spin-precession}

Taken into account the expansion in terms of spherical harmonics as argued previously, the application involving spin precession seems to be viable as well. Here, we aim at displaying the group velocity, and the fermion spin precession in order to provide additional information (parameters/data) for helping the detection of the Lorentz violation in the context of quantum gases. 

For a fermion wave packet, we can write its group velocity $\bf{\mathfrak{v}}_{g}$ as 
\ie
|{\bf{\mathfrak{v}}}_{g}| = \frac{|{\bf{p}}|}{E_{0}} + \sum_{n} (p^{2}+n m^{2}_{\psi})E_{0}^{-1-n}|{\bf{p}}|^{n-1} \times \left[a_{n}^{(d)} \mp m_{\psi}g_{n}^{(d+1)} - c_{n}^{(d)} \pm m_{\psi}H_{n}^{(d+1)}\right],
\fe
which might possibly be measured by the future experiments. Moreover, it is worth mentioning that the components $h_{g}$ and $h_{H}$ of the perturbative Hamiltonian also brings about a new phenomenon, the so-called birefringence \cite{kostelecky2016testing,kostelecky2011data,kostelecky2013fermions,kostelecky2007,kostelecky2013constraints}.
This means that, depending on the direction under consideration, our system configuration may behave differently; it entails that there can possibly exist some modifications in its respective thermal properties.

Furthermore, it can also be perceived in another remarkable way; we shall interpret such aspect as being a Larmor-like precession of the spin operator $\hat{\bf{S}}$. Thereby, its dynamics is given by
\ie
\frac{\mathrm{d}}{\mathrm{d}t} \bra{\psi_{i}}\hat{\bf{S}}\ket{\psi_{j}} = -i \bra{\psi_{i}} [h,\hat{\bf{S}}]\ket{\psi_{j}} \approx 2 (h_{g} + h_{H}) \times \bra{\psi_{i}}\hat{\bf{S}}\ket{\psi_{j}},
\fe
where we can interpret the term in the parenthesis $2 (h_{g} + h_{H})$ as precession frequency $\omega$ \cite{kostelecky2013fermions}. Therefore, based on the advantage of using our thermodynamic model of quantum gases, we propose the investigation of such frequency for the sake of probing the existence of Lorentz violation.

\pagebreak
\section{Conclusion}\label{sec:conclusion}

This work aimed at studying the thermodynamic aspects of interacting quantum gases when Lorentz violation took place. We investigated the physical consequences of taking into account both fermion and boson sectors. In order to proceed further, we have utilized the grand canonical ensemble as the starting point. Next, we obtained the so-called grand canonical partition function to address analytically all calculations of interest, namely, the particle number, the mean total energy, the entropy and the pressure. Moreover, the first three ones turned out to behave as extensive quantities although the presence of Lorentz-violating terms.  

For fermion modes, we considered a system provided by scalar, vector, pseudovector and tensor operators. Particularly, the latter two scenarios exhibited an absence of the spin degeneracy. In this way, the system turned out to acquire greater energy and particle number for spin-down particle modes in comparison with spin-down ones. Besides, pseudoscalar operator played no role at the leading order dispersion relation studied here. On the other hand, for boson particles at high temperature regime the system showed that the particle number for the vector case decreased, while increased for the tensor case. More so, we analysed the magnitude of Lorentz-violating coefficients estimating the upper bounds for the bosonic case; we proposed some possible applications in order to corroborate our results in two different contexts: \textit{phosphorene} and \textit{spin precession}.

The physical consequences of such thermal analysis of relativistic interacting quantum gases involving fermion and boson particles might possibly reveal new fingerprints of a hidden physical experimental data which might be measured by future experiments in the existence of Lorentz violation. Thereby, this theoretical proposition can lead to a toy model for further promising studies to search for any trace of Lorentz violation. In addition, being expected to go beyond the current analysis presented in this work, our procedure of treating a modified relativistic energy for an arbitrary quantum state, as long as there exist only momenta involved, may lead to further different investigations and applications depending on the scenario worked out.

As future perspectives, an analogous study for the thermodynamic functions regarding rather the nonminimal SME appears to be worthy to examine. Additionally, analyzing how the phase transition occurs for the minimal and nonminimal SME as well as investigating the implications of a system having an ensemble of antiparticles seem to be interesting open questions to be studied. These and other ideas are now under development. 


\section*{Acknowledgments}
\hspace{0.5cm}

The authors would like to express their gratitude to Conselho Nacional de Desenvolvimento Cient\'{\i}fico e Tecnol\'{o}gico (CNPq) - 142412/2018-0, Coordenação de Aperfeiçoamento de Pessoal de Nível Superior (CAPES) - Finance Code 001, and CAPES-PRINT (PRINT - PROGRAMA INSTITUCIONAL DE INTERNACIONALIZAÇÃO) - 88887.508184/2020-00 the for financial support. In addition, the authors thank L.L. Mesquita, A.Y. Petrov, and J.A. Helayël-Neto for the careful reading of this manuscript; moreover, the authors would like to express their gratitude to the anonymous referee for the suggestions and João Milton for having addressed some fruitful applications for our work. Particularly, A. A. Araújo Filho acknowledges the Facultad de Física - Universitat de València and Gonzalo J. Olmo for the kind hospitality when part of this work was made.

\appendix 

\section{Are they still extensive state quantities?}\label{sec:Therm-Limit}

Here, to corroborate our results, we proceed further for the sake of verifying the validity of the derived relations in the thermodynamic limit. Indeed, we need to take $N\rightarrow \infty $, $%
V\rightarrow \infty $ and $N/V=$ const. Being proportional to the volume, the mean particle number, the entropy and the mean energy turn out to be extensive quantities in the ordinary case. Nevertheless, knowing whether the Lorentz violation removes such extensive property or not is an intriguing question to be checked. With this purpose, we proceed making the substitution in the following way
\begin{equation}
\sum_{r}\rightarrow \frac{gV}{\left( 2\pi \right) ^{3}}\int d^{3}\boldsymbol{%
p},
\end{equation}%
where $g$ is the degeneracy factor. Now, let us make some comments. Taking the thermodynamic limit is only reasonably supported when $u^{\prime}\left( \bar{n}\right)$ and
$u\left( \bar{n}\right)$ do not depend upon the volume. 
Additionally, this also entails that the
particle number density $\bar{n}=\bar{N}/V$ must not depend upon it. Now, let us verify this assumption starting with%
\begin{equation}
\left. \frac{\partial \bar{n}}{\partial V}\right\vert _{\mu ,T}=\left. \frac{%
\partial }{\partial V}\left( \frac{\bar{N}}{V}\right) \right\vert _{\mu ,T}=%
\frac{1}{V}\left( \left. \frac{\partial \bar{N}}{\partial V}\right\vert
_{\mu ,T}-\frac{\bar{N}}{V}\right),
\end{equation}%
which yields%
\begin{eqnarray}
\left. \frac{\partial \bar{N}}{\partial V}\right\vert _{\mu ,T} &=&\frac{gV}{%
\left( 2\pi \right) ^{3}}\int d^{3}\boldsymbol{p}\frac{1}{\exp \left\{ \beta %
\left[ \epsilon _{r}\left( \boldsymbol{p}\right) +\delta _{r}\left(
\boldsymbol{p}\right) +u^{\prime }\left( \bar{n}\right) -\mu \right]
\right\} +\chi }  \notag \\
&&+\frac{gV}{\left( 2\pi \right) ^{3}}\int d^{3}\boldsymbol{p}\left\{ -\left[
\frac{1}{\exp \left\{ \beta \left[ \epsilon _{r}\left( \boldsymbol{p}\right)
+\delta _{r}\left( \boldsymbol{p}\right) +u^{\prime }\left( \bar{n}\right)
-\mu \right] \right\} +\chi }\right] ^{2}\right\} \times   \notag \\
&&\times \exp \left\{ \beta \left[ \epsilon _{r}\left( \boldsymbol{p}\right)
+\delta _{r}\left( \boldsymbol{p}\right) +u^{\prime }\left( \bar{n}\right)
-\mu \right] \right\} \beta \left. \frac{\partial u^{\prime }\left( \bar{n}%
\right) }{\partial V}\right\vert _{\mu ,T}.
\end{eqnarray}%
In this sense, we identify the first term as $\bar{N}/V$; the second term we rewrite using%
\begin{equation}
\left. \frac{\partial u^{\prime }\left( \bar{n}\right) }{\partial V}%
\right\vert _{\mu ,T}=u^{\prime \prime }\left( \bar{n}\right) \left. \frac{%
\partial \bar{n}}{\partial V}\right\vert _{\mu ,T}=\frac{u^{\prime \prime
}\left( \bar{n}\right) }{V}\left( \left. \frac{\partial \bar{N}}{\partial V}%
\right\vert _{\mu ,T}-\frac{\bar{N}}{V}\right) ,
\end{equation}%
as%
\begin{equation*}
-\sum_{r}\bar{n}_{r}\frac{\beta }{V}u^{\prime \prime }\left( \bar{n}\right)
\left( \left. \frac{\partial \bar{N}}{\partial V}\right\vert _{\mu ,T}-\frac{%
\bar{N}}{V}\right) \exp \left\{ \beta \left[ \epsilon _{r}\left( \boldsymbol{%
p}\right) +\delta _{r}\left( \boldsymbol{p}\right) +u^{\prime }\left( \bar{n}%
\right) -\mu \right] \right\},
\end{equation*}%
and to finish we obtain%
\begin{eqnarray}
\left. \frac{\partial \bar{n}}{\partial V}\right\vert _{\mu ,T} &=&\frac{1}{V%
}\left( \left. \frac{\partial \bar{N}}{\partial V}\right\vert _{\mu ,T}-%
\frac{\bar{N}}{V}\right)   \notag \\
&=&\frac{1}{V}\left( -\sum_{r}\bar{n}_{r}\frac{\beta }{V}u^{\prime \prime
}\left( \bar{n}\right) \exp \left\{ \beta \left[ \epsilon _{r}\left(
\boldsymbol{p}\right) +\delta _{r}\left( \boldsymbol{p}\right) +u^{\prime
}\left( \bar{n}\right) -\mu \right] \right\} \right) \times   \notag \\
&&\times \left( \left. \frac{\partial \bar{N}}{\partial V}\right\vert _{\mu
,T}-\frac{\bar{N}}{V}\right) .
\end{eqnarray}%
Additionally, this relation is verified if
\begin{equation}
1=-\sum_{r}\bar{n}_{r}\frac{\beta }{V}\exp \left\{ \beta \left[ \epsilon
_{r}\left( \boldsymbol{p}\right) +\delta _{r}\left( \boldsymbol{p}\right)
+u^{\prime }\left( \bar{n}\right) -\mu \right] \right\} u^{\prime \prime
}\left( \bar{n}\right) .
\end{equation}%
Nevertheless, in a general case, this is not true for the reason that $u\left( n\right)$ is an absolutely arbitrary interaction potential density. As a matter of fact, it must hold that%
\begin{equation}
\left. \frac{\partial \bar{N}}{\partial V}\right\vert _{\mu ,T}=\frac{\bar{N}%
}{V},
\end{equation}%
i.e., $\left. \frac{\partial \bar{n}}{\partial V}\right\vert _{\mu ,T}$ must vanish. Finally, we should notice that, since $\delta _{r}\left( \boldsymbol{p}%
\right)$ is a function only of $\boldsymbol{p}$, it does not mess up the extensive property. Therefore, for such thermal properties, even in the presence of Lorentz violation, the extensive characteristic of the system is maintained as well. 

\section{Numerical analyses}\label{sec:num-cal}

Here, we provide such Appendix to exhibit a concise explanation for the numerical calculations encountered throughout this manuscript. We show the thermal quantities, namely, the energy, the mean particle number and the entropy per volume, for different values of $\beta$. Besides, $\mathcal{E}$, $\mathfrak{N}$ and $\mathfrak{S}$ are quantities representing the energy, the mean particle number as well as the entropy per volume respectively. The outputs for fermions and bosons modes are displayed as follows:

\begin{table}[tbh]
  \centering
  \scalebox{0.57}{
\begin{tabular}{|c|c|c|c|c|c|c|c|c|c|c|c|c|}
\hline\hline
 $\beta$ & $\mathcal{E}$ & $\delta \mathcal{E}_{1}$ & $\delta \mathcal{E}_{2}$ & $\delta \mathcal{E}_{3}$ & $\mathfrak{N}$ & $\delta \mathfrak{N}_{1}$ & $\delta \mathfrak{N}_{2}$ & $\delta \mathfrak{N}_{3}$ & $\mathfrak{S}$ & $\delta \mathfrak{S}_{1}$ & $\delta \mathfrak{S}_{2}$ & $\delta \mathfrak{S}_{3}$\\ \hline\hline
0.1 & 3231.95 & 0.104598 & 0.00104598 & 0.0000104598 & 104.543 & -0.0000416498 & -4.16498 $\times 10^{-7}$ & -4.16498 $\times 10^{-9}$ & 734.947 & 0.00522714 & 0.0000522714 & 5.22714 $10^{-7}$  \\ \hline
0.2 & 112.314 & 0.00736482 & 0.0000736482 & 7.36482 $\times 10^{-7}$ & 7.34833 & -1.99713 $\times 10^{-7}$ & -1.99713  $\times 10^{-8}$ & -1.99713 $\times 10^{-10}$ & 51.5659 & 0.000734833 & 7.34833 $\times 10^{-6}$  & 7.34833 $\times 10^{-8}$\\ \hline
0.3 & 12.2594 & 0.00121608 & 0.0000121608 & 1.21608 $\times 10^{-7}$ & 1.20978 & -1.58032 $\times 10^{-7}$  & -1.58032 $\times 10^{-9}$ & -1.58032 $\times 10^{-11}$& 8.48027 & 0.000181467  & 1.81467 $\times 10^{-6}$& 1.81467 $\times 10^{-8}$\\ \hline
0.4 & 2.13516 & 0.000284015  & 2.84015 $\times 10^{-6}$ & 2.84015 $\times 10^{-8}$ & 0.28138 & -1.50433 $\times 10^{-8}$  & -1.50433 $\times 10^{-10}$& -1.50433 $\times 10^{-12}$& 1.97117 & 0.0000562759 & 5.62759 $\times 10^{-7}$& 5.62759 $\times 10^{-9}$\\ \hline
0.5 & 0.480237 & 0.0000801838 & 8.01838 $\times 10^{-7}$ & 8.01838 $\times 10^{-9}$ & 0.0790277 & -1.5763 $\times 10^{-9}$ & -1.5763 $\times 10^{-11}$ & -1.5763 $\times 10^{-13}$ & 0.553427 & 0.0000197569 & 1.97569 $\times 10^{-7}$ & 1.97569 $\times 10^{-9}$\\ \hline 
0.6 & 0.126973 & 0.000025527 & 2.5527 $\times 10^{-7}$& 2.5527 $\times 10^{-9}$ & 0.0250043 & -1.75433 $\times 10^{-10}$ & -1.75433 $\times 10^{-12}$& -1.75433 $\times 10^{-14}$ & 0.17507 & 7.50128 $\times 10^{-6}$ & 7.50128 $\times 10^{-8}$& 7.50128 $\times 10^{-10}$\\ \hline 
0.7 & 0.0375352 & 8.83081 $\times 10^{-6}$& 8.83081 $\times 10^{-8}$& 8.83081 $\times 10^{-10}$ & 0.00858928 & -2.03663 $\times 10^{-11}$& -2.03663 $\times 10^{-13}$& -2.03663 $\times 10^{-15}$& 0.0601324 & 3.00625 $\times 10^{-6}$& 3.00625 $\times 10^{-8}$& 3.00625 $\times 10^{-10}$\\ \hline
0.8 & 0.0120401 & 3.24695 $\times 10^{-6}$& 3.24695 $\times 10^{-8}$& 3.24695 $\times 10^{-10}$ & 0.00313348 & -2.4398 $\times 10^{-12}$& -2.4398 $\times 10^{-14}$& -2.4398 $\times 10^{-16}$& 0.0219358 & 1.25339 $\times 10^{-6}$& 1.25339 $\times 10^{-8}$ & 1.25339 $\times 10^{-10}$\\ \hline
0.9 & 0.0041105 & 1.25082 $\times 10^{-6}$ & 1.25082 $\times 10^{-8}$ & 1.25082 $\times 10^{-10}$ & 0.00119682 & -2.99456 $\times 10^{-13}$& -2.99456 $\times 10^{-15}$& -2.99456 $\times 10^{-17}$& 0.00838577 & 3.70143 $\times 10^{-6}$& 3.70143 $\times 10^{-8}$& 3.70143 $\times 10^{-10}$\\ \hline
1.0 & 0.00147396 & 4.99909 $\times 10^{-7}$& 4.99909 $\times 10^{-9}$& 4.99909 $\times 10^{-11}$& 0.000473931 & -3.74665 $\times 10^{-14}$& -3.74665 $\times 10^{-16}$& -3.74665 $\times 10^{-18}$& 0.00331758 & 2.36965 $\times 10^{-7}$& 2.36965 $\times 10^{-9}$& 2.36965 $\times 10^{-11}$\\ \hline\hline
\end{tabular}}
\caption{The scalar operator concerning the fermion sector. }\label{Tab:VScalar}
\end{table}
\begin{table}[tbh]
  \centering
  \scalebox{0.57}{
\begin{tabular}{|c|c|c|c|c|c|c|c|c|c|c|c|c|}
\hline\hline
 $\beta$ & $\mathcal{E}$ & $\delta \mathcal{E}_{1}$ & $\delta \mathcal{E}_{2}$ & $\delta \mathcal{E}_{3}$ & $\mathfrak{N}$ & $\delta \mathfrak{N}_{1}$ & $\delta \mathfrak{N}_{2}$ & $\delta \mathfrak{N}_{3}$ & $\mathfrak{S}$ & $\delta \mathfrak{S}_{1}$ & $\delta \mathfrak{S}_{2}$ & $\delta \mathfrak{S}_{3}$\\ \hline\hline
0.1 & 3231.95 & 3.23195 & 0.0323195 & 0.000323195 & 104.543 & -0.00030446 & -3.0446 $\times 10^{-6}$ & -3.0446 $\times 10^{-8}$& 734.947 & 0.107691 & 0.00107691 & 0.0000107691  \\ \hline
0.2 & 112.314 & 0.112314 & 0.00112314 & 0.0000112314 & 7.34833 & -7.03273 $\times 10^{-6}$  & -7.03273 $\times 10^{-8}$ & -7.03273 $\times 10^{-10}$ & 51.5659 & 0.00747586 & 0.0000747586  & 7.47586 $\times 10^{-7}$\\ \hline
0.3 & 12.2594 & 0.0122594 & 0.000122594 & 1.22594 $\times 10^{-6}$ & 1.20978 & -3.62361 $\times 10^{-7}$  & -3.62361 $\times 10^{-9}$ & -3.62361 $\times 10^{-11}$& 8.48027 & 0.00122158 & 0.0000122158 & 1.22158 $\times 10^{-7}$\\ \hline
0.4 & 2.13516 & 0.00213516  & 0.0000213516 & 2.13516 $\times 10^{-7}$ & 0.28138 & -2.55114 $\times 10^{-8}$  & -2.55114 $\times 10^{-10}$& -2.55114 $\times 10^{-12}$& 1.97117 & 0.000282896  & 2.82896 $\times 10^{-6}$ & 2.82896 $\times 10^{-8}$\\ \hline
0.5 & 0.480237 & 0.000480237 & 4.80237 $\times 10^{-6}$ & 4.80237 $\times 10^{-8}$ & 0.0790277 & -2.12168 $\times 10^{-9}$ & -2.12168 $\times 10^{-11}$ & -2.12168 $\times 10^{-13}$ & 0.553427 & 0.0000792604 & 7.92604 $\times 10^{-7}$ & 7.92604 $\times 10^{-9}$\\ \hline 
0.6 & 0.126973 & 0.000126973 & 1.26973 $\times 10^{-6}$& 1.26973 $\times 10^{-8}$ & 0.0250043 & -1.95919 $\times 10^{-10}$ & -1.95919 $\times 10^{-12}$& -1.95919 $\times 10^{-14}$ & 0.17507 & 0.0000250442 & 2.50442 $\times 10^{-7}$& 2.50442 $\times 10^{-9}$\\ \hline 
0.7 & 0.0375352 & 0.0000375352 & 3.75352 $\times 10^{-7}$& 3.75352 $\times 10^{-9}$ & 0.00858928 & -1.94491 $\times 10^{-11}$& -1.94491 $\times 10^{-13}$& -1.94491 $\times 10^{-15}$& 0.0601324 & 8.59669 $\times 10^{-6}$& 8.59669 $\times 10^{-8}$& 8.59669 $\times 10^{-10}$\\ \hline
0.8 & 0.0120401 & 0.0000120401 & 1.20401 $\times 10^{-7}$& 1.20401 $\times 10^{-9}$ & 0.00313348 & -2.0361 $\times 10^{-12}$& -2.0361 $\times 10^{-14}$& -2.0361 $\times 10^{-16}$& 0.0219358 & 3.13493 $\times 10^{-6}$& 3.13493 $\times 10^{-8}$ & 3.13493 $\times 10^{-10}$\\ \hline
0.9 & 0.0041105 & 4.1105 $\times 10^{-6}$ & 4.1105 $\times 10^{-8}$ & 4.1105 $\times 10^{-10}$ & 0.00119682 & -2.21989 $\times 10^{-13}$& -2.21989 $\times 10^{-15}$& -2.21989 $\times 10^{-17}$& 0.00837801 & 1.19711 $\times 10^{-6}$& 1.19711 $\times 10^{-8}$& 1.19711 $\times 10^{-10}$\\ \hline
1.0 & 0.00147396 & 1.47396 $\times 10^{-6}$& 1.47396 $\times 10^{-8}$& 1.47396 $\times 10^{-10}$& 0.000473931 & -2.49878 $\times 10^{-14}$& -2.49878 $\times 10^{-16}$& -2.49878 $\times 10^{-18}$& 0.00331758 & 4.739949 $\times 10^{-7}$& 4.73994 $\times 10^{-9}$& 4.73994 $\times 10^{-11}$\\ \hline\hline
\end{tabular}}
\caption{The vector operator concerning the fermion sector.}\label{Tab:VVector}
\end{table}

\pagebreak

\begin{table}[tbh]
  \centering
  \scalebox{0.57}{
\begin{tabular}{|c|c|c|c|c|c|c|c|c|c|c|c|c|}
\hline\hline
 $\beta$ & $\mathcal{E}$ & $\delta \mathcal{E}_{1}$ & $\delta \mathcal{E}_{2}$ & $\delta \mathcal{E}_{3}$ & $\mathfrak{N}$ & $\delta \mathfrak{N}_{1}$ & $\delta \mathfrak{N}_{2}$ & $\delta \mathfrak{N}_{3}$ & $\mathfrak{S}$ & $\delta \mathfrak{S}_{1}$ & $\delta \mathfrak{S}_{2}$ & $\delta \mathfrak{S}_{3}$\\ \hline\hline
0.1 & 3231.95 & -0.209193 & -0.00209193 & -0.0000209193 & 104.543 & 0.0000829947 & 8.29947 $\times 10^{-7}$ & 8.29947 $\times 10^{-9}$& 734.947 & -0.010449 & -0.00010449 & -1.0449 $\times 10^{-6}$\\ \hline
0.2 & 112.314 & -0.014728 & -0.00014728 & -1.4728 $\times 10^{-6}$ & 7.34833 & 3.93882$\times 10^{-6}$ & 3.93882 $\times 10^{-8}$ & 3.93882 $\times 10^{-10}$ & 51.5659 & -0.00146669 &  -0.0000146669  & -1.46669$\times 10^{-7}$\\ \hline
0.3 & 12.2594 & -0.00243122 & -0.0000243122 & -2.43122 $\times 10^{-7}$ & 1.20978 & 3.07031 $\times 10^{-7}$  & 3.07031 $\times 10^{-9}$ & 3.07031 $\times 10^{-11}$& 8.48027 &  -0.000361314  & -3.61314 $\times 10^{-6}$ & -3.61314 $\times 10^{-8}$\\ \hline
0.4 & 2.13516 & -0.000567521  & -5.67521 $\times 10^{-6}$ & -5.67521 $\times 10^{-8}$ & 0.28138 & 2.87129 $\times 10^{-8}$  & 2.87129 $\times 10^{-10}$& 2.87129 $\times 10^{-12}$& 1.97117 & -0.000111689  & -1.11689 $\times 10^{-6}$& -1.11689 $\times 10^{-8}$\\ \hline
0.5 & 0.480237 & -0.000160098 & -1.60098 $\times 10^{-6}$ & -1.60098 $\times 10^{-8}$ & 0.0790277 & 2.95182 $\times 10^{-9}$ & 2.95182 $\times 10^{-11}$ & 2.95182 $\times 10^{-13}$ & 0.553427 & -0.000039061 & -3.9061 $\times 10^{-7}$ & -3.9061 $\times 10^{-9}$\\ \hline 
0.6 & 0.126973 & -0.0000509137 & -5.09137 $\times 10^{-7}$ & -5.09137 $\times 10^{-9}$ & 0.0250043 & 3.22144 $\times 10^{-10}$ & 3.22144 $\times 10^{-12}$& 3.22144 $\times 10^{-14}$ & 0.17507 & -0.0000147665 & -1.47665 $\times 10^{-7}$& -1.47665 $\times 10^{-9}$\\ \hline 
0.7 & 0.0375352 & -0.000017589& -1.7589 $\times 10^{-7}$& -1.7589 $\times 10^{-9}$ & 0.00858928 & 3.66682 $\times 10^{-11}$& 3.66682 $\times 10^{-13}$& 3.66682 $\times 10^{-15}$& 0.0601324 & -5.8899 $\times 10^{-6}$& -5.8899 $\times 10^{-8}$& -5.8899 $\times 10^{-10}$\\ \hline
0.8 & 0.0120401 & -6.45643 $\times 10^{-6}$ & -6.45643 $\times 10^{-8}$& -6.45643 $\times 10^{-10}$ & 0.00313348 & 4.30756 $\times 10^{-12}$& 4.30756 $\times 10^{-14}$& 4.30756 $\times 10^{-16}$& 0.0219358 & -2.44322 $\times 10^{-6}$& -2.44322 $\times 10^{-8}$ & -2.44322 $\times 10^{-10}$\\ \hline
0.9 & 0.0041105 & -2.48233 $\times 10^{-6}$ & -2.48233 $\times 10^{-8}$ & -2.48233 $\times 10^{-10}$ & 0.00119682 & 5.18591 $\times 10^{-13}$& 5.18591 $\times 10^{-15}$& 5.18591 $\times 10^{-17}$& 0.00837801 & -1.04421 $\times 10^{-6}$& -1.04421 $\times 10^{-8}$& -1.04421 $\times 10^{-10}$\\ \hline
1.0 & 0.00147396 & -9.8986 $\times 10^{-7}$& -9.8986 $\times 10^{-9}$& -9.8986 $\times 10^{-11}$& 0.000473931 & 6.36654 $\times 10^{-14}$& 6.36654 $\times 10^{-16}$& 6.36654 $\times 10^{-18}$& 0.00331758 & -4.56876 $\times 10^{-7}$& -4.56876 $\times 10^{-9}$& -4.56876 $\times 10^{-11}$\\
\hline\hline
\end{tabular}}
\caption{The pseudovector operator for fermions.}\label{Tab:VPseudovector}
\end{table}
\begin{table}[tbh]
  \centering
  \scalebox{0.57}{
\begin{tabular}{|c|c|c|c|c|c|c|c|c|c|c|c|c|}
\hline\hline
 $\beta$ & $\mathcal{E}$ & $\delta \mathcal{E}_{1}$ & $\delta \mathcal{E}_{2}$ & $\delta \mathcal{E}_{3}$ & $\mathfrak{N}$ & $\delta \mathfrak{N}_{1}$ & $\delta \mathfrak{N}_{2}$ & $\delta \mathfrak{N}_{3}$ & $\mathfrak{S}$ & $\delta \mathfrak{S}_{1}$ & $\delta \mathfrak{S}_{2}$ & $\delta \mathfrak{S}_{3}$\\ \hline\hline
0.1 & 3231.95 & -9.6971 & -0.096971 & -0.00096971 & 104.543 & -0.000916101 & 9.16101 $\times 10^{-6}$ & 9.16101 $\times 10^{-8}$& 734.947 & -0.323195 & -0.00323195 & -0.0000323195\\ \hline
0.2 & 112.314 & -0.33712 & -0.0033712 & -0.000033712 & 7.34833 & 0.0000213616 & 2.13616 $\times 10^{-7}$ & 2.13616 $\times 10^{-9}$ & 51.5659 & -0.0224627 & -0.000224627 & -2.24627 $\times 10^{-6}$\\ \hline
0.3 & 12.2594 & -0.0368236 & -0.000368236 & -3.68236 $\times 10^{-6}$ & 1.20978 & 1.11777 $\times 10^{-6}$  & 1.11777 $\times 10^{-8}$ & 1.11777 $\times 10^{-10}$& 8.48027 & -0.00367781  & -0.0000367781 & -3.67781 $\times 10^{-7}$\\ \hline
0.4 & 2.13516 & -0.00641989  & -0.0000641989 & -6.41989 $\times 10^{-7}$ & 0.28138 & 8.02973 $\times 10^{-8}$  & 8.02973 $\times 10^{-10}$& 8.02973 $\times 10^{-12}$& 1.97117 & -0.000854066  & -8.54066 $\times 10^{-6}$& -8.54066 $\times 10^{-8}$\\ \hline
0.5 & 0.480237 & -0.00144586 & -0.0000144586 & -1.44586 $\times 10^{-7}$ & 0.0790277 & 6.83726 $\times 10^{-9}$ & 6.83726 $\times 10^{-11}$ & 6.83726 $\times 10^{-13}$ & 0.553427 & -0.000240119 & -2.40119 $\times 10^{-6}$ & -2.40119 $\times 10^{-8}$\\ \hline 
0.6 & 0.126973 & -0.000382907 & -3.82907 $\times 10^{-6}$& -3.82907 $\times 10^{-8}$ & 0.0250043 & 6.47963 $\times 10^{-10}$ & 6.47963 $\times 10^{-12}$& 6.47963 $\times 10^{-14}$ & 0.17507 & -0.0000761841 & -7.61841 $\times 10^{-7}$& -7.61841 $\times 10^{-9}$\\ \hline 
0.7 & 0.0375352 & -0.000113414 & -1.13414 $\times 10^{-6}$& -1.13414 $\times 10^{-8}$ & 0.00858928 & 6.61241 $\times 10^{-11}$& 6.61241 $\times 10^{-13}$& 6.61241 $\times 10^{-15}$& 0.0601324 & -0.0000262746 & -2.62746 $\times 10^{-7}$& -2.62746 $\times 10^{-9}$\\ \hline
0.8 & 0.0120401 & -0.0000364624 & -3.64624 $\times 10^{-7}$&  -3.64624 $\times 10^{-9}$ & 0.00313348 & 7.12401 $\times 10^{-12}$& 7.12401 $\times 10^{-14}$& 7.12401 $\times 10^{-16}$& 0.0219358 & -9.63211 $\times 10^{-6}$& -9.63211 $\times 10^{-8}$ & -9.63211 $\times 10^{-10}$\\ \hline
0.9 & 0.0041105 & -0.0000124805 & -1.24805 $\times 10^{-7}$ & -1.24805 $\times 10^{-9}$ & 0.00119682 & 7.99893 $\times 10^{-13}$& 7.99893 $\times 10^{-15}$& 7.99893 $\times 10^{-17}$& 0.00837801 & -3.69945 $\times 10^{-6}$& -3.69945 $\times 10^{-8}$& -3.69945 $\times 10^{-10}$\\ \hline
1.0 & 0.00147396 &-4.48836 $\times 10^{-6}$& -4.48836 $\times 10^{-8}$& -4.48836 $\times 10^{-10}$& 0.000473931 & 9.27651 $\times 10^{-14}$& 9.27651 $\times 10^{-16}$& 9.27651 $\times 10^{-18}$& 0.00331758 & -1.47396 $\times 10^{-6}$& -1.47396 $\times 10^{-8}$& -1.47396 $\times 10^{-10}$\\ \hline\hline
\end{tabular}}
\caption{The tensor operator for fermions.}\label{Tab:Vtensor}
\end{table}
\begin{table}[h]
  \centering
  \scalebox{0.57}{
\begin{tabular}{|c|c|c|c|c|c|c|c|c|c|c|c|c|}
\hline\hline
 $\beta$ & $\mathcal{E}$ & $\delta \mathcal{E}_{1}$ & $\delta \mathcal{E}_{2}$ & $\delta \mathcal{E}_{3}$ & $\mathfrak{N}$ & $\delta \mathfrak{N}_{1}$ & $\delta \mathfrak{N}_{2}$ & $\delta \mathfrak{N}_{3}$ & $\mathfrak{S}$ & $\delta \mathfrak{S}_{1}$ & $\delta \mathfrak{S}_{2}$ & $\delta \mathfrak{S}_{3}$\\ \hline\hline
0.1 & 3465.22 & -0.12049 & -0.0012049 & -0.000012049 & 120.39 & 0.000110936 & 1.10936 $\times 10^{-6}$ & 1.10936 $\times 10^{-8}$ & 837.793 & -0.0060195 & -0.000060195 & -6.0195 $\times 10^{-7}$\\ \hline
0.2 & 116.632 & -0.00794586 & 0.0000794586 & -794586$\times 10^{-7}$ & 7.92365 & 3.31996  $\times 10^{-6}$ & 3.31996  $\times 10^{-8}$ & 3.31996  $\times 10^{-10}$ & 55.3038 & -0.000792365 & -7.92365  $\times 10^{-6}$ & -7.92365  $\times 10^{-8}$\\ \hline
0.3 & 12.5142 & -0.00126748 & -0.0000126748 & -126748$\times 10^{-7}$ & 1.26011 & 2.07175 $\times 10^{-7}$  & 2.07175 $\times 10^{-9}$ & 2.07175 $\times 10^{-11}$ & 8.80739 & -0.000189017  & -1.89017 $\times 10^{-6}$ & -1.89017 $\times 10^{-8}$\\ \hline
0.4 & 2.15927 & -0.000290522  & -2.90522$\times 10^{-6}$& -2.90522 $\times 10^{-8}$ & 0.287658 & 1.73861 $\times 10^{-8}$  & 1.73861 $\times 10^{-10}$& 1.73861 $\times 10^{-12}$ & 2.01198 & -0.0000575316  & -5.75316 $\times 10^{-7}$& -5.75316 $\times 10^{-9}$\\ \hline
0.5 & 0.483186 & -0.0000811842 & -8.11842 $\times 10^{-7}$ & -8.11842 $\times 10^{-9}$ & 0.0799761 & 1.70289 $\times 10^{-9}$ & 1.70289 $\times 10^{-11}$ & 1.70289 $\times 10^{-13}$& 0.559591 & -0.000019994 & -1.9994 $\times 10^{-7}$ & -1.9994 $\times 10^{-9}$\\ \hline
0.6 & 0.127398 & -0.0000257007 & -2.57007 $\times 10^{-7}$& -2.57007 $\times 10^{-9}$ & 0.0251657 & 1.82797 $\times 10^{-10}$ & 1.82797 $\times 10^{-12}$& 1.82797 $\times 10^{-14}$ & 0.176119 & -7.5497 $\times 10^{-6}$ & -7.5497 $\times 10^{-8}$& -7.5497 $\times 10^{-10}$\\ \hline
0.7 & 0.0376033 & -8.8636 $\times 10^{-6}$& -8.8636 $\times 10^{-8}$& -8.8636 $\times 10^{-10}$ & 0.00861908 & 2.08159 $\times 10^{-11}$& 2.08159 $\times 10^{-13}$& 2.08159 $\times 10^{-15}$& 0.060326 & -3.01668 $\times 10^{-6}$& -3.01668 $\times 10^{-8}$& -3.01668 $\times 10^{-10}$\\ \hline
0.8 & 0.012052 & -3.25352 $\times 10^{-6}$& -3.25352 $\times 10^{-8}$ & -3.25352 $\times 10^{-10}$ & 0.00313931 & 2.4682 $\times 10^{-12}$& 2.4682 $\times 10^{-14}$ & 2.4682 $\times 10^{-16}$ & 0.0219737 & -1.25572 $\times 10^{-6}$& -1.25572 $\times 10^{-8}$ & -1.25572 $\times 10^{-10}$\\ \hline
0.9 & 0.00411269 & -1.2522 $\times 10^{-6}$& -1.2522 $\times 10^{-8}$& -1.2522 $\times 10^{-10}$ & 0.00119801 & 3.01294 $\times 10^{-13}$& 3.01294 $\times 10^{-15}$& 3.01294 $\times 10^{-17}$& 0.00838577 & -5.39104 $\times 10^{-7}$& -5.39104 $\times 10^{-9}$& -5.39104 $\times 10^{-11}$\\ \hline
1.0 & 0.00147439 & -5.00208 $\times 10^{-7}$& -5.00208 $\times 10^{-9}$& -5.00208 $\times 10^{-11}$ & 0.000474184 & 3.75879 $\times 10^{-14}$& 3.75879 $\times 10^{-16}$& 3.75879 $\times 10^{-18}$& 0.00331922 & -2.37092 $\times 10^{-7}$& -2.37092 $\times 10^{-9}$& -2.37092 $\times 10^{-11}$\\ \hline\hline
\end{tabular}}
\caption{The vector operator in the boson sector.}\label{Tab:allboson1}
\end{table}

\pagebreak

\begin{table}[h]
  \centering
  \scalebox{0.57}{
\begin{tabular}{|c|c|c|c|c|c|c|c|c|c|c|c|c|}
\hline\hline
 $\beta$ & $\mathcal{E}$ & $\delta \mathcal{E}_{1}$ & $\delta \mathcal{E}_{2}$ & $\delta \mathcal{E}_{3}$ & $\mathfrak{N}$ & $\delta \mathfrak{N}_{1}$ & $\delta \mathfrak{N}_{2}$ & $\delta \mathfrak{N}_{3}$ & $\mathfrak{S}$ & $\delta \mathfrak{S}_{1}$ & $\delta \mathfrak{S}_{2}$ & $\delta \mathfrak{S}_{3}$\\
  \hline\hline
0.1 & 3465.22 & 10.3973 & 0.103973 & 0.00103973 & 120.39 & -0.00189399 & -0.0000189399 & -1.89399 $\times 10^{-7}$& 837.793 & 0.346522 & 0.00346522 & 0.0000346522\\ \hline
0.2 & 116.632 & 0.350106 & 0.00350106 & 0.0000350106 & 7.92365 & -0.0000313526 & -3.13526  $\times 10^{-7}$ & -3.13526 $\times 10^{-9}$ & 55.3038 & 0.0233264 & 0.000233264  & 2.33264 $\times 10^{-6}$\\ \hline
0.3 & 12.5142 & 0.037592 & 0.00037592 & 3.7592$\times 10^{-6}$ & 1.26011 & -1.37484 $\times 10^{-6}$  & -1.37484 $\times 10^{-8}$ & -1.37484 $\times 10^{-10}$& 8.80739 & 0.00375425  & 0.0000375425 & 3.75425 $\times 10^{-7}$\\ \hline
0.4 & 2.15927 & 0.00649289  & -0.0000649289& 6.49289 $\times 10^{-7}$ & 0.287658 & -8.98141 $\times 10^{-8}$  & -8.98141 $\times 10^{-10}$& -8.98141 $\times 10^{-12}$& 2.01198 & 0.000863707  & 8.63707 $\times 10^{-6}$& 8.63707 $\times 10^{-8}$\\ \hline
0.5 & 0.483186 & 0.00145484 & 0.0000145484 & 1.45484 $\times 10^{-7}$ & 0.0799761 & -7.26379 $\times 10^{-9}$ & -7.26379 $\times 10^{-11}$ & -7.26379 $\times 10^{-13}$ & 0.559591 & 0.000241593 & 2.41593 $\times 10^{-6}$ & 2.41593 $\times 10^{-8}$\\ \hline 
0.6 & 0.127398 & 0.000384207 & 3.84207 $\times 10^{-6}$& 3.84207 $\times 10^{-8}$ & 0.0251657 & -6.6944 $\times 10^{-10}$ & -6.6944 $\times 10^{-12}$& -6.6944 $\times 10^{-14}$ & 0.176119 & 0.0000764385 & 7.64385 $\times 10^{-7}$& 7.64385 $\times 10^{-9}$\\ \hline 
0.7 & 0.0376033 & -8.8636 $\times 10^{-6}$& -8.8636 $\times 10^{-8}$& -8.8636 $\times 10^{-10}$ & 0.00861908 & 2.08159 $\times 10^{-11}$& 2.08159 $\times 10^{-13}$& 2.08159 $\times 10^{-15}$& 0.060326 & -3.01668 $\times 10^{-6}$& -3.01668 $\times 10^{-8}$& -3.01668 $\times 10^{-10}$\\ \hline
0.8 & 0.0376033 & 0.000113624 & 1.13624 $\times 10^{-6}$& 1.13624 $\times 10^{-8}$ & 0.00861908 & -6.7293 $\times 10^{-11}$& -6.7293 $\times 10^{-13}$& -6.7293 $\times 10^{-15}$& 0.0219737 & 9.64159 $\times 10^{-6}$& 9.64159 $\times 10^{-8}$ & 9.64159 $\times 10^{-10}$\\ \hline
0.9 & 0.00411269 & 0.0000124874 & 1.24874 $\times 10^{-7}$ & 1.24874 $\times 10^{-9}$ & 0.00119801 & -8.03919 $\times 10^{-13}$& -8.03919 $\times 10^{-15}$& -8.03919 $\times 10^{-17}$& 0.00838577 & 3.70143 $\times 10^{-6}$& 3.70143 $\times 10^{-8}$& 3.70143 $\times 10^{-10}$\\ \hline
1.0 & 0.00147439 & 4.48971 $\times 10^{-6}$& 4.48971 $\times 10^{-8}$& 4.48971 $\times 10^{-10}$& 0.000474184 & -9.30139 $\times 10^{-14}$& -9.30139 $\times 10^{-16}$& -9.30139 $\times 10^{-18}$& 0.00331922 & 1.47439 $\times 10^{-6}$& 1.47439 $\times 10^{-8}$& -2.37092 $\times 10^{-10}$\\ \hline\hline
\end{tabular}}
\caption{The tensor operator for the boson sector.}\label{Tab:allboson2}
\end{table}



\bibliographystyle{apsrev4-1}
\bibliography{main}

\begin{thebibliography}{116}%
\makeatletter
\providecommand \@ifxundefined [1]{%
 \@ifx{#1\undefined}
}%
\providecommand \@ifnum [1]{%
 \ifnum #1\expandafter \@firstoftwo
 \else \expandafter \@secondoftwo
 \fi
}%
\providecommand \@ifx [1]{%
 \ifx #1\expandafter \@firstoftwo
 \else \expandafter \@secondoftwo
 \fi
}%
\providecommand \natexlab [1]{#1}%
\providecommand \enquote  [1]{``#1''}%
\providecommand \bibnamefont  [1]{#1}%
\providecommand \bibfnamefont [1]{#1}%
\providecommand \citenamefont [1]{#1}%
\providecommand \href@noop [0]{\@secondoftwo}%
\providecommand \href [0]{\begingroup \@sanitize@url \@href}%
\providecommand \@href[1]{\@@startlink{#1}\@@href}%
\providecommand \@@href[1]{\endgroup#1\@@endlink}%
\providecommand \@sanitize@url [0]{\catcode `\\12\catcode `\$12\catcode
  `\&12\catcode `\#12\catcode `\^12\catcode `\_12\catcode `\%12\relax}%
\providecommand \@@startlink[1]{}%
\providecommand \@@endlink[0]{}%
\providecommand \url  [0]{\begingroup\@sanitize@url \@url }%
\providecommand \@url [1]{\endgroup\@href {#1}{\urlprefix }}%
\providecommand \urlprefix  [0]{URL }%
\providecommand \Eprint [0]{\href }%
\providecommand \doibase [0]{http://dx.doi.org/}%
\providecommand \selectlanguage [0]{\@gobble}%
\providecommand \bibinfo  [0]{\@secondoftwo}%
\providecommand \bibfield  [0]{\@secondoftwo}%
\providecommand \translation [1]{[#1]}%
\providecommand \BibitemOpen [0]{}%
\providecommand \bibitemStop [0]{}%
\providecommand \bibitemNoStop [0]{.\EOS\space}%
\providecommand \EOS [0]{\spacefactor3000\relax}%
\providecommand \BibitemShut  [1]{\csname bibitem#1\endcsname}%
\let\auto@bib@innerbib\@empty
\bibitem [{\citenamefont {Judes}\ and\ \citenamefont {Visser}(2003)}]{STR1}%
  \BibitemOpen
  \bibfield  {author} {\bibinfo {author} {\bibfnamefont {S.}~\bibnamefont
  {Judes}}\ and\ \bibinfo {author} {\bibfnamefont {M.}~\bibnamefont {Visser}},\
  }\href@noop {} {\bibfield  {journal} {\bibinfo  {journal} {Phys. Rev. D}\
  }\textbf {\bibinfo {volume} {68}},\ \bibinfo {pages} {045001} (\bibinfo
  {year} {2003})}\BibitemShut {NoStop}%
\bibitem [{\citenamefont {Robertson}(1949)}]{STR2}%
  \BibitemOpen
  \bibfield  {author} {\bibinfo {author} {\bibfnamefont {H.~P.}\ \bibnamefont
  {Robertson}},\ }\href@noop {} {\bibfield  {journal} {\bibinfo  {journal}
  {Rev. Mod. Phys.}\ }\textbf {\bibinfo {volume} {21}},\ \bibinfo {pages} {378}
  (\bibinfo {year} {1949})}\BibitemShut {NoStop}%
\bibitem [{\citenamefont {Myers}\ and\ \citenamefont {Pospelov}(2003)}]{STR3}%
  \BibitemOpen
  \bibfield  {author} {\bibinfo {author} {\bibfnamefont {R.~C.}\ \bibnamefont
  {Myers}}\ and\ \bibinfo {author} {\bibfnamefont {M.}~\bibnamefont
  {Pospelov}},\ }\href@noop {} {\bibfield  {journal} {\bibinfo  {journal}
  {Phys. Rev. Lett.}\ }\textbf {\bibinfo {volume} {90}},\ \bibinfo {pages}
  {211601} (\bibinfo {year} {2003})}\BibitemShut {NoStop}%
\bibitem [{\citenamefont {Bertolami}\ and\ \citenamefont {Rosa}(2005)}]{STR4}%
  \BibitemOpen
  \bibfield  {author} {\bibinfo {author} {\bibfnamefont {O.}~\bibnamefont
  {Bertolami}}\ and\ \bibinfo {author} {\bibfnamefont {J.~G.}\ \bibnamefont
  {Rosa}},\ }\href@noop {} {\bibfield  {journal} {\bibinfo  {journal} {Phys.
  Rev. D}\ }\textbf {\bibinfo {volume} {71}},\ \bibinfo {pages} {097901}
  (\bibinfo {year} {2005})}\BibitemShut {NoStop}%
\bibitem [{\citenamefont {Reyes}\ \emph {et~al.}(2008)\citenamefont {Reyes},
  \citenamefont {Urrutia},\ and\ \citenamefont {Vergara}}]{STR5}%
  \BibitemOpen
  \bibfield  {author} {\bibinfo {author} {\bibfnamefont {C.~M.}\ \bibnamefont
  {Reyes}}, \bibinfo {author} {\bibfnamefont {L.~F.}\ \bibnamefont {Urrutia}},
  \ and\ \bibinfo {author} {\bibfnamefont {J.~D.}\ \bibnamefont {Vergara}},\
  }\href@noop {} {\bibfield  {journal} {\bibinfo  {journal} {Phys. Rev. D}\
  }\textbf {\bibinfo {volume} {78}},\ \bibinfo {pages} {125011} (\bibinfo
  {year} {2008})}\BibitemShut {NoStop}%
\bibitem [{\citenamefont {Mattingly}(2008)}]{STR6}%
  \BibitemOpen
  \bibfield  {author} {\bibinfo {author} {\bibfnamefont {D.}~\bibnamefont
  {Mattingly}},\ }\href@noop {} {\bibfield  {journal} {\bibinfo  {journal}
  {arXiv preprint arXiv:0802.1561}\ } (\bibinfo {year} {2008})}\BibitemShut
  {NoStop}%
\bibitem [{\citenamefont {Rubtsov}\ \emph {et~al.}(2014)\citenamefont
  {Rubtsov}, \citenamefont {Satunin},\ and\ \citenamefont {Sibiryakov}}]{STR7}%
  \BibitemOpen
  \bibfield  {author} {\bibinfo {author} {\bibfnamefont {G.}~\bibnamefont
  {Rubtsov}}, \bibinfo {author} {\bibfnamefont {P.}~\bibnamefont {Satunin}}, \
  and\ \bibinfo {author} {\bibfnamefont {S.}~\bibnamefont {Sibiryakov}},\
  }\href@noop {} {\bibfield  {journal} {\bibinfo  {journal} {CPT and Lorentz
  Symmetry}\ ,\ \bibinfo {pages} {192–195}} (\bibinfo {year}
  {2014})}\BibitemShut {NoStop}%
\bibitem [{\citenamefont {Liberati}(2013)}]{liberati2013}%
  \BibitemOpen
  \bibfield  {author} {\bibinfo {author} {\bibfnamefont {S.}~\bibnamefont
  {Liberati}},\ }\href@noop {} {\bibfield  {journal} {\bibinfo  {journal}
  {Classical and Quantum Gravity}\ }\textbf {\bibinfo {volume} {30}},\ \bibinfo
  {pages} {133001} (\bibinfo {year} {2013})}\BibitemShut {NoStop}%
\bibitem [{\citenamefont {Tasson}(2014)}]{tasson2014}%
  \BibitemOpen
  \bibfield  {author} {\bibinfo {author} {\bibfnamefont {J.~D.}\ \bibnamefont
  {Tasson}},\ }\href@noop {} {\bibfield  {journal} {\bibinfo  {journal}
  {Reports on Progress in Physics}\ }\textbf {\bibinfo {volume} {77}},\
  \bibinfo {pages} {062901} (\bibinfo {year} {2014})}\BibitemShut {NoStop}%
\bibitem [{\citenamefont {Hees}\ \emph {et~al.}(2016)\citenamefont {Hees},
  \citenamefont {Bailey}, \citenamefont {Bourgoin}, \citenamefont {Bars},
  \citenamefont {Guerlin}, \citenamefont {Poncin-Lafitte} \emph
  {et~al.}}]{hees2016}%
  \BibitemOpen
  \bibfield  {author} {\bibinfo {author} {\bibfnamefont {A.}~\bibnamefont
  {Hees}}, \bibinfo {author} {\bibfnamefont {Q.~G.}\ \bibnamefont {Bailey}},
  \bibinfo {author} {\bibfnamefont {A.}~\bibnamefont {Bourgoin}}, \bibinfo
  {author} {\bibfnamefont {P.-L.}\ \bibnamefont {Bars}}, \bibinfo {author}
  {\bibfnamefont {C.}~\bibnamefont {Guerlin}}, \bibinfo {author} {\bibfnamefont
  {L.}~\bibnamefont {Poncin-Lafitte}},  \emph {et~al.},\ }\href@noop {}
  {\bibfield  {journal} {\bibinfo  {journal} {Universe}\ }\textbf {\bibinfo
  {volume} {2}},\ \bibinfo {pages} {30} (\bibinfo {year} {2016})}\BibitemShut
  {NoStop}%
\bibitem [{\citenamefont {Rovelli}(2004)}]{rovelli2004}%
  \BibitemOpen
  \bibfield  {author} {\bibinfo {author} {\bibfnamefont {C.}~\bibnamefont
  {Rovelli}},\ }\href@noop {} {\emph {\bibinfo {title} {Quantum gravity}}}\
  (\bibinfo  {publisher} {Cambridge university press},\ \bibinfo {year}
  {2004})\BibitemShut {NoStop}%
\bibitem [{\citenamefont {Kosteleck\'y}\ and\ \citenamefont
  {Samuel}(1989{\natexlab{a}})}]{New1}%
  \BibitemOpen
  \bibfield  {author} {\bibinfo {author} {\bibfnamefont {V.~A.}\ \bibnamefont
  {Kosteleck\'y}}\ and\ \bibinfo {author} {\bibfnamefont {S.}~\bibnamefont
  {Samuel}},\ }\href@noop {} {\bibfield  {journal} {\bibinfo  {journal} {Phys.
  Rev. D}\ }\textbf {\bibinfo {volume} {39}},\ \bibinfo {pages} {683} (\bibinfo
  {year} {1989}{\natexlab{a}})}\BibitemShut {NoStop}%
\bibitem [{\citenamefont {Kosteleck\'y}\ and\ \citenamefont
  {Samuel}(1989{\natexlab{b}})}]{New2}%
  \BibitemOpen
  \bibfield  {author} {\bibinfo {author} {\bibfnamefont {V.~A.}\ \bibnamefont
  {Kosteleck\'y}}\ and\ \bibinfo {author} {\bibfnamefont {S.}~\bibnamefont
  {Samuel}},\ }\href@noop {} {\bibfield  {journal} {\bibinfo  {journal} {Phys.
  Rev. Lett.}\ }\textbf {\bibinfo {volume} {63}},\ \bibinfo {pages} {224}
  (\bibinfo {year} {1989}{\natexlab{b}})}\BibitemShut {NoStop}%
\bibitem [{\citenamefont {Kosteleck\'y}\ and\ \citenamefont
  {Samuel}(1989{\natexlab{c}})}]{New3}%
  \BibitemOpen
  \bibfield  {author} {\bibinfo {author} {\bibfnamefont {V.~A.}\ \bibnamefont
  {Kosteleck\'y}}\ and\ \bibinfo {author} {\bibfnamefont {S.}~\bibnamefont
  {Samuel}},\ }\href@noop {} {\bibfield  {journal} {\bibinfo  {journal} {Phys.
  Rev. D}\ }\textbf {\bibinfo {volume} {40}},\ \bibinfo {pages} {1886}
  (\bibinfo {year} {1989}{\natexlab{c}})}\BibitemShut {NoStop}%
\bibitem [{\citenamefont {Kostelecký}\ and\ \citenamefont
  {Potting}(1991)}]{New4}%
  \BibitemOpen
  \bibfield  {author} {\bibinfo {author} {\bibfnamefont {V.~A.}\ \bibnamefont
  {Kostelecký}}\ and\ \bibinfo {author} {\bibfnamefont {R.}~\bibnamefont
  {Potting}},\ }\href@noop {} {\bibfield  {journal} {\bibinfo  {journal}
  {Nuclear Physics B}\ }\textbf {\bibinfo {volume} {359}},\ \bibinfo {pages}
  {545 } (\bibinfo {year} {1991})}\BibitemShut {NoStop}%
\bibitem [{\citenamefont {Kosteleck\'y}\ and\ \citenamefont
  {Potting}()}]{New5}%
  \BibitemOpen
  \bibfield  {author} {\bibinfo {author} {\bibfnamefont {V.~A.}\ \bibnamefont
  {Kosteleck\'y}}\ and\ \bibinfo {author} {\bibfnamefont {R.}~\bibnamefont
  {Potting}},\ }\href@noop {} {\bibfield  {journal} {\bibinfo  {journal} {Phys.
  Rev. D}\ }\textbf {\bibinfo {volume} {51}},\ \bibinfo {pages}
  {3923}}\BibitemShut {NoStop}%
\bibitem [{\citenamefont {Gambini}\ and\ \citenamefont {Pullin}(1999)}]{New6}%
  \BibitemOpen
  \bibfield  {author} {\bibinfo {author} {\bibfnamefont {R.}~\bibnamefont
  {Gambini}}\ and\ \bibinfo {author} {\bibfnamefont {J.}~\bibnamefont
  {Pullin}},\ }\href@noop {} {\bibfield  {journal} {\bibinfo  {journal} {Phys.
  Rev. D}\ }\textbf {\bibinfo {volume} {59}},\ \bibinfo {pages} {124021}
  (\bibinfo {year} {1999})}\BibitemShut {NoStop}%
\bibitem [{\citenamefont {Bojowald}\ \emph {et~al.}(2005)\citenamefont
  {Bojowald}, \citenamefont {Morales-T\'ecotl},\ and\ \citenamefont
  {Sahlmann}}]{New7}%
  \BibitemOpen
  \bibfield  {author} {\bibinfo {author} {\bibfnamefont {M.}~\bibnamefont
  {Bojowald}}, \bibinfo {author} {\bibfnamefont {H.~A.}\ \bibnamefont
  {Morales-T\'ecotl}}, \ and\ \bibinfo {author} {\bibfnamefont
  {H.}~\bibnamefont {Sahlmann}},\ }\href@noop {} {\bibfield  {journal}
  {\bibinfo  {journal} {Phys. Rev. D}\ }\textbf {\bibinfo {volume} {71}},\
  \bibinfo {pages} {084012} (\bibinfo {year} {2005})}\BibitemShut {NoStop}%
\bibitem [{\citenamefont {Amelino-Camelia}\ and\ \citenamefont
  {Majid}(2000)}]{New8}%
  \BibitemOpen
  \bibfield  {author} {\bibinfo {author} {\bibfnamefont {G.}~\bibnamefont
  {Amelino-Camelia}}\ and\ \bibinfo {author} {\bibfnamefont {S.}~\bibnamefont
  {Majid}},\ }\href@noop {} {\bibfield  {journal} {\bibinfo  {journal}
  {International Journal of Modern Physics A}\ }\textbf {\bibinfo {volume}
  {15}},\ \bibinfo {pages} {4301} (\bibinfo {year} {2000})}\BibitemShut
  {NoStop}%
\bibitem [{\citenamefont {Carroll}\ \emph {et~al.}(2001)\citenamefont
  {Carroll}, \citenamefont {Harvey}, \citenamefont {Kosteleck\'y},
  \citenamefont {Lane},\ and\ \citenamefont {Okamoto}}]{New9}%
  \BibitemOpen
  \bibfield  {author} {\bibinfo {author} {\bibfnamefont {S.~M.}\ \bibnamefont
  {Carroll}}, \bibinfo {author} {\bibfnamefont {J.~A.}\ \bibnamefont {Harvey}},
  \bibinfo {author} {\bibfnamefont {V.~A.}\ \bibnamefont {Kosteleck\'y}},
  \bibinfo {author} {\bibfnamefont {C.~D.}\ \bibnamefont {Lane}}, \ and\
  \bibinfo {author} {\bibfnamefont {T.}~\bibnamefont {Okamoto}},\ }\href@noop
  {} {\bibfield  {journal} {\bibinfo  {journal} {Phys. Rev. Lett.}\ }\textbf
  {\bibinfo {volume} {87}},\ \bibinfo {pages} {141601} (\bibinfo {year}
  {2001})}\BibitemShut {NoStop}%
\bibitem [{\citenamefont {Klinkhamer}\ and\ \citenamefont
  {Rupp}(2004)}]{New10}%
  \BibitemOpen
  \bibfield  {author} {\bibinfo {author} {\bibfnamefont {F.~R.}\ \bibnamefont
  {Klinkhamer}}\ and\ \bibinfo {author} {\bibfnamefont {C.}~\bibnamefont
  {Rupp}},\ }\href@noop {} {\bibfield  {journal} {\bibinfo  {journal} {Phys.
  Rev. D}\ }\textbf {\bibinfo {volume} {70}},\ \bibinfo {pages} {045020}
  (\bibinfo {year} {2004})}\BibitemShut {NoStop}%
\bibitem [{\citenamefont {Bernadotte}\ and\ \citenamefont
  {Klinkhamer}(2007)}]{New11}%
  \BibitemOpen
  \bibfield  {author} {\bibinfo {author} {\bibfnamefont {S.}~\bibnamefont
  {Bernadotte}}\ and\ \bibinfo {author} {\bibfnamefont {F.~R.}\ \bibnamefont
  {Klinkhamer}},\ }\href@noop {} {\bibfield  {journal} {\bibinfo  {journal}
  {Phys. Rev. D}\ }\textbf {\bibinfo {volume} {75}},\ \bibinfo {pages} {024028}
  (\bibinfo {year} {2007})}\BibitemShut {NoStop}%
\bibitem [{\citenamefont {Klinkhamer}(1998)}]{New12}%
  \BibitemOpen
  \bibfield  {author} {\bibinfo {author} {\bibfnamefont {F.}~\bibnamefont
  {Klinkhamer}},\ }\href@noop {} {\bibfield  {journal} {\bibinfo  {journal}
  {Nuclear Physics B}\ }\textbf {\bibinfo {volume} {535}},\ \bibinfo {pages}
  {233 } (\bibinfo {year} {1998})}\BibitemShut {NoStop}%
\bibitem [{\citenamefont {Klinkhamer}(2000)}]{New13}%
  \BibitemOpen
  \bibfield  {author} {\bibinfo {author} {\bibfnamefont {F.}~\bibnamefont
  {Klinkhamer}},\ }\href@noop {} {\bibfield  {journal} {\bibinfo  {journal}
  {Nuclear Physics B}\ }\textbf {\bibinfo {volume} {578}},\ \bibinfo {pages}
  {277 } (\bibinfo {year} {2000})}\BibitemShut {NoStop}%
\bibitem [{\citenamefont {Klinkhamer}\ and\ \citenamefont
  {Schimmel}(2002)}]{New14}%
  \BibitemOpen
  \bibfield  {author} {\bibinfo {author} {\bibfnamefont {F.}~\bibnamefont
  {Klinkhamer}}\ and\ \bibinfo {author} {\bibfnamefont {J.}~\bibnamefont
  {Schimmel}},\ }\href@noop {} {\bibfield  {journal} {\bibinfo  {journal}
  {Nuclear Physics B}\ }\textbf {\bibinfo {volume} {639}},\ \bibinfo {pages}
  {241 } (\bibinfo {year} {2002})}\BibitemShut {NoStop}%
\bibitem [{\citenamefont {Ghosh}\ and\ \citenamefont
  {Klinkhamer}(2018)}]{New15}%
  \BibitemOpen
  \bibfield  {author} {\bibinfo {author} {\bibfnamefont {K.}~\bibnamefont
  {Ghosh}}\ and\ \bibinfo {author} {\bibfnamefont {F.}~\bibnamefont
  {Klinkhamer}},\ }\href@noop {} {\bibfield  {journal} {\bibinfo  {journal}
  {Nuclear Physics B}\ }\textbf {\bibinfo {volume} {926}},\ \bibinfo {pages}
  {335 } (\bibinfo {year} {2018})}\BibitemShut {NoStop}%
\bibitem [{\citenamefont {Ho\ifmmode~\check{r}\else
  \v{r}\fi{}ava}(2009)}]{New16}%
  \BibitemOpen
  \bibfield  {author} {\bibinfo {author} {\bibfnamefont {P.}~\bibnamefont
  {Ho\ifmmode~\check{r}\else \v{r}\fi{}ava}},\ }\href@noop {} {\bibfield
  {journal} {\bibinfo  {journal} {Phys. Rev. D}\ }\textbf {\bibinfo {volume}
  {79}},\ \bibinfo {pages} {084008} (\bibinfo {year} {2009})}\BibitemShut
  {NoStop}%
\bibitem [{\citenamefont {Cognola}\ \emph {et~al.}(2016)\citenamefont
  {Cognola}, \citenamefont {Myrzakulov}, \citenamefont {Sebastiani},
  \citenamefont {Vagnozzi},\ and\ \citenamefont {Zerbini}}]{sv1}%
  \BibitemOpen
  \bibfield  {author} {\bibinfo {author} {\bibfnamefont {G.}~\bibnamefont
  {Cognola}}, \bibinfo {author} {\bibfnamefont {R.}~\bibnamefont {Myrzakulov}},
  \bibinfo {author} {\bibfnamefont {L.}~\bibnamefont {Sebastiani}}, \bibinfo
  {author} {\bibfnamefont {S.}~\bibnamefont {Vagnozzi}}, \ and\ \bibinfo
  {author} {\bibfnamefont {S.}~\bibnamefont {Zerbini}},\ }\href@noop {}
  {\bibfield  {journal} {\bibinfo  {journal} {Classical and quantum gravity}\
  }\textbf {\bibinfo {volume} {33}},\ \bibinfo {pages} {225014} (\bibinfo
  {year} {2016})}\BibitemShut {NoStop}%
\bibitem [{\citenamefont {Casalino}\ \emph {et~al.}(2018)\citenamefont
  {Casalino}, \citenamefont {Rinaldi}, \citenamefont {Sebastiani},\ and\
  \citenamefont {Vagnozzi}}]{sv2}%
  \BibitemOpen
  \bibfield  {author} {\bibinfo {author} {\bibfnamefont {A.}~\bibnamefont
  {Casalino}}, \bibinfo {author} {\bibfnamefont {M.}~\bibnamefont {Rinaldi}},
  \bibinfo {author} {\bibfnamefont {L.}~\bibnamefont {Sebastiani}}, \ and\
  \bibinfo {author} {\bibfnamefont {S.}~\bibnamefont {Vagnozzi}},\ }\href@noop
  {} {\bibfield  {journal} {\bibinfo  {journal} {Classical and Quantum
  Gravity}\ }\textbf {\bibinfo {volume} {36}},\ \bibinfo {pages} {017001}
  (\bibinfo {year} {2018})}\BibitemShut {NoStop}%
\bibitem [{\citenamefont {Colladay}\ and\ \citenamefont
  {Kosteleck{\`y}}(1998{\natexlab{a}})}]{k1}%
  \BibitemOpen
  \bibfield  {author} {\bibinfo {author} {\bibfnamefont {D.}~\bibnamefont
  {Colladay}}\ and\ \bibinfo {author} {\bibfnamefont {V.~A.}\ \bibnamefont
  {Kosteleck{\`y}}},\ }\href@noop {} {\bibfield  {journal} {\bibinfo  {journal}
  {Physical Review D}\ }\textbf {\bibinfo {volume} {58}},\ \bibinfo {pages}
  {116002} (\bibinfo {year} {1998}{\natexlab{a}})}\BibitemShut {NoStop}%
\bibitem [{\citenamefont {Colladay}\ and\ \citenamefont
  {Kosteleck{\`y}}(1998{\natexlab{b}})}]{k2}%
  \BibitemOpen
  \bibfield  {author} {\bibinfo {author} {\bibfnamefont {D.}~\bibnamefont
  {Colladay}}\ and\ \bibinfo {author} {\bibfnamefont {V.~A.}\ \bibnamefont
  {Kosteleck{\`y}}},\ }\href@noop {} {\bibfield  {journal} {\bibinfo  {journal}
  {Physical Review D}\ }\textbf {\bibinfo {volume} {58}},\ \bibinfo {pages}
  {116002} (\bibinfo {year} {1998}{\natexlab{b}})}\BibitemShut {NoStop}%
\bibitem [{\citenamefont {Kosteleck{\`y}}\ and\ \citenamefont
  {Samuel}(1989{\natexlab{a}})}]{k4}%
  \BibitemOpen
  \bibfield  {author} {\bibinfo {author} {\bibfnamefont {V.~A.}\ \bibnamefont
  {Kosteleck{\`y}}}\ and\ \bibinfo {author} {\bibfnamefont {S.}~\bibnamefont
  {Samuel}},\ }\href@noop {} {\bibfield  {journal} {\bibinfo  {journal}
  {Physical Review Letters}\ }\textbf {\bibinfo {volume} {63}},\ \bibinfo
  {pages} {224} (\bibinfo {year} {1989}{\natexlab{a}})}\BibitemShut {NoStop}%
\bibitem [{\citenamefont {Kosteleck{\`y}}\ and\ \citenamefont
  {Samuel}(1989{\natexlab{b}})}]{k5}%
  \BibitemOpen
  \bibfield  {author} {\bibinfo {author} {\bibfnamefont {V.~A.}\ \bibnamefont
  {Kosteleck{\`y}}}\ and\ \bibinfo {author} {\bibfnamefont {S.}~\bibnamefont
  {Samuel}},\ }\href@noop {} {\bibfield  {journal} {\bibinfo  {journal}
  {Physical Review D}\ }\textbf {\bibinfo {volume} {39}},\ \bibinfo {pages}
  {683} (\bibinfo {year} {1989}{\natexlab{b}})}\BibitemShut {NoStop}%
\bibitem [{\citenamefont {Kosteleck{\`y}}\ and\ \citenamefont
  {Potting}(1996)}]{k7}%
  \BibitemOpen
  \bibfield  {author} {\bibinfo {author} {\bibfnamefont {V.~A.}\ \bibnamefont
  {Kosteleck{\`y}}}\ and\ \bibinfo {author} {\bibfnamefont {R.}~\bibnamefont
  {Potting}},\ }\href@noop {} {\bibfield  {journal} {\bibinfo  {journal}
  {Physics Letters B}\ }\textbf {\bibinfo {volume} {381}},\ \bibinfo {pages}
  {89} (\bibinfo {year} {1996})}\BibitemShut {NoStop}%
\bibitem [{\citenamefont {Kosteleck{\`y}}(2004)}]{kostelecky2004}%
  \BibitemOpen
  \bibfield  {author} {\bibinfo {author} {\bibfnamefont {V.~A.}\ \bibnamefont
  {Kosteleck{\`y}}},\ }\href@noop {} {\bibfield  {journal} {\bibinfo  {journal}
  {Physical Review D}\ }\textbf {\bibinfo {volume} {69}},\ \bibinfo {pages}
  {105009} (\bibinfo {year} {2004})}\BibitemShut {NoStop}%
\bibitem [{\citenamefont {Mewes}(2019)}]{mewes2019}%
  \BibitemOpen
  \bibfield  {author} {\bibinfo {author} {\bibfnamefont {M.}~\bibnamefont
  {Mewes}},\ }\href@noop {} {\bibfield  {journal} {\bibinfo  {journal}
  {Physical Review D}\ }\textbf {\bibinfo {volume} {99}},\ \bibinfo {pages}
  {104062} (\bibinfo {year} {2019})}\BibitemShut {NoStop}%
\bibitem [{\citenamefont {Maluf}\ \emph {et~al.}(2019)\citenamefont {Maluf},
  \citenamefont {Ara{\'u}jo~Filho}, \citenamefont {Cruz},\ and\ \citenamefont
  {Almeida}}]{adailton2}%
  \BibitemOpen
  \bibfield  {author} {\bibinfo {author} {\bibfnamefont {R.}~\bibnamefont
  {Maluf}}, \bibinfo {author} {\bibfnamefont {A.}~\bibnamefont
  {Ara{\'u}jo~Filho}}, \bibinfo {author} {\bibfnamefont {W.}~\bibnamefont
  {Cruz}}, \ and\ \bibinfo {author} {\bibfnamefont {C.}~\bibnamefont
  {Almeida}},\ }\href@noop {} {\bibfield  {journal} {\bibinfo  {journal} {EPL
  (Europhysics Letters)}\ }\textbf {\bibinfo {volume} {124}},\ \bibinfo {pages}
  {61001} (\bibinfo {year} {2019})}\BibitemShut {NoStop}%
\bibitem [{\citenamefont {Kosteleck{\`y}}\ and\ \citenamefont
  {Lehnert}(2001)}]{kostelecky2001}%
  \BibitemOpen
  \bibfield  {author} {\bibinfo {author} {\bibfnamefont {V.~A.}\ \bibnamefont
  {Kosteleck{\`y}}}\ and\ \bibinfo {author} {\bibfnamefont {R.}~\bibnamefont
  {Lehnert}},\ }\href@noop {} {\bibfield  {journal} {\bibinfo  {journal}
  {Physical Review D}\ }\textbf {\bibinfo {volume} {63}},\ \bibinfo {pages}
  {065008} (\bibinfo {year} {2001})}\BibitemShut {NoStop}%
\bibitem [{\citenamefont {Shore}(2005)}]{f1}%
  \BibitemOpen
  \bibfield  {author} {\bibinfo {author} {\bibfnamefont {G.~M.}\ \bibnamefont
  {Shore}},\ }\href@noop {} {\bibfield  {journal} {\bibinfo  {journal} {Nuclear
  Physics B}\ }\textbf {\bibinfo {volume} {717}},\ \bibinfo {pages} {86}
  (\bibinfo {year} {2005})}\BibitemShut {NoStop}%
\bibitem [{\citenamefont {Colladay}\ and\ \citenamefont
  {Kosteleck{\`y}}(2001)}]{f2}%
  \BibitemOpen
  \bibfield  {author} {\bibinfo {author} {\bibfnamefont {D.}~\bibnamefont
  {Colladay}}\ and\ \bibinfo {author} {\bibfnamefont {V.~A.}\ \bibnamefont
  {Kosteleck{\`y}}},\ }\href@noop {} {\bibfield  {journal} {\bibinfo  {journal}
  {Physics Letters B}\ }\textbf {\bibinfo {volume} {511}},\ \bibinfo {pages}
  {209} (\bibinfo {year} {2001})}\BibitemShut {NoStop}%
\bibitem [{\citenamefont {Kharlanov}\ and\ \citenamefont
  {Zhukovsky}(2007)}]{f3}%
  \BibitemOpen
  \bibfield  {author} {\bibinfo {author} {\bibfnamefont {O.}~\bibnamefont
  {Kharlanov}}\ and\ \bibinfo {author} {\bibfnamefont {V.~C.}\ \bibnamefont
  {Zhukovsky}},\ }\href@noop {} {\bibfield  {journal} {\bibinfo  {journal}
  {Journal of mathematical physics}\ }\textbf {\bibinfo {volume} {48}},\
  \bibinfo {pages} {092302} (\bibinfo {year} {2007})}\BibitemShut {NoStop}%
\bibitem [{\citenamefont {Bluhm}\ \emph {et~al.}(2000)\citenamefont {Bluhm},
  \citenamefont {Kosteleck{\`y}},\ and\ \citenamefont {Lane}}]{f4}%
  \BibitemOpen
  \bibfield  {author} {\bibinfo {author} {\bibfnamefont {R.}~\bibnamefont
  {Bluhm}}, \bibinfo {author} {\bibfnamefont {V.~A.}\ \bibnamefont
  {Kosteleck{\`y}}}, \ and\ \bibinfo {author} {\bibfnamefont {C.~D.}\
  \bibnamefont {Lane}},\ }\href@noop {} {\bibfield  {journal} {\bibinfo
  {journal} {Physical Review Letters}\ }\textbf {\bibinfo {volume} {84}},\
  \bibinfo {pages} {1098} (\bibinfo {year} {2000})}\BibitemShut {NoStop}%
\bibitem [{\citenamefont {Kruglov}(2012)}]{f5}%
  \BibitemOpen
  \bibfield  {author} {\bibinfo {author} {\bibfnamefont {S.}~\bibnamefont
  {Kruglov}},\ }\href@noop {} {\bibfield  {journal} {\bibinfo  {journal}
  {Physics Letters B}\ }\textbf {\bibinfo {volume} {718}},\ \bibinfo {pages}
  {228} (\bibinfo {year} {2012})}\BibitemShut {NoStop}%
\bibitem [{\citenamefont {Reis}\ and\ \citenamefont {Schreck}(2017)}]{f6}%
  \BibitemOpen
  \bibfield  {author} {\bibinfo {author} {\bibfnamefont {J.~A. A.~S.}\
  \bibnamefont {Reis}}\ and\ \bibinfo {author} {\bibfnamefont {M.}~\bibnamefont
  {Schreck}},\ }\href@noop {} {\bibfield  {journal} {\bibinfo  {journal} {Phys.
  Rev. D}\ }\textbf {\bibinfo {volume} {95}},\ \bibinfo {pages} {075016}
  (\bibinfo {year} {2017})}\BibitemShut {NoStop}%
\bibitem [{\citenamefont {Schreck}(2017)}]{f7}%
  \BibitemOpen
  \bibfield  {author} {\bibinfo {author} {\bibfnamefont {M.}~\bibnamefont
  {Schreck}},\ }\href@noop {} {\bibfield  {journal} {\bibinfo  {journal} {Phys.
  Rev. D}\ }\textbf {\bibinfo {volume} {96}},\ \bibinfo {pages} {095026}
  (\bibinfo {year} {2017})}\BibitemShut {NoStop}%
\bibitem [{\citenamefont {Adam}\ and\ \citenamefont {Klinkhamer}(2001)}]{e1}%
  \BibitemOpen
  \bibfield  {author} {\bibinfo {author} {\bibfnamefont {C.}~\bibnamefont
  {Adam}}\ and\ \bibinfo {author} {\bibfnamefont {F.~R.}\ \bibnamefont
  {Klinkhamer}},\ }\href@noop {} {\bibfield  {journal} {\bibinfo  {journal}
  {Nuclear Physics B}\ }\textbf {\bibinfo {volume} {607}},\ \bibinfo {pages}
  {247} (\bibinfo {year} {2001})}\BibitemShut {NoStop}%
\bibitem [{\citenamefont {Andrianov}\ and\ \citenamefont {Soldati}(1998)}]{e2}%
  \BibitemOpen
  \bibfield  {author} {\bibinfo {author} {\bibfnamefont {A.~A.}\ \bibnamefont
  {Andrianov}}\ and\ \bibinfo {author} {\bibfnamefont {R.}~\bibnamefont
  {Soldati}},\ }\href@noop {} {\bibfield  {journal} {\bibinfo  {journal}
  {Physics Letters B}\ }\textbf {\bibinfo {volume} {435}},\ \bibinfo {pages}
  {449} (\bibinfo {year} {1998})}\BibitemShut {NoStop}%
\bibitem [{\citenamefont {Andrianov}\ \emph {et~al.}(1998)\citenamefont
  {Andrianov}, \citenamefont {Soldati},\ and\ \citenamefont {Sorbo}}]{e3}%
  \BibitemOpen
  \bibfield  {author} {\bibinfo {author} {\bibfnamefont {A.~A.}\ \bibnamefont
  {Andrianov}}, \bibinfo {author} {\bibfnamefont {R.}~\bibnamefont {Soldati}},
  \ and\ \bibinfo {author} {\bibfnamefont {L.}~\bibnamefont {Sorbo}},\
  }\href@noop {} {\bibfield  {journal} {\bibinfo  {journal} {Physical Review
  D}\ }\textbf {\bibinfo {volume} {59}},\ \bibinfo {pages} {025002} (\bibinfo
  {year} {1998})}\BibitemShut {NoStop}%
\bibitem [{\citenamefont {Belich}\ \emph {et~al.}(2013)\citenamefont {Belich},
  \citenamefont {Bernald}, \citenamefont {Gaete},\ and\ \citenamefont
  {Helay{\"e}l-Neto}}]{e4}%
  \BibitemOpen
  \bibfield  {author} {\bibinfo {author} {\bibfnamefont {H.}~\bibnamefont
  {Belich}}, \bibinfo {author} {\bibfnamefont {L.}~\bibnamefont {Bernald}},
  \bibinfo {author} {\bibfnamefont {P.}~\bibnamefont {Gaete}}, \ and\ \bibinfo
  {author} {\bibfnamefont {J.}~\bibnamefont {Helay{\"e}l-Neto}},\ }\href@noop
  {} {\bibfield  {journal} {\bibinfo  {journal} {The European Physical Journal
  C}\ }\textbf {\bibinfo {volume} {73}},\ \bibinfo {pages} {2632} (\bibinfo
  {year} {2013})}\BibitemShut {NoStop}%
\bibitem [{\citenamefont {Scarpelli}\ \emph {et~al.}(2003)\citenamefont
  {Scarpelli}, \citenamefont {Belich}, \citenamefont {Boldo},\ and\
  \citenamefont {Helayel-Neto}}]{e5}%
  \BibitemOpen
  \bibfield  {author} {\bibinfo {author} {\bibfnamefont {A.~B.}\ \bibnamefont
  {Scarpelli}}, \bibinfo {author} {\bibfnamefont {H.}~\bibnamefont {Belich}},
  \bibinfo {author} {\bibfnamefont {J.}~\bibnamefont {Boldo}}, \ and\ \bibinfo
  {author} {\bibfnamefont {J.}~\bibnamefont {Helayel-Neto}},\ }\href@noop {}
  {\bibfield  {journal} {\bibinfo  {journal} {Physical Review D}\ }\textbf
  {\bibinfo {volume} {67}},\ \bibinfo {pages} {085021} (\bibinfo {year}
  {2003})}\BibitemShut {NoStop}%
\bibitem [{\citenamefont {Alfaro}\ \emph {et~al.}(2010)\citenamefont {Alfaro},
  \citenamefont {Andrianov}, \citenamefont {Cambiaso}, \citenamefont
  {Giacconi},\ and\ \citenamefont {Soldati}}]{e6}%
  \BibitemOpen
  \bibfield  {author} {\bibinfo {author} {\bibfnamefont {J.}~\bibnamefont
  {Alfaro}}, \bibinfo {author} {\bibfnamefont {A.}~\bibnamefont {Andrianov}},
  \bibinfo {author} {\bibfnamefont {M.}~\bibnamefont {Cambiaso}}, \bibinfo
  {author} {\bibfnamefont {P.}~\bibnamefont {Giacconi}}, \ and\ \bibinfo
  {author} {\bibfnamefont {R.}~\bibnamefont {Soldati}},\ }\href@noop {}
  {\bibfield  {journal} {\bibinfo  {journal} {International Journal of Modern
  Physics A}\ }\textbf {\bibinfo {volume} {25}},\ \bibinfo {pages} {3271}
  (\bibinfo {year} {2010})}\BibitemShut {NoStop}%
\bibitem [{\citenamefont {Casana}\ \emph
  {et~al.}(2018{\natexlab{a}})\citenamefont {Casana}, \citenamefont {Ferreira},
  \citenamefont {Lisboa-Santos}, \citenamefont {dos Santos},\ and\
  \citenamefont {Schreck}}]{e7}%
  \BibitemOpen
  \bibfield  {author} {\bibinfo {author} {\bibfnamefont {R.}~\bibnamefont
  {Casana}}, \bibinfo {author} {\bibfnamefont {M.~M.}\ \bibnamefont
  {Ferreira}}, \bibinfo {author} {\bibfnamefont {L.}~\bibnamefont
  {Lisboa-Santos}}, \bibinfo {author} {\bibfnamefont {F.~E.~P.}\ \bibnamefont
  {dos Santos}}, \ and\ \bibinfo {author} {\bibfnamefont {M.}~\bibnamefont
  {Schreck}},\ }\href@noop {} {\bibfield  {journal} {\bibinfo  {journal} {Phys.
  Rev. D}\ }\textbf {\bibinfo {volume} {97}},\ \bibinfo {pages} {115043}
  (\bibinfo {year} {2018}{\natexlab{a}})}\BibitemShut {NoStop}%
\bibitem [{\citenamefont {Ferreira}\ \emph {et~al.}(2019)\citenamefont
  {Ferreira}, \citenamefont {Lisboa-Santos}, \citenamefont {Maluf},\ and\
  \citenamefont {Schreck}}]{e8}%
  \BibitemOpen
  \bibfield  {author} {\bibinfo {author} {\bibfnamefont {M.~M.}\ \bibnamefont
  {Ferreira}}, \bibinfo {author} {\bibfnamefont {L.}~\bibnamefont
  {Lisboa-Santos}}, \bibinfo {author} {\bibfnamefont {R.~V.}\ \bibnamefont
  {Maluf}}, \ and\ \bibinfo {author} {\bibfnamefont {M.}~\bibnamefont
  {Schreck}},\ }\href@noop {} {\bibfield  {journal} {\bibinfo  {journal} {Phys.
  Rev. D}\ }\textbf {\bibinfo {volume} {100}},\ \bibinfo {pages} {055036}
  (\bibinfo {year} {2019})}\BibitemShut {NoStop}%
\bibitem [{\citenamefont {Reis}\ \emph {et~al.}(2019)\citenamefont {Reis},
  \citenamefont {Ferreira},\ and\ \citenamefont {Schreck}}]{e9}%
  \BibitemOpen
  \bibfield  {author} {\bibinfo {author} {\bibfnamefont {J.~A. A.~S.}\
  \bibnamefont {Reis}}, \bibinfo {author} {\bibfnamefont {M.~M.}\ \bibnamefont
  {Ferreira}}, \ and\ \bibinfo {author} {\bibfnamefont {M.}~\bibnamefont
  {Schreck}},\ }\href@noop {} {\bibfield  {journal} {\bibinfo  {journal} {Phys.
  Rev. D}\ }\textbf {\bibinfo {volume} {100}},\ \bibinfo {pages} {095026}
  (\bibinfo {year} {2019})}\BibitemShut {NoStop}%
\bibitem [{\citenamefont {Kosteleck{\`y}}\ and\ \citenamefont
  {Mewes}(2001)}]{c1}%
  \BibitemOpen
  \bibfield  {author} {\bibinfo {author} {\bibfnamefont {V.~A.}\ \bibnamefont
  {Kosteleck{\`y}}}\ and\ \bibinfo {author} {\bibfnamefont {M.}~\bibnamefont
  {Mewes}},\ }\href@noop {} {\bibfield  {journal} {\bibinfo  {journal}
  {Physical Review Letters}\ }\textbf {\bibinfo {volume} {87}},\ \bibinfo
  {pages} {251304} (\bibinfo {year} {2001})}\BibitemShut {NoStop}%
\bibitem [{\citenamefont {Klinkhamer}\ and\ \citenamefont {Risse}(2008)}]{c2}%
  \BibitemOpen
  \bibfield  {author} {\bibinfo {author} {\bibfnamefont {F.}~\bibnamefont
  {Klinkhamer}}\ and\ \bibinfo {author} {\bibfnamefont {M.}~\bibnamefont
  {Risse}},\ }\href@noop {} {\bibfield  {journal} {\bibinfo  {journal}
  {Physical Review D}\ }\textbf {\bibinfo {volume} {77}},\ \bibinfo {pages}
  {016002} (\bibinfo {year} {2008})}\BibitemShut {NoStop}%
\bibitem [{\citenamefont {Altschul}(2007)}]{c3}%
  \BibitemOpen
  \bibfield  {author} {\bibinfo {author} {\bibfnamefont {B.}~\bibnamefont
  {Altschul}},\ }\href@noop {} {\bibfield  {journal} {\bibinfo  {journal}
  {Physical review letters}\ }\textbf {\bibinfo {volume} {98}},\ \bibinfo
  {pages} {041603} (\bibinfo {year} {2007})}\BibitemShut {NoStop}%
\bibitem [{\citenamefont {Schreck}(2012)}]{c4}%
  \BibitemOpen
  \bibfield  {author} {\bibinfo {author} {\bibfnamefont {M.}~\bibnamefont
  {Schreck}},\ }\href@noop {} {\bibfield  {journal} {\bibinfo  {journal}
  {Physical Review D}\ }\textbf {\bibinfo {volume} {86}},\ \bibinfo {pages}
  {065038} (\bibinfo {year} {2012})}\BibitemShut {NoStop}%
\bibitem [{\citenamefont {Kosteleck{\`y}}\ and\ \citenamefont
  {Mewes}(2009{\natexlab{a}})}]{kostelecky2009electro}%
  \BibitemOpen
  \bibfield  {author} {\bibinfo {author} {\bibfnamefont {V.~A.}\ \bibnamefont
  {Kosteleck{\`y}}}\ and\ \bibinfo {author} {\bibfnamefont {M.}~\bibnamefont
  {Mewes}},\ }\href@noop {} {\bibfield  {journal} {\bibinfo  {journal}
  {Physical Review D}\ }\textbf {\bibinfo {volume} {80}},\ \bibinfo {pages}
  {015020} (\bibinfo {year} {2009}{\natexlab{a}})}\BibitemShut {NoStop}%
\bibitem [{\citenamefont {Kosteleck{\`y}}\ and\ \citenamefont
  {Mewes}(2012)}]{kostelecky2012neutrinos}%
  \BibitemOpen
  \bibfield  {author} {\bibinfo {author} {\bibfnamefont {V.~A.}\ \bibnamefont
  {Kosteleck{\`y}}}\ and\ \bibinfo {author} {\bibfnamefont {M.}~\bibnamefont
  {Mewes}},\ }\href@noop {} {\bibfield  {journal} {\bibinfo  {journal}
  {Physical Review D}\ }\textbf {\bibinfo {volume} {85}},\ \bibinfo {pages}
  {096005} (\bibinfo {year} {2012})}\BibitemShut {NoStop}%
\bibitem [{\citenamefont {Kosteleck{\`y}}\ and\ \citenamefont
  {Mewes}(2013{\natexlab{a}})}]{kostelecky2013fermions}%
  \BibitemOpen
  \bibfield  {author} {\bibinfo {author} {\bibfnamefont {V.~A.}\ \bibnamefont
  {Kosteleck{\`y}}}\ and\ \bibinfo {author} {\bibfnamefont {M.}~\bibnamefont
  {Mewes}},\ }\href@noop {} {\bibfield  {journal} {\bibinfo  {journal}
  {Physical Review D}\ }\textbf {\bibinfo {volume} {88}},\ \bibinfo {pages}
  {096006} (\bibinfo {year} {2013}{\natexlab{a}})}\BibitemShut {NoStop}%
\bibitem [{\citenamefont {Ding}\ and\ \citenamefont
  {Kosteleck\'y}(2016)}]{kostelecky2016Penning}%
  \BibitemOpen
  \bibfield  {author} {\bibinfo {author} {\bibfnamefont {Y.}~\bibnamefont
  {Ding}}\ and\ \bibinfo {author} {\bibfnamefont {V.~A.}\ \bibnamefont
  {Kosteleck\'y}},\ }\href@noop {} {\bibfield  {journal} {\bibinfo  {journal}
  {Phys. Rev. D}\ }\textbf {\bibinfo {volume} {94}},\ \bibinfo {pages} {056008}
  (\bibinfo {year} {2016})}\BibitemShut {NoStop}%
\bibitem [{\citenamefont {Kosteleck\'y}\ and\ \citenamefont
  {Li}(2019)}]{kostelecky2019gauge}%
  \BibitemOpen
  \bibfield  {author} {\bibinfo {author} {\bibfnamefont {V.~A.}\ \bibnamefont
  {Kosteleck\'y}}\ and\ \bibinfo {author} {\bibfnamefont {Z.}~\bibnamefont
  {Li}},\ }\href@noop {} {\bibfield  {journal} {\bibinfo  {journal} {Phys. Rev.
  D}\ }\textbf {\bibinfo {volume} {99}},\ \bibinfo {pages} {056016} (\bibinfo
  {year} {2019})}\BibitemShut {NoStop}%
\bibitem [{\citenamefont {Casana}\ \emph
  {et~al.}(2013{\natexlab{a}})\citenamefont {Casana}, \citenamefont
  {Ferreira~Jr}, \citenamefont {Passos}, \citenamefont {dos Santos},\ and\
  \citenamefont {Silva}}]{d1}%
  \BibitemOpen
  \bibfield  {author} {\bibinfo {author} {\bibfnamefont {R.}~\bibnamefont
  {Casana}}, \bibinfo {author} {\bibfnamefont {M.}~\bibnamefont {Ferreira~Jr}},
  \bibinfo {author} {\bibfnamefont {E.}~\bibnamefont {Passos}}, \bibinfo
  {author} {\bibfnamefont {F.}~\bibnamefont {dos Santos}}, \ and\ \bibinfo
  {author} {\bibfnamefont {E.}~\bibnamefont {Silva}},\ }\href@noop {}
  {\bibfield  {journal} {\bibinfo  {journal} {Physical Review D}\ }\textbf
  {\bibinfo {volume} {87}},\ \bibinfo {pages} {047701} (\bibinfo {year}
  {2013}{\natexlab{a}})}\BibitemShut {NoStop}%
\bibitem [{\citenamefont {Casana}\ \emph
  {et~al.}(2013{\natexlab{b}})\citenamefont {Casana}, \citenamefont
  {Ferreira~Jr}, \citenamefont {Maluf},\ and\ \citenamefont {dos Santos}}]{d2}%
  \BibitemOpen
  \bibfield  {author} {\bibinfo {author} {\bibfnamefont {R.}~\bibnamefont
  {Casana}}, \bibinfo {author} {\bibfnamefont {M.}~\bibnamefont {Ferreira~Jr}},
  \bibinfo {author} {\bibfnamefont {R.}~\bibnamefont {Maluf}}, \ and\ \bibinfo
  {author} {\bibfnamefont {F.}~\bibnamefont {dos Santos}},\ }\href@noop {}
  {\bibfield  {journal} {\bibinfo  {journal} {Physics Letters B}\ }\textbf
  {\bibinfo {volume} {726}},\ \bibinfo {pages} {815} (\bibinfo {year}
  {2013}{\natexlab{b}})}\BibitemShut {NoStop}%
\bibitem [{\citenamefont {Belich}\ \emph {et~al.}(2009)\citenamefont {Belich},
  \citenamefont {Colatto}, \citenamefont {Costa-Soares}, \citenamefont
  {Helay{\"e}l-Neto},\ and\ \citenamefont {Orlando}}]{d3}%
  \BibitemOpen
  \bibfield  {author} {\bibinfo {author} {\bibfnamefont {H.}~\bibnamefont
  {Belich}}, \bibinfo {author} {\bibfnamefont {L.}~\bibnamefont {Colatto}},
  \bibinfo {author} {\bibfnamefont {T.}~\bibnamefont {Costa-Soares}}, \bibinfo
  {author} {\bibfnamefont {J.}~\bibnamefont {Helay{\"e}l-Neto}}, \ and\
  \bibinfo {author} {\bibfnamefont {M.}~\bibnamefont {Orlando}},\ }\href@noop
  {} {\bibfield  {journal} {\bibinfo  {journal} {The European Physical Journal
  C}\ }\textbf {\bibinfo {volume} {62}},\ \bibinfo {pages} {425} (\bibinfo
  {year} {2009})}\BibitemShut {NoStop}%
\bibitem [{\citenamefont {Schreck}(2014)}]{schreck2014}%
  \BibitemOpen
  \bibfield  {author} {\bibinfo {author} {\bibfnamefont {M.}~\bibnamefont
  {Schreck}},\ }\href@noop {} {\bibfield  {journal} {\bibinfo  {journal}
  {Physical Review D}\ }\textbf {\bibinfo {volume} {90}},\ \bibinfo {pages}
  {085025} (\bibinfo {year} {2014})}\BibitemShut {NoStop}%
\bibitem [{\citenamefont {Cuzinatto}\ \emph {et~al.}(2011)\citenamefont
  {Cuzinatto}, \citenamefont {De~Melo}, \citenamefont {Medeiros},\ and\
  \citenamefont {Pompeia}}]{cuzinatto2011}%
  \BibitemOpen
  \bibfield  {author} {\bibinfo {author} {\bibfnamefont {R.}~\bibnamefont
  {Cuzinatto}}, \bibinfo {author} {\bibfnamefont {C.}~\bibnamefont {De~Melo}},
  \bibinfo {author} {\bibfnamefont {L.}~\bibnamefont {Medeiros}}, \ and\
  \bibinfo {author} {\bibfnamefont {P.}~\bibnamefont {Pompeia}},\ }\href@noop
  {} {\bibfield  {journal} {\bibinfo  {journal} {International Journal of
  Modern Physics A}\ }\textbf {\bibinfo {volume} {26}},\ \bibinfo {pages}
  {3641} (\bibinfo {year} {2011})}\BibitemShut {NoStop}%
\bibitem [{\citenamefont {Casana}\ \emph
  {et~al.}(2018{\natexlab{b}})\citenamefont {Casana}, \citenamefont
  {Ferreira~Jr}, \citenamefont {Lisboa-Santos}, \citenamefont {dos Santos},\
  and\ \citenamefont {Schreck}}]{casana2018}%
  \BibitemOpen
  \bibfield  {author} {\bibinfo {author} {\bibfnamefont {R.}~\bibnamefont
  {Casana}}, \bibinfo {author} {\bibfnamefont {M.~M.}\ \bibnamefont
  {Ferreira~Jr}}, \bibinfo {author} {\bibfnamefont {L.}~\bibnamefont
  {Lisboa-Santos}}, \bibinfo {author} {\bibfnamefont {F.~E.}\ \bibnamefont {dos
  Santos}}, \ and\ \bibinfo {author} {\bibfnamefont {M.}~\bibnamefont
  {Schreck}},\ }\href@noop {} {\bibfield  {journal} {\bibinfo  {journal}
  {Physical Review D}\ }\textbf {\bibinfo {volume} {97}},\ \bibinfo {pages}
  {115043} (\bibinfo {year} {2018}{\natexlab{b}})}\BibitemShut {NoStop}%
\bibitem [{\citenamefont {Filho}\ and\ \citenamefont {Maluf}(2020)}]{adailton}%
  \BibitemOpen
  \bibfield  {author} {\bibinfo {author} {\bibfnamefont {A.~A.}\ \bibnamefont
  {Filho}}\ and\ \bibinfo {author} {\bibfnamefont {R.}~\bibnamefont {Maluf}},\
  }\href@noop {} {\bibfield  {journal} {\bibinfo  {journal} {arXiv preprint
  arXiv:2003.02380}\ } (\bibinfo {year} {2020})}\BibitemShut {NoStop}%
\bibitem [{\citenamefont {Anacleto}\ \emph {et~al.}(2018)\citenamefont
  {Anacleto}, \citenamefont {Brito}, \citenamefont {Maciel}, \citenamefont
  {Mohammadi}, \citenamefont {Passos}, \citenamefont {Santos},\ and\
  \citenamefont {Santos}}]{anacleto2018}%
  \BibitemOpen
  \bibfield  {author} {\bibinfo {author} {\bibfnamefont {M.}~\bibnamefont
  {Anacleto}}, \bibinfo {author} {\bibfnamefont {F.}~\bibnamefont {Brito}},
  \bibinfo {author} {\bibfnamefont {E.}~\bibnamefont {Maciel}}, \bibinfo
  {author} {\bibfnamefont {A.}~\bibnamefont {Mohammadi}}, \bibinfo {author}
  {\bibfnamefont {E.}~\bibnamefont {Passos}}, \bibinfo {author} {\bibfnamefont
  {W.}~\bibnamefont {Santos}}, \ and\ \bibinfo {author} {\bibfnamefont
  {J.}~\bibnamefont {Santos}},\ }\href@noop {} {\bibfield  {journal} {\bibinfo
  {journal} {Physics Letters B}\ }\textbf {\bibinfo {volume} {785}},\ \bibinfo
  {pages} {191} (\bibinfo {year} {2018})}\BibitemShut {NoStop}%
\bibitem [{\citenamefont {Borges}\ \emph {et~al.}(2019)\citenamefont {Borges},
  \citenamefont {Barone}, \citenamefont {de~Melo},\ and\ \citenamefont
  {Barone}}]{borges2019}%
  \BibitemOpen
  \bibfield  {author} {\bibinfo {author} {\bibfnamefont {L.}~\bibnamefont
  {Borges}}, \bibinfo {author} {\bibfnamefont {F.}~\bibnamefont {Barone}},
  \bibinfo {author} {\bibfnamefont {C.}~\bibnamefont {de~Melo}}, \ and\
  \bibinfo {author} {\bibfnamefont {F.}~\bibnamefont {Barone}},\ }\href@noop {}
  {\bibfield  {journal} {\bibinfo  {journal} {Nuclear Physics B}\ }\textbf
  {\bibinfo {volume} {944}},\ \bibinfo {pages} {114634} (\bibinfo {year}
  {2019})}\BibitemShut {NoStop}%
\bibitem [{\citenamefont {Colladay}\ and\ \citenamefont {McDonald}(2004)}]{t1}%
  \BibitemOpen
  \bibfield  {author} {\bibinfo {author} {\bibfnamefont {D.}~\bibnamefont
  {Colladay}}\ and\ \bibinfo {author} {\bibfnamefont {P.}~\bibnamefont
  {McDonald}},\ }\href@noop {} {\bibfield  {journal} {\bibinfo  {journal}
  {Physical Review D}\ }\textbf {\bibinfo {volume} {70}},\ \bibinfo {pages}
  {125007} (\bibinfo {year} {2004})}\BibitemShut {NoStop}%
\bibitem [{\citenamefont {Colladay}\ and\ \citenamefont {McDonald}(2006)}]{t2}%
  \BibitemOpen
  \bibfield  {author} {\bibinfo {author} {\bibfnamefont {D.}~\bibnamefont
  {Colladay}}\ and\ \bibinfo {author} {\bibfnamefont {P.}~\bibnamefont
  {McDonald}},\ }\href@noop {} {\bibfield  {journal} {\bibinfo  {journal}
  {Physical Review D}\ }\textbf {\bibinfo {volume} {73}},\ \bibinfo {pages}
  {105006} (\bibinfo {year} {2006})}\BibitemShut {NoStop}%
\bibitem [{\citenamefont {Castellanos}\ and\ \citenamefont
  {L{\"a}mmerzahl}(2012)}]{t3}%
  \BibitemOpen
  \bibfield  {author} {\bibinfo {author} {\bibfnamefont {E.}~\bibnamefont
  {Castellanos}}\ and\ \bibinfo {author} {\bibfnamefont {C.}~\bibnamefont
  {L{\"a}mmerzahl}},\ }\href@noop {} {\bibfield  {journal} {\bibinfo  {journal}
  {Modern Physics Letters A}\ }\textbf {\bibinfo {volume} {27}},\ \bibinfo
  {pages} {1250181} (\bibinfo {year} {2012})}\BibitemShut {NoStop}%
\bibitem [{\citenamefont {Castellanos}\ and\ \citenamefont
  {Camacho}(2009)}]{t4}%
  \BibitemOpen
  \bibfield  {author} {\bibinfo {author} {\bibfnamefont {E.}~\bibnamefont
  {Castellanos}}\ and\ \bibinfo {author} {\bibfnamefont {A.}~\bibnamefont
  {Camacho}},\ }\href@noop {} {\bibfield  {journal} {\bibinfo  {journal}
  {General Relativity and Gravitation}\ }\textbf {\bibinfo {volume} {41}},\
  \bibinfo {pages} {2677} (\bibinfo {year} {2009})}\BibitemShut {NoStop}%
\bibitem [{\citenamefont {Gomes}\ \emph {et~al.}(2010)\citenamefont {Gomes},
  \citenamefont {Mariz}, \citenamefont {Nascimento}, \citenamefont {Petrov},
  \citenamefont {Santos},\ and\ \citenamefont {da~Silva}}]{t5}%
  \BibitemOpen
  \bibfield  {author} {\bibinfo {author} {\bibfnamefont {M.}~\bibnamefont
  {Gomes}}, \bibinfo {author} {\bibfnamefont {T.}~\bibnamefont {Mariz}},
  \bibinfo {author} {\bibfnamefont {J.}~\bibnamefont {Nascimento}}, \bibinfo
  {author} {\bibfnamefont {A.~Y.}\ \bibnamefont {Petrov}}, \bibinfo {author}
  {\bibfnamefont {A.}~\bibnamefont {Santos}}, \ and\ \bibinfo {author}
  {\bibfnamefont {A.}~\bibnamefont {da~Silva}},\ }\href@noop {} {\bibfield
  {journal} {\bibinfo  {journal} {Physical Review D}\ }\textbf {\bibinfo
  {volume} {81}},\ \bibinfo {pages} {045013} (\bibinfo {year}
  {2010})}\BibitemShut {NoStop}%
\bibitem [{\citenamefont {Casana}\ \emph {et~al.}(2009)\citenamefont {Casana},
  \citenamefont {Ferreira~Jr}, \citenamefont {Rodrigues},\ and\ \citenamefont
  {Silva}}]{t6}%
  \BibitemOpen
  \bibfield  {author} {\bibinfo {author} {\bibfnamefont {R.}~\bibnamefont
  {Casana}}, \bibinfo {author} {\bibfnamefont {M.~M.}\ \bibnamefont
  {Ferreira~Jr}}, \bibinfo {author} {\bibfnamefont {J.~S.}\ \bibnamefont
  {Rodrigues}}, \ and\ \bibinfo {author} {\bibfnamefont {M.~R.}\ \bibnamefont
  {Silva}},\ }\href@noop {} {\bibfield  {journal} {\bibinfo  {journal}
  {Physical Review D}\ }\textbf {\bibinfo {volume} {80}},\ \bibinfo {pages}
  {085026} (\bibinfo {year} {2009})}\BibitemShut {NoStop}%
\bibitem [{\citenamefont {Kaniadakis}(2005)}]{t7}%
  \BibitemOpen
  \bibfield  {author} {\bibinfo {author} {\bibfnamefont {G.}~\bibnamefont
  {Kaniadakis}},\ }\href@noop {} {\bibfield  {journal} {\bibinfo  {journal}
  {Physical Review E}\ }\textbf {\bibinfo {volume} {72}},\ \bibinfo {pages}
  {036108} (\bibinfo {year} {2005})}\BibitemShut {NoStop}%
\bibitem [{\citenamefont {Casana}\ and\ \citenamefont {da~Silva}(2015)}]{t8}%
  \BibitemOpen
  \bibfield  {author} {\bibinfo {author} {\bibfnamefont {R.}~\bibnamefont
  {Casana}}\ and\ \bibinfo {author} {\bibfnamefont {K.~A.~T.}\ \bibnamefont
  {da~Silva}},\ }\href@noop {} {\bibfield  {journal} {\bibinfo  {journal}
  {Modern Physics Letters A}\ }\textbf {\bibinfo {volume} {30}},\ \bibinfo
  {pages} {1550037} (\bibinfo {year} {2015})}\BibitemShut {NoStop}%
\bibitem [{\citenamefont {Colladay}\ and\ \citenamefont
  {McDonald}(2005)}]{Colladay1t}%
  \BibitemOpen
  \bibfield  {author} {\bibinfo {author} {\bibfnamefont {D.}~\bibnamefont
  {Colladay}}\ and\ \bibinfo {author} {\bibfnamefont {P.}~\bibnamefont
  {McDonald}},\ }in\ \href@noop {} {\emph {\bibinfo {booktitle} {CPT and
  Lorentz Symmetry}}}\ (\bibinfo  {publisher} {World Scientific},\ \bibinfo
  {year} {2005})\ pp.\ \bibinfo {pages} {264--269}\BibitemShut {NoStop}%
\bibitem [{\citenamefont {Kosteleck{\`y}}\ and\ \citenamefont
  {Mewes}(2007)}]{kostelecky2007}%
  \BibitemOpen
  \bibfield  {author} {\bibinfo {author} {\bibfnamefont {V.~A.}\ \bibnamefont
  {Kosteleck{\`y}}}\ and\ \bibinfo {author} {\bibfnamefont {M.}~\bibnamefont
  {Mewes}},\ }\href@noop {} {\bibfield  {journal} {\bibinfo  {journal}
  {Physical review letters}\ }\textbf {\bibinfo {volume} {99}},\ \bibinfo
  {pages} {011601} (\bibinfo {year} {2007})}\BibitemShut {NoStop}%
\bibitem [{\citenamefont {Kosteleck{\`y}}\ and\ \citenamefont
  {Mewes}(2009{\natexlab{b}})}]{kostelecky2009}%
  \BibitemOpen
  \bibfield  {author} {\bibinfo {author} {\bibfnamefont {V.~A.}\ \bibnamefont
  {Kosteleck{\`y}}}\ and\ \bibinfo {author} {\bibfnamefont {M.}~\bibnamefont
  {Mewes}},\ }\href@noop {} {\bibfield  {journal} {\bibinfo  {journal}
  {Physical Review D}\ }\textbf {\bibinfo {volume} {80}},\ \bibinfo {pages}
  {015020} (\bibinfo {year} {2009}{\natexlab{b}})}\BibitemShut {NoStop}%
\bibitem [{\citenamefont {Cambiaso}\ \emph {et~al.}(2012)\citenamefont
  {Cambiaso}, \citenamefont {Lehnert},\ and\ \citenamefont
  {Potting}}]{cambiaso2012}%
  \BibitemOpen
  \bibfield  {author} {\bibinfo {author} {\bibfnamefont {M.}~\bibnamefont
  {Cambiaso}}, \bibinfo {author} {\bibfnamefont {R.}~\bibnamefont {Lehnert}}, \
  and\ \bibinfo {author} {\bibfnamefont {R.}~\bibnamefont {Potting}},\
  }\href@noop {} {\bibfield  {journal} {\bibinfo  {journal} {Physical Review
  D}\ }\textbf {\bibinfo {volume} {85}},\ \bibinfo {pages} {085023} (\bibinfo
  {year} {2012})}\BibitemShut {NoStop}%
\bibitem [{\citenamefont {Mewes}(2012)}]{mewes2012}%
  \BibitemOpen
  \bibfield  {author} {\bibinfo {author} {\bibfnamefont {M.}~\bibnamefont
  {Mewes}},\ }\href@noop {} {\bibfield  {journal} {\bibinfo  {journal}
  {Physical Review D}\ }\textbf {\bibinfo {volume} {85}},\ \bibinfo {pages}
  {116012} (\bibinfo {year} {2012})}\BibitemShut {NoStop}%
\bibitem [{\citenamefont {Edwards}\ and\ \citenamefont
  {Kosteleck{\`y}}(2018)}]{EDWARDS2018}%
  \BibitemOpen
  \bibfield  {author} {\bibinfo {author} {\bibfnamefont {B.~R.}\ \bibnamefont
  {Edwards}}\ and\ \bibinfo {author} {\bibfnamefont {V.~A.}\ \bibnamefont
  {Kosteleck{\`y}}},\ }\href@noop {} {\bibfield  {journal} {\bibinfo  {journal}
  {Physics Letters B}\ }\textbf {\bibinfo {volume} {786}},\ \bibinfo {pages}
  {319} (\bibinfo {year} {2018})}\BibitemShut {NoStop}%
\bibitem [{\citenamefont {Passos}\ and\ \citenamefont
  {Petrov}(2008)}]{passos2008}%
  \BibitemOpen
  \bibfield  {author} {\bibinfo {author} {\bibfnamefont {E.}~\bibnamefont
  {Passos}}\ and\ \bibinfo {author} {\bibfnamefont {A.~Y.}\ \bibnamefont
  {Petrov}},\ }\href@noop {} {\bibfield  {journal} {\bibinfo  {journal}
  {Physics Letters B}\ }\textbf {\bibinfo {volume} {662}},\ \bibinfo {pages}
  {441} (\bibinfo {year} {2008})}\BibitemShut {NoStop}%
\bibitem [{\citenamefont {Gomes}\ and\ \citenamefont
  {Nascimento}(2010)}]{gomes201}%
  \BibitemOpen
  \bibfield  {author} {\bibinfo {author} {\bibfnamefont {M.}~\bibnamefont
  {Gomes}}\ and\ \bibinfo {author} {\bibfnamefont {J.}~\bibnamefont
  {Nascimento}},\ }\href@noop {} {\bibfield  {journal} {\bibinfo  {journal}
  {Phys. Rev. D}\ }\textbf {\bibinfo {volume} {81}},\ \bibinfo {pages} {045018}
  (\bibinfo {year} {2010})}\BibitemShut {NoStop}%
\bibitem [{\citenamefont {Khoa}\ \emph {et~al.}(1997)\citenamefont {Khoa},
  \citenamefont {Satchler},\ and\ \citenamefont
  {Von~Oertzen}}]{khoa1997nuclear}%
  \BibitemOpen
  \bibfield  {author} {\bibinfo {author} {\bibfnamefont {D.~T.}\ \bibnamefont
  {Khoa}}, \bibinfo {author} {\bibfnamefont {G.}~\bibnamefont {Satchler}}, \
  and\ \bibinfo {author} {\bibfnamefont {W.}~\bibnamefont {Von~Oertzen}},\
  }\href@noop {} {\bibfield  {journal} {\bibinfo  {journal} {Physical Review
  C}\ }\textbf {\bibinfo {volume} {56}},\ \bibinfo {pages} {954} (\bibinfo
  {year} {1997})}\BibitemShut {NoStop}%
\bibitem [{\citenamefont {Barnett}\ \emph {et~al.}(1996)\citenamefont
  {Barnett}, \citenamefont {Carone}, \citenamefont {Groom}, \citenamefont
  {Trippe}, \citenamefont {Wohl}, \citenamefont {Armstrong}, \citenamefont
  {Gee}, \citenamefont {Wagman}, \citenamefont {James}, \citenamefont {Mangano}
  \emph {et~al.}}]{barnett1996}%
  \BibitemOpen
  \bibfield  {author} {\bibinfo {author} {\bibfnamefont {R.~M.}\ \bibnamefont
  {Barnett}}, \bibinfo {author} {\bibfnamefont {C.}~\bibnamefont {Carone}},
  \bibinfo {author} {\bibfnamefont {D.}~\bibnamefont {Groom}}, \bibinfo
  {author} {\bibfnamefont {T.}~\bibnamefont {Trippe}}, \bibinfo {author}
  {\bibfnamefont {C.}~\bibnamefont {Wohl}}, \bibinfo {author} {\bibfnamefont
  {B.}~\bibnamefont {Armstrong}}, \bibinfo {author} {\bibfnamefont
  {P.}~\bibnamefont {Gee}}, \bibinfo {author} {\bibfnamefont {G.}~\bibnamefont
  {Wagman}}, \bibinfo {author} {\bibfnamefont {F.}~\bibnamefont {James}},
  \bibinfo {author} {\bibfnamefont {M.}~\bibnamefont {Mangano}},  \emph
  {et~al.},\ }\href@noop {} {\bibfield  {journal} {\bibinfo  {journal}
  {Physical Review D}\ }\textbf {\bibinfo {volume} {54}},\ \bibinfo {pages} {1}
  (\bibinfo {year} {1996})}\BibitemShut {NoStop}%
\bibitem [{\citenamefont {Eidelman}\ \emph {et~al.}(2004)\citenamefont
  {Eidelman}, \citenamefont {Hayes}, \citenamefont {Olive}, \citenamefont
  {Aguilar-Benitez}, \citenamefont {Amsler}, \citenamefont {Asner},
  \citenamefont {Babu}, \citenamefont {Barnett}, \citenamefont {Beringer},
  \citenamefont {Burchat} \emph {et~al.}}]{eidelman2004}%
  \BibitemOpen
  \bibfield  {author} {\bibinfo {author} {\bibfnamefont {S.}~\bibnamefont
  {Eidelman}}, \bibinfo {author} {\bibfnamefont {K.}~\bibnamefont {Hayes}},
  \bibinfo {author} {\bibfnamefont {K.~e.}\ \bibnamefont {Olive}}, \bibinfo
  {author} {\bibfnamefont {M.}~\bibnamefont {Aguilar-Benitez}}, \bibinfo
  {author} {\bibfnamefont {C.}~\bibnamefont {Amsler}}, \bibinfo {author}
  {\bibfnamefont {D.}~\bibnamefont {Asner}}, \bibinfo {author} {\bibfnamefont
  {K.}~\bibnamefont {Babu}}, \bibinfo {author} {\bibfnamefont {R.}~\bibnamefont
  {Barnett}}, \bibinfo {author} {\bibfnamefont {J.}~\bibnamefont {Beringer}},
  \bibinfo {author} {\bibfnamefont {P.}~\bibnamefont {Burchat}},  \emph
  {et~al.},\ }\href@noop {} {\bibfield  {journal} {\bibinfo  {journal} {Physics
  letters B}\ }\textbf {\bibinfo {volume} {592}},\ \bibinfo {pages} {1}
  (\bibinfo {year} {2004})}\BibitemShut {NoStop}%
\bibitem [{\citenamefont {Lalazissis}\ \emph {et~al.}(2005)\citenamefont
  {Lalazissis}, \citenamefont {Nik{\v{s}}i{\'c}}, \citenamefont {Vretenar},\
  and\ \citenamefont {Ring}}]{lalazissis2005new}%
  \BibitemOpen
  \bibfield  {author} {\bibinfo {author} {\bibfnamefont {G.}~\bibnamefont
  {Lalazissis}}, \bibinfo {author} {\bibfnamefont {T.}~\bibnamefont
  {Nik{\v{s}}i{\'c}}}, \bibinfo {author} {\bibfnamefont {D.}~\bibnamefont
  {Vretenar}}, \ and\ \bibinfo {author} {\bibfnamefont {P.}~\bibnamefont
  {Ring}},\ }\href@noop {} {\bibfield  {journal} {\bibinfo  {journal} {Physical
  Review C}\ }\textbf {\bibinfo {volume} {71}},\ \bibinfo {pages} {024312}
  (\bibinfo {year} {2005})}\BibitemShut {NoStop}%
\bibitem [{\citenamefont {Humphries}\ \emph {et~al.}(1972)\citenamefont
  {Humphries}, \citenamefont {James},\ and\ \citenamefont
  {Luckhurst}}]{humphries1972}%
  \BibitemOpen
  \bibfield  {author} {\bibinfo {author} {\bibfnamefont {R.}~\bibnamefont
  {Humphries}}, \bibinfo {author} {\bibfnamefont {P.}~\bibnamefont {James}}, \
  and\ \bibinfo {author} {\bibfnamefont {G.}~\bibnamefont {Luckhurst}},\
  }\href@noop {} {\bibfield  {journal} {\bibinfo  {journal} {Journal of the
  Chemical Society, Faraday Transactions 2: Molecular and Chemical Physics}\
  }\textbf {\bibinfo {volume} {68}},\ \bibinfo {pages} {1031} (\bibinfo {year}
  {1972})}\BibitemShut {NoStop}%
\bibitem [{\citenamefont {Klein}(1969)}]{klein1969}%
  \BibitemOpen
  \bibfield  {author} {\bibinfo {author} {\bibfnamefont {M.~W.}\ \bibnamefont
  {Klein}},\ }\href@noop {} {\bibfield  {journal} {\bibinfo  {journal}
  {Physical Review}\ }\textbf {\bibinfo {volume} {188}},\ \bibinfo {pages}
  {933} (\bibinfo {year} {1969})}\BibitemShut {NoStop}%
\bibitem [{\citenamefont {Wojtowicz}\ and\ \citenamefont
  {Rayl}(1968)}]{wojtowicz}%
  \BibitemOpen
  \bibfield  {author} {\bibinfo {author} {\bibfnamefont {P.~J.}\ \bibnamefont
  {Wojtowicz}}\ and\ \bibinfo {author} {\bibfnamefont {M.}~\bibnamefont
  {Rayl}},\ }\href@noop {} {\bibfield  {journal} {\bibinfo  {journal} {Physical
  Review Letters}\ }\textbf {\bibinfo {volume} {20}},\ \bibinfo {pages} {1489}
  (\bibinfo {year} {1968})}\BibitemShut {NoStop}%
\bibitem [{\citenamefont {Ter~Haar}\ and\ \citenamefont
  {Lines}(1962)}]{ter1962molecular}%
  \BibitemOpen
  \bibfield  {author} {\bibinfo {author} {\bibfnamefont {D.}~\bibnamefont
  {Ter~Haar}}\ and\ \bibinfo {author} {\bibfnamefont {M.}~\bibnamefont
  {Lines}},\ }\href@noop {} {\bibfield  {journal} {\bibinfo  {journal}
  {Philosophical Transactions of the Royal Society of London. Series A,
  Mathematical and Physical Sciences}\ }\textbf {\bibinfo {volume} {254}},\
  \bibinfo {pages} {521} (\bibinfo {year} {1962})}\BibitemShut {NoStop}%
\bibitem [{\citenamefont {Ara{\'u}jo-Filho}\ \emph {et~al.}(2017)\citenamefont
  {Ara{\'u}jo-Filho}, \citenamefont {Silva}, \citenamefont {Righi},
  \citenamefont {da~Silva}, \citenamefont {Silva}, \citenamefont {Caetano},\
  and\ \citenamefont {Freire}}]{araujo2017}%
  \BibitemOpen
  \bibfield  {author} {\bibinfo {author} {\bibfnamefont {A.~A.}\ \bibnamefont
  {Ara{\'u}jo-Filho}}, \bibinfo {author} {\bibfnamefont {F.~L.}\ \bibnamefont
  {Silva}}, \bibinfo {author} {\bibfnamefont {A.}~\bibnamefont {Righi}},
  \bibinfo {author} {\bibfnamefont {M.~B.}\ \bibnamefont {da~Silva}}, \bibinfo
  {author} {\bibfnamefont {B.~P.}\ \bibnamefont {Silva}}, \bibinfo {author}
  {\bibfnamefont {E.~W.}\ \bibnamefont {Caetano}}, \ and\ \bibinfo {author}
  {\bibfnamefont {V.~N.}\ \bibnamefont {Freire}},\ }\href@noop {} {\bibfield
  {journal} {\bibinfo  {journal} {Journal of Solid State Chemistry}\ }\textbf
  {\bibinfo {volume} {250}},\ \bibinfo {pages} {68} (\bibinfo {year}
  {2017})}\BibitemShut {NoStop}%
\bibitem [{\citenamefont {Silva}\ \emph {et~al.}(2018)\citenamefont {Silva},
  \citenamefont {Filho}, \citenamefont {da~Silva}, \citenamefont {Balzuweit},
  \citenamefont {Bantignies}, \citenamefont {Caetano}, \citenamefont {Moreira},
  \citenamefont {Freire},\ and\ \citenamefont {Righi}}]{silva2018}%
  \BibitemOpen
  \bibfield  {author} {\bibinfo {author} {\bibfnamefont {F.~L. R.~e.}\
  \bibnamefont {Silva}}, \bibinfo {author} {\bibfnamefont {A.~A.~A.}\
  \bibnamefont {Filho}}, \bibinfo {author} {\bibfnamefont {M.~B.}\ \bibnamefont
  {da~Silva}}, \bibinfo {author} {\bibfnamefont {K.}~\bibnamefont {Balzuweit}},
  \bibinfo {author} {\bibfnamefont {J.-L.}\ \bibnamefont {Bantignies}},
  \bibinfo {author} {\bibfnamefont {E.~W.~S.}\ \bibnamefont {Caetano}},
  \bibinfo {author} {\bibfnamefont {R.~L.}\ \bibnamefont {Moreira}}, \bibinfo
  {author} {\bibfnamefont {V.~N.}\ \bibnamefont {Freire}}, \ and\ \bibinfo
  {author} {\bibfnamefont {A.}~\bibnamefont {Righi}},\ }\href@noop {}
  {\bibfield  {journal} {\bibinfo  {journal} {Journal of Raman Spectroscopy}\
  }\textbf {\bibinfo {volume} {49}},\ \bibinfo {pages} {538} (\bibinfo {year}
  {2018})}\BibitemShut {NoStop}%
\bibitem [{\citenamefont {Greiner}\ \emph {et~al.}(2012)\citenamefont
  {Greiner}, \citenamefont {Neise},\ and\ \citenamefont
  {St{\"o}cker}}]{greiner2012}%
  \BibitemOpen
  \bibfield  {author} {\bibinfo {author} {\bibfnamefont {W.}~\bibnamefont
  {Greiner}}, \bibinfo {author} {\bibfnamefont {L.}~\bibnamefont {Neise}}, \
  and\ \bibinfo {author} {\bibfnamefont {H.}~\bibnamefont {St{\"o}cker}},\
  }\href@noop {} {\emph {\bibinfo {title} {Thermodynamics and statistical
  mechanics}}}\ (\bibinfo  {publisher} {Springer Science \& Business Media},\
  \bibinfo {year} {2012})\BibitemShut {NoStop}%
\bibitem [{\citenamefont {Salinas}(1997)}]{salinas1997}%
  \BibitemOpen
  \bibfield  {author} {\bibinfo {author} {\bibfnamefont {S.~R.}\ \bibnamefont
  {Salinas}},\ }\href@noop {} {\emph {\bibinfo {title} {Introdu{\c{c}}{\~a}o a
  f{\'\i}sica estat{\'\i}stica vol. 09}}}\ (\bibinfo  {publisher} {Edusp},\
  \bibinfo {year} {1997})\BibitemShut {NoStop}%
\bibitem [{\citenamefont {Pathria}\ and\ \citenamefont {Beale}()}]{pathria32}%
  \BibitemOpen
  \bibfield  {author} {\bibinfo {author} {\bibfnamefont {R.}~\bibnamefont
  {Pathria}}\ and\ \bibinfo {author} {\bibfnamefont {P.~D.}\ \bibnamefont
  {Beale}},\ }\href@noop {} {\bibfield  {journal} {\bibinfo  {journal} {Butter
  worth}\ }\textbf {\bibinfo {volume} {32}}}\BibitemShut {NoStop}%
\bibitem [{\citenamefont {AA~Filho}(2020)}]{adailton3}%
  \BibitemOpen
  \bibfield  {author} {\bibinfo {author} {\bibfnamefont {A.}~\bibnamefont
  {AA~Filho}},\ }\href@noop {} {\bibfield  {journal} {\bibinfo  {journal}
  {arXiv preprint arXiv:2004.07799}\ } (\bibinfo {year} {2020})}\BibitemShut
  {NoStop}%
\bibitem [{\citenamefont {Oliveira}\ and\ \citenamefont
  {Ara{\'u}jo~Filho}(2020)}]{oliveira2020}%
  \BibitemOpen
  \bibfield  {author} {\bibinfo {author} {\bibfnamefont {R.}~\bibnamefont
  {Oliveira}}\ and\ \bibinfo {author} {\bibfnamefont {A.}~\bibnamefont
  {Ara{\'u}jo~Filho}},\ }\href@noop {} {\bibfield  {journal} {\bibinfo
  {journal} {The European Physical Journal Plus}\ }\textbf {\bibinfo {volume}
  {135}},\ \bibinfo {pages} {99} (\bibinfo {year} {2020})}\BibitemShut
  {NoStop}%
\bibitem [{\citenamefont {Oliveira}\ \emph {et~al.}(2019)\citenamefont
  {Oliveira}, \citenamefont {Ara{\'u}jo~Filho}, \citenamefont {Lima},
  \citenamefont {Maluf},\ and\ \citenamefont {Almeida}}]{oliveira2019}%
  \BibitemOpen
  \bibfield  {author} {\bibinfo {author} {\bibfnamefont {R.~R.}\ \bibnamefont
  {Oliveira}}, \bibinfo {author} {\bibfnamefont {A.~A.}\ \bibnamefont
  {Ara{\'u}jo~Filho}}, \bibinfo {author} {\bibfnamefont {F.~C.}\ \bibnamefont
  {Lima}}, \bibinfo {author} {\bibfnamefont {R.~V.}\ \bibnamefont {Maluf}}, \
  and\ \bibinfo {author} {\bibfnamefont {C.~A.}\ \bibnamefont {Almeida}},\
  }\href@noop {} {\bibfield  {journal} {\bibinfo  {journal} {The European
  Physical Journal Plus}\ }\textbf {\bibinfo {volume} {134}},\ \bibinfo {pages}
  {495} (\bibinfo {year} {2019})}\BibitemShut {NoStop}%
\bibitem [{\citenamefont {Pacheco}\ \emph {et~al.}(2014)\citenamefont
  {Pacheco}, \citenamefont {Maluf}, \citenamefont {Almeida},\ and\
  \citenamefont {Landim}}]{pacheco2014}%
  \BibitemOpen
  \bibfield  {author} {\bibinfo {author} {\bibfnamefont {M.}~\bibnamefont
  {Pacheco}}, \bibinfo {author} {\bibfnamefont {R.}~\bibnamefont {Maluf}},
  \bibinfo {author} {\bibfnamefont {C.}~\bibnamefont {Almeida}}, \ and\
  \bibinfo {author} {\bibfnamefont {R.}~\bibnamefont {Landim}},\ }\href@noop {}
  {\bibfield  {journal} {\bibinfo  {journal} {EPL (Europhysics Letters)}\
  }\textbf {\bibinfo {volume} {108}},\ \bibinfo {pages} {10005} (\bibinfo
  {year} {2014})}\BibitemShut {NoStop}%
\bibitem [{\citenamefont {Casana}\ \emph {et~al.}(2008)\citenamefont {Casana},
  \citenamefont {Ferreira~Jr},\ and\ \citenamefont {Rodrigues}}]{casana2008}%
  \BibitemOpen
  \bibfield  {author} {\bibinfo {author} {\bibfnamefont {R.}~\bibnamefont
  {Casana}}, \bibinfo {author} {\bibfnamefont {M.~M.}\ \bibnamefont
  {Ferreira~Jr}}, \ and\ \bibinfo {author} {\bibfnamefont {J.~S.}\ \bibnamefont
  {Rodrigues}},\ }\href@noop {} {\bibfield  {journal} {\bibinfo  {journal}
  {Physical Review D}\ }\textbf {\bibinfo {volume} {78}},\ \bibinfo {pages}
  {125013} (\bibinfo {year} {2008})}\BibitemShut {NoStop}%
\bibitem [{\citenamefont {Cohen}\ and\ \citenamefont {Glashow}(2006)}]{VSR}%
  \BibitemOpen
  \bibfield  {author} {\bibinfo {author} {\bibfnamefont {A.~G.}\ \bibnamefont
  {Cohen}}\ and\ \bibinfo {author} {\bibfnamefont {S.~L.}\ \bibnamefont
  {Glashow}},\ }\href@noop {} {\bibfield  {journal} {\bibinfo  {journal}
  {Physical review letters}\ }\textbf {\bibinfo {volume} {97}},\ \bibinfo
  {pages} {021601} (\bibinfo {year} {2006})}\BibitemShut {NoStop}%
\bibitem [{\citenamefont {Kolomeitsev}\ and\ \citenamefont
  {Voskresensky}(2018)}]{Data1}%
  \BibitemOpen
  \bibfield  {author} {\bibinfo {author} {\bibfnamefont {E.}~\bibnamefont
  {Kolomeitsev}}\ and\ \bibinfo {author} {\bibfnamefont {D.}~\bibnamefont
  {Voskresensky}},\ }\href@noop {} {\bibfield  {journal} {\bibinfo  {journal}
  {Nuclear Physics A}\ }\textbf {\bibinfo {volume} {973}},\ \bibinfo {pages}
  {89} (\bibinfo {year} {2018})}\BibitemShut {NoStop}%
\bibitem [{\citenamefont {Lee}\ \emph {et~al.}(1988)\citenamefont {Lee},
  \citenamefont {Yang},\ and\ \citenamefont {Parr}}]{lee1988development}%
  \BibitemOpen
  \bibfield  {author} {\bibinfo {author} {\bibfnamefont {C.}~\bibnamefont
  {Lee}}, \bibinfo {author} {\bibfnamefont {W.}~\bibnamefont {Yang}}, \ and\
  \bibinfo {author} {\bibfnamefont {R.~G.}\ \bibnamefont {Parr}},\ }\href@noop
  {} {\bibfield  {journal} {\bibinfo  {journal} {Physical Review B}\ }\textbf
  {\bibinfo {volume} {37}},\ \bibinfo {pages} {785} (\bibinfo {year}
  {1988})}\BibitemShut {NoStop}%
\bibitem [{\citenamefont {Casida}\ \emph {et~al.}(1998)\citenamefont {Casida},
  \citenamefont {Jamorski}, \citenamefont {Casida},\ and\ \citenamefont
  {Salahub}}]{aa1}%
  \BibitemOpen
  \bibfield  {author} {\bibinfo {author} {\bibfnamefont {M.~E.}\ \bibnamefont
  {Casida}}, \bibinfo {author} {\bibfnamefont {C.}~\bibnamefont {Jamorski}},
  \bibinfo {author} {\bibfnamefont {K.~C.}\ \bibnamefont {Casida}}, \ and\
  \bibinfo {author} {\bibfnamefont {D.~R.}\ \bibnamefont {Salahub}},\
  }\href@noop {} {\bibfield  {journal} {\bibinfo  {journal} {The Journal of
  Chemical Physics}\ }\textbf {\bibinfo {volume} {108}},\ \bibinfo {pages}
  {4439} (\bibinfo {year} {1998})}\BibitemShut {NoStop}%
\bibitem [{\citenamefont {Segall}\ \emph {et~al.}(2002)\citenamefont {Segall},
  \citenamefont {Lindan}, \citenamefont {Probert}, \citenamefont {Pickard},
  \citenamefont {Hasnip}, \citenamefont {Clark},\ and\ \citenamefont
  {Payne}}]{aa2}%
  \BibitemOpen
  \bibfield  {author} {\bibinfo {author} {\bibfnamefont {M.~D.}\ \bibnamefont
  {Segall}}, \bibinfo {author} {\bibfnamefont {P.~J.~D.}\ \bibnamefont
  {Lindan}}, \bibinfo {author} {\bibfnamefont {M.~J.~a.}\ \bibnamefont
  {Probert}}, \bibinfo {author} {\bibfnamefont {C.~J.}\ \bibnamefont
  {Pickard}}, \bibinfo {author} {\bibfnamefont {P.~J.}\ \bibnamefont {Hasnip}},
  \bibinfo {author} {\bibfnamefont {S.}~\bibnamefont {Clark}}, \ and\ \bibinfo
  {author} {\bibfnamefont {M.}~\bibnamefont {Payne}},\ }\href@noop {}
  {\bibfield  {journal} {\bibinfo  {journal} {Journal of Physics: Condensed
  Matter}\ }\textbf {\bibinfo {volume} {14}},\ \bibinfo {pages} {2717}
  (\bibinfo {year} {2002})}\BibitemShut {NoStop}%
\bibitem [{\citenamefont {Langreth}\ and\ \citenamefont {Mehl}(1983)}]{aa3}%
  \BibitemOpen
  \bibfield  {author} {\bibinfo {author} {\bibfnamefont {D.~C.}\ \bibnamefont
  {Langreth}}\ and\ \bibinfo {author} {\bibfnamefont {M.~J.}\ \bibnamefont
  {Mehl}},\ }\href@noop {} {\bibfield  {journal} {\bibinfo  {journal} {Physical
  Review B}\ }\textbf {\bibinfo {volume} {28}},\ \bibinfo {pages} {1809}
  (\bibinfo {year} {1983})}\BibitemShut {NoStop}%
\bibitem [{\citenamefont {Cunha}\ \emph {et~al.}(2020)\citenamefont {Cunha},
  \citenamefont {da~Costa}, \citenamefont {Felix}, \citenamefont {Chaves},\
  and\ \citenamefont {Pereira~Jr}}]{cunha2020}%
  \BibitemOpen
  \bibfield  {author} {\bibinfo {author} {\bibfnamefont {S.~M.}\ \bibnamefont
  {Cunha}}, \bibinfo {author} {\bibfnamefont {D.~R.}\ \bibnamefont {da~Costa}},
  \bibinfo {author} {\bibfnamefont {L.~C.}\ \bibnamefont {Felix}}, \bibinfo
  {author} {\bibfnamefont {A.}~\bibnamefont {Chaves}}, \ and\ \bibinfo {author}
  {\bibfnamefont {J.~M.}\ \bibnamefont {Pereira~Jr}},\ }\href@noop {}
  {\bibfield  {journal} {\bibinfo  {journal} {Physical Review B}\ }\textbf
  {\bibinfo {volume} {102}},\ \bibinfo {pages} {045427} (\bibinfo {year}
  {2020})}\BibitemShut {NoStop}%
\bibitem [{\citenamefont {Kosteleck{\`y}}\ and\ \citenamefont
  {Mewes}(2016)}]{kostelecky2016testing}%
  \BibitemOpen
  \bibfield  {author} {\bibinfo {author} {\bibfnamefont {V.~A.}\ \bibnamefont
  {Kosteleck{\`y}}}\ and\ \bibinfo {author} {\bibfnamefont {M.}~\bibnamefont
  {Mewes}},\ }\href@noop {} {\bibfield  {journal} {\bibinfo  {journal} {Physics
  Letters B}\ }\textbf {\bibinfo {volume} {757}},\ \bibinfo {pages} {510}
  (\bibinfo {year} {2016})}\BibitemShut {NoStop}%
\bibitem [{\citenamefont {Kosteleck{\`y}}\ and\ \citenamefont
  {Russell}(2011)}]{kostelecky2011data}%
  \BibitemOpen
  \bibfield  {author} {\bibinfo {author} {\bibfnamefont {V.~A.}\ \bibnamefont
  {Kosteleck{\`y}}}\ and\ \bibinfo {author} {\bibfnamefont {N.}~\bibnamefont
  {Russell}},\ }\href@noop {} {\bibfield  {journal} {\bibinfo  {journal}
  {Reviews of Modern Physics}\ }\textbf {\bibinfo {volume} {83}},\ \bibinfo
  {pages} {11} (\bibinfo {year} {2011})}\BibitemShut {NoStop}%
\bibitem [{\citenamefont {Kosteleck{\`y}}\ and\ \citenamefont
  {Mewes}(2013{\natexlab{b}})}]{kostelecky2013constraints}%
  \BibitemOpen
  \bibfield  {author} {\bibinfo {author} {\bibfnamefont {V.~A.}\ \bibnamefont
  {Kosteleck{\`y}}}\ and\ \bibinfo {author} {\bibfnamefont {M.}~\bibnamefont
  {Mewes}},\ }\href@noop {} {\bibfield  {journal} {\bibinfo  {journal}
  {Physical review letters}\ }\textbf {\bibinfo {volume} {110}},\ \bibinfo
  {pages} {201601} (\bibinfo {year} {2013}{\natexlab{b}})}\BibitemShut
  {NoStop}%
\end{thebibliography}%

\end{document}